\documentclass[10pt]{article}
\usepackage{fullpage,citesort,epsfig,graphics,amsbsy}
\usepackage{psfrag}
\usepackage[normal,small]{caption}
\newcommand{\beq}{\begin{equation}}
\newcommand{\eeq}{\end{equation}}
\newcommand{\bea}{\begin{eqnarray}}
\newcommand{\eea}{\end{eqnarray}}

\newcommand{\be}{\begin{equation}}
\newcommand{\ee}{\end{equation}}
\newcommand{\bq}{\begin{eqnarray}}
\newcommand{\eq}{\end{eqnarray}}

\newcommand{\ie}{{\it i.e.\ }}
\def\math{\mathsurround=0pt }
\def\leftrightarrowfill{$\math \mathord\leftarrow \mkern-6mu \cleaders\hbox{$\mkern-2mu \mathord- \mkern-2mu$}\hfill
 \mkern-6mu \mathord\rightarrow$}
\def\overleftrightarrow#1{\vbox{\ialign{##\crcr
     \leftrightarrowfill\crcr\noalign{\kern-1pt\nointerlineskip}
     $\hfil\displaystyle{#1}\hfil$\crcr}}}

\newcommand{\bfs}{\boldsymbol}
\let\l=\lambda

 \def\bd{\begin{document}} \def\ed{\end{document}}
\def\ds{\documentstyle} \let\fr=\frac \let\bl=\bigl \let\br=\bigr
\let\Br=\Bigr \let\Bl=\Bigl
\let\bm=\bibitem
\let\na=\nabla
\let\pa=\partial \let\ov=\overline
\def\ft#1#2{{\textstyle{{\scriptstyle #1}\over {\scriptstyle #2}}}}
\def\fft#1#2{{#1 \over #2}}
\def\vp{\varphi}
\def\sst#1{{\scriptscriptstyle #1}}
\def\oneone{\rlap 1\mkern4mu{\rm l}}
\def\td{\tilde}
\def\wtd{\widetilde}
\def\dalemb#1#2{{\vbox{\hrule height .#2pt
        \hbox{\vrule width.#2pt height#1pt \kern#1pt
                \vrule width.#2pt}
        \hrule height.#2pt}}}
\def\square{\mathord{\dalemb{6.8}{7}\hbox{\hskip1pt}}}
\def\wtd{\widetilde}
\def\R{\rlap{\rm I}\mkern3mu{\rm R}}
\def\im{{\rm i}}
\def\tilg{\tilde{g}}
\def\tilF{\tilde{F}}
\def\tilA{\tilde{A}}
\def\varf{\varphi}
\def\tilf{\tilde{\phi}}
\def\tilh{\tilde{h}}
\def\rme{{\rm e}}
\def\ep{\epsilon}
\def\0{{(0)}}
\def\9{{(9)}}
\def\8{{(8)}}
\def\7{{(7)}}
\def\6{{(6)}}
\def\5{{(5)}}
\def\4{{(4)}}
\def\3{{(3)}}
\def\2{{(2)}}
\def\1{{(1)}}
\newcommand{\trace}{{\rm Tr}}
\newcommand{\ub}{\overline{U}}
\newcommand{\vb}{\overline{V}}
\newcommand{\uh}{\widehat{U}}
\newcommand{\vh}{\widehat{V}}
\newcommand{\ubh}{\overline{\widehat{U}}}
\newcommand{\vbh}{\overline{\widehat{V}}}
\newcommand{\lb}{\bar{\l}}
\newcommand{\Fb}{\overline{F}}
\newcommand{\Fh}{\widehat{F}}
\newcommand{\Fbh}{\overline{\widehat{F}}}
\newcommand{\Ab}{\overline{A}}
\newcommand{\Ah}{\widehat{A}}
\newcommand{\Abh}{\overline{\widehat{A}}}
\newcommand{\Gb}{\overline{G}}
\newcommand{\Gh}{\widehat{G}}
\newcommand{\Gbh}{\overline{\widehat{G}}}
\newcommand{\Pb}{\overline{P}}
\newcommand{\Ph}{\widehat{P}}
\newcommand{\Pbh}{\overline{\widehat{P}}}
\newcommand{\Qb}{\overline{Q}}
\newcommand{\Qh}{\widehat{Q}}
\newcommand{\Qbh}{\overline{\widehat{Q}}}
\newcommand{\Bb}{\overline{B}}
\newcommand{\Bh}{\widehat{B}}
\newcommand{\Bbh}{\overline{\widehat{B}}}
\newcommand{\fhns}{\hat{F}^{\rm (NS)}}
\newcommand{\fhrr}{\hat{F}^{\rm (RR)}}
\newcommand{\ahns}{\hat{A}^{\rm (NS)}}
\newcommand{\ahrr}{\hat{A}^{\rm (RR)}}
\newcommand{\hhrr}{\hat{H}^{\rm (RR)}}
\newcommand{\hchi}{\hat{\chi}}
\newcommand{\hphi}{\hat{\phi}}
\newcommand{\htau}{\hat{\tau}}
\newcommand{\cG}{{\cal G}}
\newcommand{\cGb}{\overline{{\cal G}}}
\newcommand{\cH}{{\cal H}}
\newcommand{\cP}{{\cal P}}
\newcommand{\cPb}{\overline{{\cal P}}}
\newcommand{\cQ}{{\cal Q}}
\newcommand{\cQb}{\overline{{\cal Q}}}
\newcommand{\cM}{{\cal M}}
\newcommand{\cN}{{\cal N}}
\newcommand{\cO}{{\cal O}}
\newcommand{\cD}{{\cal D}}
\newcommand{\cL}{{\cal L}}
\newcommand{\cA}{{\cal A}}
\newcommand{\cB}{{\cal B}}
\newcommand{\hg}{\hat{g}}
\newcommand{\cE}{{\cal E}}
\newcommand{\vpp}{\mbox{$\langle{\scriptstyle++}\rangle$}}
\newcommand{\vmp}{\mbox{$\langle{\scriptstyle-+}\rangle$}}
\newcommand{\vppp}{\mbox{$\langle{\scriptstyle+++}\rangle$}}
\newcommand{\vmpp}{\mbox{$\langle{\scriptstyle-++}\rangle$}}
\newcommand{\vpmp}{\mbox{$\langle{\scriptstyle+-+}\rangle$}}

\begin{document}
\setlength{\captionmargin}{20pt}
\begin{titlepage}
\begin{flushright}
UFIFT-HEP-05-16\\
hep-th/0507213
\end{flushright}

\vskip 3cm

\begin{center}
\begin{Large}
{\bf Notes on One-loop Calculations in Light-cone Gauge\footnote{Supported 
in part by the Department
of Energy under Grant No. DE-FG02-97ER-41029. 
}}
\end{Large}

\vskip 2cm
{\large 
 Charles B. Thorn\footnote{E-mail  address: {\tt thorn@phys.ufl.edu}}
}
\vskip0.20cm
{\it Institute for Fundamental Theory\\
Department of Physics, University of Florida,
Gainesville FL 32611}

%(\today)

\vskip 1.0cm
\end{center}

\begin{abstract}\noindent
Loop calculations in light-cone gauge must confront many technical
complexities. We offer here a compendium of detailed 
light-cone calculations in Yang-Mills theories (with no matter fields).
We consistently regulate the  $p^+=0$ singularities through
discretization of the $p^+$ component of momentum. Although it is
more cumbersome than the Mandelstam-Leibbrandt prescription, this
choice has the virtue of employing only positive norm states,
retaining manifest unitarity. 
Some of the results
given here are useful for the forthcoming paper \cite{chakrabartiqt},
specifically the results for the gluon
self energy and one-loop vertex corrections. 
\end{abstract}
\vfill
\end{titlepage}
\section{Introduction}

In these notes we carry out several one loop calculations
using lightcone gauge and employing 
a novel regularization of Feynman diagrams motivated
by the light-cone worldsheet picture of planar diagrams
\cite{bardakcit,thornsheet}.
For its rigorous 
definition the worldsheet formalism relies on a discretization
of $\sigma,\tau$, and hence of $T_i,p_i^+$ the light cone 
times and longitudinal momenta associated with the
various propagators of the diagram. On the other hand,
conventional Feynman diagrams require continuous $T_i,p_i^+$.
The ultraviolet divergences of quantum field theory
correspond in lightcone variables to infinities due to
integration at large transverse momentum. These transverse
momentum infinities will
get entangled with, and will spoil, the continuum limit of
the $T_i,p_i^+$ unless they are regulated independently of these 
longitudinal variables \cite{glazek,thorngote,thornscalar}.
The requirement that this transverse regulator be local on the
worldsheet then dictates that it be applied only to the
boundary values ${\bfs q}_i$ of the worldsheet fields ${\bfs q}(\sigma,\tau)$. 
Such a cutoff is local
in both $\sigma,\tau$ because it only need be applied
at the beginning or end of an internal boundary (because ${\bfs q}$
satisfies Dirichlet boundary conditions), i.e.  at a point on the worldsheet. 
It is particularly convenient for
our analysis to simply impose a Gaussian cutoff, i.e. to
insert in the integrand the factor $e^{-\delta\sum{\bfs q}^2_i}$
\cite{bardakcitimp,bardakci,thorngote}.
This factor can be directly interpreted as a local modification of the
worldsheet action.

With $\delta>0$ and fixed, the rigorously defined world sheet path integral
for each multi-loop planar
diagram can be explicitly evaluated on the worldsheet lattice \cite{gilest} 
and then the continuum limit of the $T_i$ for the various
propagators can be safely taken. In gauge theories in lightcone gauge it
is necessary to keep $p^+$ discrete until the end, taking the
continuum limit only for physical quantities. This is because
 $p^+=0$ divergences will remain in unphysical intermediate
quantities\footnote{Another approach is the Mandelstam-Leibbrandt
principal value prescription, which retains continuous $p^+$
but gives up manifest unitarity. We prefer retaining unitarity.}.
The result, essentially 
by construction, reduces to
one of the standard representations of the Feynman diagram
in momentum space with the regulator factor $e^{-\delta\sum{\bfs q}^2_i}$
inserted in the integrand. Because the $\delta$
regulator is in place and $p^+$ is discrete, 
these integrals are manifestly finite.
In Section 2 we describe how this reduction takes place. 

After this reduction, there remains an almost conventional
analysis of the renormalization procedure in the context of this
somewhat novel regulator. The novelty stems from the
fact that the $q_i$'s, the variables subject to the cut-off, 
are not the momenta 
flowing through the propagators. Rather, they are ``dual-momentum'' 
variables, 
one assigned to each loop. There is also a set of external
dual-momenta $q^e_i$, one assigned to each region between
external lines.
The momentum flowing through the propagator
that separates loop $i$ from loop $j$ is the difference $q_i-q_j$.
Thus the regulator breaks a ``translation'' symmetry
$q^e_i\to q^e_i+a$ enjoyed by the {\it bare} unregulated 
diagram\footnote{Because the regulator only 
cuts off the transverse components
of $q$, the translation symmetry in the longitudinal momenta remains 
unbroken.}. Because of this broken symmetry, with $\delta>0$
the $n$-point function depends on $n$ dual-momenta rather than on $n-1$ 
actual momenta. Formally the limit $\delta\to0$ should restore
the symmetry and the amplitudes should become independent of
one of the dual-momenta. Because of ultraviolet
divergences, the introduction of counter-terms is necessary to ensure 
that this happens.

In Section 3 we describe the properties of this regularization in
detail. In section 4 we discuss the self-energy and
its renormalization through one loop by direct calculation.
The one loop three point vertex is calculated in Section 5,
and the correct asymptotically free behavior is confirmed.
Finally, in section 6 we include some analysis of the
box diagrams with maximal and next to maximal helicity
violation.

\section{From Lightcone to the Schwinger Representation}
\label{sec2}

By construction, the evaluation of the
worldsheet path integral representing a specific
planar Feynman diagram produces a certain discretized version of
the usual multi-loop integral. Each propagator appears
in its mixed ${\bfs p},p^+>0,x^+$ representation
\bea
\int {dp^-\over2\pi}e^{-ix^+p^-}{-i\over p^2+\mu_0^2-i\epsilon}
&=&{\theta(x^+)\over 2p^+}e^{-ix^+({\bfs p}^2+\mu_0^2)/2p^+}\nonumber\\
&\to&{\theta(\tau)\over 2p^+}e^{-\tau({\bfs p}^2+\mu_0^2)/2p^+}.
\eea
The Feynman integration is 
over all independent $\tau_i,p^+_i,{\bfs p}_i$. However the
worldsheet lattice formalism produces instead sums over
discretized $\tau_i=k_ia,p^+_i=l_im$, while keeping the ${\bfs p}_i$
integrals continuous. However, in the presence of the regulator $\delta>0$
described in the introduction, one can safely replace all
of the discretized sums by continuous integrals.

We would like to now show that, for cubic scalar vertices,
 these perhaps unfamiliar lightcone
multi-loop integrals are identical to the covariant
Feynman integrals in which each propagator is written in a
Schwinger representation. 
\bea
{1\over p^2+\mu^2}=\int_0^\infty dT e^{-T(p^2+\mu^2)}.
\label{schwingerrep}
\eea
Indeed, it is straightforward to show that the number of
independent $\tau_i,p^+_i$ in the diagram's
lightcone representation is
precisely equal to the number of $T_i$ in the diagram's 
Schwinger representation. If one explicitly carries out the
Gaussian integrals in the two representations by completing the
square the remaining integrals in the two representations
will be of the same dimensionality. The integrands are very
similar except that the determinant prefactor from the lightcone
is raised to the $(D-2)/2$ power compared to the $D/2$ power
in the Schwinger representation. One can make
the exponentials in the
integrands identical by changing integration variables
from the $\tau_i,p^+_i$ to appropriate $T_i$. It then turns out
that the Jacobian for this change of variables supplies the
missing determinant factors.

To understand why this happens, just consider the transform
to light-cone representation of the Schwinger representation:
\bea
-i\int {dp^-\over2\pi}e^{-ix^+p^-}\int idT e^{-iT({\bf p}^2
+\mu_0^2-2p^+p^--i\epsilon)}
&=&-i\int idT\delta(x^+-2p^+T) e^{-iT({\bf p}^2+\mu_0^2)}\nonumber\\
&\to&\int dT\delta(\tau-2p^+T) e^{-T({\bf p}^2+\mu_0^2)}.
\eea
From this result we see that the appropriate change of
variables is $T=\tau/2p^+$. It is interesting and satisfying
that the passage to imaginary $x^+$ in the lightcone
representation is completely equivalent to writing the
Schwinger representation with a real exponential, which
of course is only meaningful after the Wick rotation to
Euclidean space.

For the rest of the discussion of renormalization we need 
no longer refer to the explicit worldsheet representation.
We only have to write the usual covariant rules using
dual momenta $q_i$, and insert the regulator factor
$e^{-\delta\sum_i{\bfs q}_i^2}$. Once we have established
the form of the counter-terms required for renormalization
we shall return to give their worldsheet representation
at the end of the article.

\section{Regularization}
Draw an arbitrary planar diagram so that its lines divide the
plane into different regions, the external lines
all going off to infinity. Then the external lines
bound infinite regions, and the finite regions fill each
loop of the multi-loop diagram.  For each loop introduce
a momentum $q^\mu_i$, assigned to the loop's region. Then each
propagator line separates two regions, say $i_1$ and $i_2$,
and the propagator's momentum is then taken to be
$q_{i_1}-q_{i_2}$, and momentum is automatically conserved.
We regulate each diagram by including in the integrand the
factor $e^{-\delta\sum_{i=1}^L{\boldsymbol q}_i^2}$. Since we are
using a light-cone world sheet we only cut off the transverse
momentum integrals, because we want to preserve
longitudinal boost invariance\footnote{One 
could easily extent the cutoff to the
longitudinal variables, but then the light-cone interpretation
would be obscured.}. This regularization sacrifices full Lorentz
invariance, but respects the $O(D-2)$ rotational invariance in transverse
space as well as the longitudinal boost invariance.
The transverse boost invariances generated by $M^{\pm i}$ are broken, 
and it will require counter-terms to restore 
them in the renormalizable case.interpretation. 

Without loss of generality, we can and do restrict attention to
proper (\ie connected one particle irreducible) diagrams,
with propagators removed from external legs. Such diagrams
never have tadpole sub-diagrams, which would be problematic
for the lightcone description (because $p^+>0$), though not for a covariant
description. The only 1PIR diagram involving a tadpole is
the one-point function itself, $\langle\Phi\rangle$. 
It is true that the lightcone description
has no convenient representation of the one point function.
However, in a covariant description, the only effect of 
tadpoles as sub-diagrams in larger (improper) diagrams is pure
mass renormalization, which means their effect can be absorbed
in an additive constant in the self-energy counter-term.
In this article we assume that this is always done and therefore
drop tadpoles completely. Then we can freely
pass back and forth between light-cone and covariant descriptions, 
as long as we refrain from considering
the one-point function itself. Since the one-point function cannot
be directly measured in any case, this is no limitation
on the lightcone description. If needed, the value of the
one-point function can be related via the field equations 
to $\langle\Phi^2\rangle$,
which in turn can be extracted from the high momentum
limit of the two point function.

It is convenient to employ the Schwinger representation
of each propagator (\ref{schwingerrep}):
\bea
{1\over p^2+\mu^2}=\int_0^\infty dT e^{-T(p^2+\mu^2)}\nonumber
\eea
which enables the execution of all loop momentum integrals
by completing the square in the exponents of the Gaussian integrals.
To describe this for an $L$ loop diagram, assemble the
loop momenta in an $L$ dimensional vector $q$ and call $M_0$ the 
$L\times L$ symmetric matrix that describes the quadratic
terms in the $q_i$, so the exponent reads
\bea
-q^T\cdot(M_0+\delta)q+v^T\cdot q +q^T\cdot v -B
\label{diagrambilinear}
\eea
where the $L$-vector $v$ describes the couplings to the momenta
assigned to the external regions and $B$ is a bilinear form in those
external momenta. It is understood that $\delta\neq0$ only for the
transverse components. Then the result of the loop integrations is just
\bea
&&{\pi^L\over\det M_0}\left({\pi^L\over\det(\delta+ M_0)}\right)^{(D-2)/2}
\exp\left\{-B+v^T\cdot {1\over\delta+M_0}v\right\}=\nonumber\\
&&{\pi^L\over\det M_0}\left({\pi^L\over\det(\delta+ M_0)}\right)^{(D-2)/2}
\exp\left\{-B+v^T\cdot {1\over M_0}v
-\delta {\boldsymbol v}^T{1\over M_0}\cdot{1\over\delta+M_0}
{\boldsymbol v}\right\}
\label{diagramint}
\eea
We see that the shift of $M_0$ by $\delta$ regulates the integration
region near the zeroes of the determinant, which is the source
of ultraviolet divergences in the diagram. The first two terms
in the exponent are manifestly Lorentz invariant
and are precisely what they would be in the 
unregulated theory. The last term in the exponent breaks Lorentz invariance
because it depends explicitly on the transverse momentum
components. If we could argue that it were negligible (as $\delta\to0$), 
we could assert from the
known proofs of renormalizability that all divergences 
as $\delta\to0$ could
be covariantly absorbed in the renormalization of mass $\mu$ and coupling $g$
to all orders in perturbation theory.

The term in question 
is nominally of order $O(\delta)$ but since it also
depends on the $T_i$'s we must check this estimate more carefully.
First note that $q_0\equiv (\delta+M_0)^{-1}{\boldsymbol v}$ is in fact
the location of the minimum of a bilinear form in the $q_i$'s
that has the interpretation as the potential energy of 
$L$ particles tied to each other and to the fixed
external momenta with a bunch of springs
with spring constants $T_i>0$ and to the origin with springs of
spring constant $\delta$. It is obvious that the resulting
equilibrium has every $q_{0 i}$ within the simplex with vertices
at the origin and the external momenta. If $\delta=0$ they are
within the simplex with vertices at the external momenta. In 
either case it follows that $|{\boldsymbol q}_{0i}|$ is
uniformly bounded by the largest external momentum. Thus we can conclude
that the term in question is uniformly bounded over the
whole integration region by $L\delta |{\boldsymbol q_{ext}}|_{max}^2$. 
Thus the $O(\delta)$ estimate is rigorous.

Even so, Lorentz non-covariance can survive due to ultraviolet divergences
of degree $1/\delta$ or worse which can overwhelm
the $O(\delta)$ suppression. Fortunately, in a renormalizable theory
we can isolate where these divergences can occur and accordingly
identify the subtractions necessary to remove these contributions
which would violate Lorentz invariance. Indeed the ultraviolet
divergences in vertex parts are superficially linear in
momentum ( $1/\sqrt{\delta}$) while those in 
self-energy parts are quadratic in momentum ($O(1/\delta)$). 
Thus the Lorentz violations due to the term in the
exponent will be associated with
self-energy divergences, but of course we must follow their impact
in sub-diagrams of larger diagrams as well. That term will be negligible
in three and higher point diagrams, but there are also some $\delta$
artifacts due to the linear momentum factors in the cubic vertex,
which survive because of the latter's superficial linear divergence.
Thus we should expect the associated counter-terms to involve
at most three factors of the gauge field.

\section{Gluon Self Energy}
In order to acquaint the reader with some of the novelties
of calculations using the $\delta$ regulator, we  carry out
in this section a direct calculation or the self energy through one loop,
with an explicit separation of all divergences and 
Lorentz-violating artifacts. 
We call the bare gluon self-energy $\Pi^{ij}_0$, but it is convenient
to calculate $Z\Pi^{ij}_0$ and absorb the factor of $Z$ in the bare coupling 
by defining the renormalized coupling $g=g_0Z^{3/2}/Z_1$, where
$Z_1$ is the three vertex renormalization constant. In other words we
write down the Feynman rules in terms of renormalized mass and coupling,
canceling infinities against 
the self-energy counter-term $Z\Pi^{ij}_{\rm C.T.}$  and the three
vertex counter-term $g(Z_1-1)A^3$, which are included in
the Feynman rules, rather than absorbing them in redefinitions
of the bare parameters.

Choose the complex
basis for the gluon polarization $1,2$: $\wedge=(1+i2)/\sqrt2$,
$\vee=(1-i2)/\sqrt2$. 
The unsubtracted one-loop self-energy diagrams have the values
\bea
Z\Pi^{\wedge\vee}_0&=&Z\Pi^{\vee\wedge}_0
={g^2N_c\over4\pi^2}\int_0^\infty{dT_1dT_2
\over T_1+T_2}\left[{1\over(T_1+T_2+\delta)^2}
+{\delta^2[T_1{\bfs q}+T_2{\bfs q}^\prime]^2\over
(T_1+T_2)^2(T_1+T_2+\delta)^3}\right]\nonumber\\
&&\qquad\left(1+{(T_1+T_2)^2\over T_1^2}
+{(T_1+T_2)^2\over T_2^2}\right)
\exp\left\{-{T_1T_2\over T_1+T_2}(q-q^{\prime})^2
-\delta{(T_1{\boldsymbol q}+T_2{\boldsymbol q}^\prime)^2
\over(T_1+T_2)(T_1+T_2+\delta)}\right\}\nonumber\\
&=&
{g^2N_c\over4\pi^2}\int_0^\infty dT\int_0^1dx \left[{1\over (T+\delta)^2}
+{\delta^2[x{\bfs q}+(1-x){\bfs q}^\prime]^2\over
(T+\delta)^3}\right]\nonumber\\
&& \left(1+{1\over x^2}
+{1\over (1-x)^2}\right)
\exp\left\{-Tx(1-x)(q-q^{\prime})^2
-{\delta T\over T+\delta}(x{\boldsymbol q}+(1-x){\boldsymbol q}^\prime)^2
\right\}\\
Z\Pi^{\vee\vee}_0&=&{g^2N_c\over2\pi^2}\int_0^\infty dT\int_0^1dx 
{\delta^2[x{q^\vee}+(1-x){q}^{\prime\vee}]^2\over
(T+\delta)^3}\nonumber\\&& 
\exp\left\{-Tx(1-x)(q-q^{\prime})^2
-{\delta T\over T+\delta}(x{\boldsymbol q}+(1-x){\boldsymbol q}^\prime)^2
\right\}\to {g^2N_c\over4\pi^2}\int_0^1dx 
{[x{q^\vee}+(1-x){q}^{\prime\vee}]^2}
\eea
And $\Pi^{\wedge\wedge}_0$ is obtained from $\Pi^{\vee\vee}_0$
by the substitution $\vee\to\wedge$. Note that these two quantities
are simply quadratic polynomials in ${\bfs q},{\bfs q}^\prime$,
so a counter-term can be introduce to cancel them completely.
The quadratic divergence in $\Pi^{\wedge\vee}_0$ can be 
simply extracted with an integration
by parts. We observe that
\bea
\left[{1\over (T+\delta)^2}
+{\delta^2[x{\bfs q}+(1-x){\bfs q}^\prime]^2\over
(T+\delta)^3}\right]\exp\left\{-{\delta T\over T+\delta}
(x{\boldsymbol q}+(1-x){\boldsymbol q}^\prime)^2
\right\}&=&\nonumber\\
&&\hskip-1.5in -{\partial\over\partial T}{1\over T+\delta}
\exp\left\{-{\delta T\over T+\delta}
(x{\boldsymbol q}+(1-x){\boldsymbol q}^\prime)^2
\right\}
\eea
So we can rewrite the self energy as
\bea
Z\Pi^{\wedge\vee}_0&=&-{g^2N_c\over4\pi^2}(q-q^{\prime})^2
\int_0^1 dx\left(x(1-x)+{1-x\over x}
+{x\over 1-x}\right) I( H\delta) \nonumber\\&&
+{g^2\over4\pi^2}{1\over\delta}
\int_0^1 dx\left(1+{1\over x^2}
+{1\over (1-x)^2}\right)\\
H&\equiv&x(1-x)(q-q^{\prime})^2\\
I(H\delta)&\equiv&\int_0^\infty du {e^{-uH\delta
-u\delta(x{\boldsymbol q}+(1-x){\boldsymbol q}^\prime)^2/(1+u)}\over 1+u}
\quad{}_{\widetilde{\delta\to0}} 
\quad -\gamma-\ln(H\delta)=-\ln(H\delta e^\gamma)
\eea
where $\gamma=-\Gamma^\prime(1)/\Gamma(1)$ is Euler's constant.
Clearly the $x$ integrals diverge at the end points of integration.
These divergences are spurious artifacts of the light-cone gauge
and have nothing to do with the usual ultraviolet divergences of the
gauge theory. They must cancel in physical quantities without invoking
renormalization or counter-terms. In our approach the $x$ integration
just corresponds to integration over the location on the
worldsheet of the boundary representing the loop. On the
world sheet lattice this location is an integer $l$ with 
$x=l/M$ and $M$ is the discretized total plus momentum
entering the self energy: $q^+=mM$. The discreteness of
$p^+$ regulates the endpoint x divergences. 

Let us discuss first the fate of the quadratic $1/\delta$ divergence,
which for discrete $p^+$ reads:
\bea
{g^2\over4\pi^2}{1\over M\delta }\sum_{l=1}^{M-1}
\left(1+{M^2\over l^2}+{M^2\over(M-l)^2}\right)
\sim {g^2\over4\pi^2}{1\over\delta}
\left({\pi^2\over3}M-1+O\left({1\over M}\right)\right)
\eea
where the right side indicates the large $M$ behavior of the
sums. The term linear in $M=p^+/m$ cannot be canceled by a
gluon self mass, because it is linear in $p^+$. However, 
precisely because it is linear in $p^+$, it represents a
constant $-g^2M/(24p^+\delta)=-g^2/(24m\delta)$ 
added to the energy $p^-$ of each gluon. A constant added to $p^-$
can be interpreted as
energy associated with the boundary of the worldsheet representing
that gluon, in
other words with a boundary cosmological constant. If we
start with a nonzero boundary cosmological constant $\lambda_b$ in
zeroth order, we can tune its value to cancel the
linear terms in $p^+$ generated by loop effects. Its lowest order value is 
then
\bea
\lambda_b=+{g^2\over24m\delta}
\eea
After this cancellation,
there is left behind a constant which can be canceled by a gluon mass
counter-term $\delta\mu^2$. So to this order
\bea
\delta\mu^2={g^2\over4\pi^2\delta}
\eea 
Of course, the gluon mass is zero in tree approximation,
but since loop corrections generate a gluon mass, the tree value
must be non-zero and adjusted to cancel the loop contributions
order by order in perturbation theory. A nonzero mass at tree level 
violates gauge invariance, which means a violation of Lorentz invariance
in the completely fixed lightcone gauge. So an alternative prescription
is: in lightcone gauge, allow a nonzero gluon mass $\mu_0^2$ 
as an input parameter,
and calculate physical quantities as functions of this parameter.
Finally, choose a value of this parameter that restores Lorentz invariance.
Note that to one loop, $\mu^2=0$ requires a tachyonic gluon mass:
$\mu_0^2=-\delta\mu^2$.

Next we turn to the logarithmic divergences in the self energy.
For dimensional reasons write $H\delta=(H/\mu^2)(\mu^2\delta)$,
so as $\delta\to0$ $I(H\delta)\to \Gamma^\prime(1)-\ln(H/\mu^2)
-\ln(\mu^2\delta)$. Also call $p=q^\prime-q$ and remember that
$x=q^+/p^+$. Including the above mentioned counter-terms, we
then find
\bea
Z\Pi^{\wedge\vee}-2p^+\lambda_b+\delta\mu^2&=& 
{g^2N_c\over4\pi^2}p^2
\sum_{q^+}\left(x(1-x)-2+\left[{1\over q^+}
+{1\over p^+-q^+}\right]\right)\ln\left\{x(1-x)p^2\delta e^\gamma\right\}\\
&\sim& 
{g^2N_c\over4\pi^2}p^2
\left(-{11\over6}+\sum_{q^+}\left[{1\over q^+}
+{1\over p^+-q^+}\right]\right)\ln(\mu^2\delta)+ {\rm Finite}
\eea
where ${\rm Finite}$ means with respect to $\delta\to0$. We may
associate this log divergent contribution with a $p^+$
dependent wave function renormalization factor
\bea
Z_{p^+}=1+{g^2N_c\over4\pi^2}\left(-{11\over6}+\sum_{q^+}\left[{1\over q^+}
+{1\over p^+-q^+}\right]\right)\ln(\mu^2\delta)
\eea
Note that this $Z_{p^+}<1$ by virtue of the (divergent) $q^+$ sums,
in accordance with the requirements of unitarity.  Of course, we
really want renormalization constants to be independent of
the momenta, so we define instead the wave function renormalization
\bea
Z_3&=&1+{g^2N_c\over4\pi^2}\int_0^1dx\left(x(1-x)-2\right)
\ln\left\{x(1-x)\mu^2\delta e^\gamma\right\}\nonumber\\
&=&1+{g^2N_c\over4\pi^2}\left\{-{11\over6}
\ln\left\{\mu^2\delta e^\gamma\right\}+{67\over18}\right\}
\label{zee3}
\eea
which is {\it larger} than 1. We then leave the $p^+$ dependent part of the
log divergences in the definition of the ``renormalized''
$\Pi$ and must find that these are precisely canceled
by other contributions. Indeed we will find these canceling contributions
in the two vertex renormalizations each (internal) propagator attaches to.
Naturally, the cancellation is incomplete on the external lines.
Removing a factor of $Z_3$ from the gluon propagator corresponds to using the
``renormalized'' self energy
\bea
Z\Pi^{\wedge\vee}_R&\equiv& 
{g^2N_c\over4\pi^2}p^2\left\{
\sum_{q^+}\left[{1\over q^+}
+{1\over p^+-q^+}\right]\ln\left\{x(1-x)p^2\delta e^\gamma\right\}
-{11\over6}\ln{p^2\over\mu^2}\right\}
\eea
where we must find that the $\delta$ dependence cancels in physical quantities.
To simplify future equations, we give the anomalous quantity in braces
a name:
\bea
{\cal A}(p^2,p^+)\equiv\sum_{q^+}\left[{1\over q^+}
+{1\over p^+-q^+}\right]\ln\left\{x(1-x)p^2\delta e^\gamma\right\}.
\eea
We shall find this quantity occurring in vertex calculations. Then
\bea
Z\Pi^{\wedge\vee}_R&\equiv& 
{g^2N_c\over4\pi^2}p^2\left\{{\cal A}(p^2,p^+)
-{11\over6}\ln{p^2\over\mu^2}\right\}
\eea

\section{Three Point Function and one-loop coupling renormalization} 
\subsection{Triangle graph}
The tree vertex is
\bea
2g{p_3^+\over p_1^+p_2^+}{\bfs K}_{12}\equiv 
2g{p_3^+\over p_1^+p_2^+}(p_1^+{\bfs p}_2-p_2^+{\bfs p}_1)
\eea
In this section we turn to the 1PIR loop corrections
to the cubic vertex. 
First consider the maximal helicity violating
triangle, which must be finite because there is no
tree contribution to this amplitude:
\bea
\Gamma^{\wedge\wedge\wedge}&=&
-{g^3\over \pi^2}
\int_0^\infty {dT_1dT_2dT_3\over 
T_{13}(T_{13}+\delta)}
\exp\left\{-{\delta({\boldsymbol k}_0T_3+{\boldsymbol k}_1T_1
+{\boldsymbol k}_2 T_2)^2\over
T_{13}(T_{13}+\delta)}\right\}\nonumber\\
&&\exp\left\{
-{T_1T_3(k_1-k_0)^2
+T_1T_2(k_1-k_2)^2
+T_2T_3(k_2-k_0)^2\over T_1+T_2+T_3}
\right\}\nonumber\\
&&\prod_{j=1}^3\left\{{T_jK^\wedge\over p^+_{j-1}T_{13}}
-{\delta(T_1k_1^\wedge+T_2k_2^{\wedge}+T_3k_0^{\wedge})
\over T_{13}(\delta+T_{13})} \right\}\\
{\bfs K}&=& {\bfs K}_{12}=p^+_3{\bfs k}_1+p^+_1{\bfs k}_2
+p^+_2{\bfs k}_0
=p^+_1({\bfs k}_2-{\bfs k}_1)-p^+_2({\bfs k}_1-{\bfs k}_0) .
\eea
We see by inspection that the $\delta\to0$ limit of this
amplitude is perfectly finite.
\bea
\Gamma^{\wedge\wedge\wedge}&\sim&
-{g^3\over \pi^2}
\int_0^\infty {dT_1dT_2dT_3\over 
p^+_1p^+_2p^+_3T_{13}^5}T_1T_2T_3(K^\wedge)^3\exp\left\{
-{T_1T_3(k_1-k_0)^2
+T_1T_2(k_1-k_2)^2
+T_2T_3(k_2-k_0)^2\over T_1+T_2+T_3}
\right\}\nonumber\\
&\sim&
-{g^3\over \pi^2}{(K^\wedge)^3\over 
p^+_1p^+_2p^+_3}\int_{x+y<1}dxdy{xy(1-x-y)\over
x(1-x-y)(k_1-k_0)^2
+xy(k_1-k_2)^2
+y(1-x-y)(k_2-k_0)^2 }
\eea
The other maximal helicity violating amplitude $\Gamma^{\vee\vee\vee}$
is obtained from this result by the  substitution $K^\wedge\to K^\vee$.
For the contribution of the triangle to on-shell scattering at one
loop, two of the $p_i^2=0$, and the amplitude simplifies to
\bea
\Gamma^{\wedge\wedge\wedge}&\sim&
-{g^3\over 6\pi^2}{(K^\wedge)^3\over 
p^+_1p^+_2p^+_3 p^2}
\eea
where $p$ is the momentum of the off-shell leg.

The amplitudes, $\Gamma^{\wedge\wedge\vee}$, $\Gamma^{\vee\vee\wedge}$,
and those obtained by cyclic permutation,
are more challenging because they contain both infra-red and
ultraviolet divergences. We shall work out the first case, together
with its cyclic permutations, in complete detail.. 
The amplitude for $\Gamma^{\wedge\wedge\vee}$, 
after the shift of loop momentum ${\bfs q}$
which completes the square in the exponent, is
\bea
\Gamma^{\wedge\wedge\vee}&=&-4g^3\sum_{q^+}\int dT_1dT_2dT_3
\delta(T_{13}q^+-T_{12}p^+_1-T_2p^+_2)\int {d^2 q\over(2\pi)^3}
\exp\left\{
-{\delta({\boldsymbol k}_0T_3+{\boldsymbol k}_1T_1
+{\boldsymbol k}_2T_2)^2\over
T_{13}(T_{13}+\delta)}\right\}\nonumber\\
&&\exp\left\{-(T_{13}+\delta){\bfs q}^2
-{T_1T_3(k_1-k_0)^2
+T_1T_2(k_1-k_2)^2
+T_2T_3(k_2-k_0)^2\over T_1+T_2+T_3}
\right\}\nonumber\\
&&\bigg\{A^{\wedge\wedge\vee}_3(q+K_1)^\wedge(q+K_2)^\wedge(q+K_3)^\vee 
+A^{\wedge\wedge\vee}_2(q+K_1)^\wedge(q+K_2)^\vee(q+K_3)^\wedge \nonumber\\&&\qquad\qquad
+A^{\wedge\wedge\vee}_1(q+K_1)^\vee(q+K_2)^\wedge
(q+K_3)^\wedge\bigg\}
\label{baretriangle}
\eea
where for brevity we have defined\footnote{The amplitudes for the other
spin configurations,
$\Gamma^{\vee\wedge\wedge}$ and $\Gamma^{\wedge\vee\wedge}$, are obtained
by modifying the $A^{\wedge\wedge\vee}_i$ appropriately:
\bea
A^{\wedge\wedge\vee}_1&\to&A^{\vee\wedge\wedge}_1={q^{+2}\over(p^+_1-q^+)^2}
+{(p^+_1-q^+)^2\over q^{+2}}\qquad\qquad A_2\to A^{\vee\wedge\wedge}_2
={p_2^{+2}p_1^{+2}
\over(q^+-p_1^+)^2(p^+_1+p^+_2-q^+)^2}\nonumber\\
A^{\wedge\wedge\vee}_3&\to& A^{\vee\wedge\wedge}_3 = {p_3^{+2}p_1^{+2}
\over q^{+2}(p^+_1+p^+_2-q^+)^2}\\
A^{\wedge\wedge\vee}_1&\to&
A^{\wedge\vee\wedge}_1={p_1^{+2}p_2^{+2}\over q^{+2}(p^+_1-q^+)^2}
\qquad\qquad A^{\wedge\wedge\vee}_2\to A^{\wedge\vee\wedge}_2={
(q^+-p_1^+)^2\over(p^+_1+p^+_2-q^+)^2}
+{(p^+_1+p^+_2-q^+)^2\over(q^+-p_1^+)^2}\nonumber\\
A^{\wedge\wedge\vee}_3&\to& A^{\wedge\vee\wedge}_3 = {p_3^{+2}p_2^{+2}
\over q^{+2}(p^+_1+p^+_2-q^+)^2}
\eea}
\bea 
A^{\wedge\wedge\vee}_1&=&{p_1^{+2}p_3^{+2}\over(q^+-p_1^+)^2q^{+2}},
\qquad\qquad A^{\wedge\wedge\vee}_2={p_2^{+2}p_3^{+2}
\over(q^+-p_1^+)^2(p^+_1+p^+_2-q^+)^2}\nonumber\\
A^{\wedge\wedge\vee}_3&=&{q^{+2}\over(p^+_1+p^+_2-q^+)^2}
+{(p^+_1+p^+_2-q^+)^2\over q^{+2}}
\eea
By $p^+$ conservation at least one of the $p_i^+$ is positive and at least
one is negative. For definiteness we choose $p_1^+>0$ and $p_3^+<0$. Then
$p_2^+$ could have either sign. We shall work out in detail the
case $p_2^+>0$. We can then obtain the results for the case
$p_2^+<0$ by the following argument. Consider the expression for the amplitude
with $p_1\to p^\prime_1=-p_3$, $p_2\to p^\prime_2=-p_2$, 
$p_3\to p^\prime_3=-p_1$, ${\bfs k}_0\to{\bfs k}^\prime_0={\bfs k}_0$,
${\bfs k}_1\to{\bfs k}^\prime_1={\bfs k}_2$, 
${\bfs k}_2\to{\bfs k}^\prime_2={\bfs k}_1$. For clarity we also
identify the new Schwinger parameters as $T^\prime_1=T_2$, $T^\prime_2=T_1$,
and $T^\prime_3=T_3$. Then, by inspection we find that
\bea
\Gamma^{\wedge\wedge\vee}(p_i,{\bfs k}_i)&=&
\Gamma^{\vee\wedge\wedge}(p^\prime_i,{\bfs k}^\prime_i),\qquad
\Gamma^{\vee\wedge\wedge}(p_i,{\bfs k}_i)=
\Gamma^{\wedge\wedge\vee}(p^\prime_i,{\bfs k}^\prime_i),\qquad
\Gamma^{\wedge\vee\wedge}(p_i,{\bfs k}_i)=
\Gamma^{\wedge\vee\wedge}(p^\prime_i,{\bfs k}^\prime_i)
\label{signflips}
\eea
We see that $p_1^+>0, p_2^+<0, p_3^+<0$ implies 
$p^{\prime+}_1>0, p^{\prime+}_2>0, p^{\prime+}_3<0$, so we can read off the
answer for $p_2^+<0$ from the result for $p_2^{\prime+}>0$.

In view of the divergences at $q^+=0,p_1^+,p_1^++p_2^+$, we have kept the
integral over $q^+$ in discretized form, where setting $q^+=lm$ and
$p^+_i=M_im$, $\sum_{q^+}$ means $m\sum_{l=1,l\neq M_1}^{M_1+M_2-1}$.
Also the $K_i$ are given by
\bea
{\bfs K}_1&=&{T_2\over T_{13}p^+_1}{\bfs K}-{\delta(T_1{\bfs k}_1
+T_2{\bfs k}_2+T_3{\bfs k}_0)\over T_{13}(\delta+T_{13})}\\
{\bfs K}_2&=&{T_3\over T_{13}p^+_2}{\bfs K}-{\delta(T_1{\bfs k}_1
+T_2{\bfs k}_2+T_3{\bfs k}_0)\over T_{13}(\delta+T_{13})}\\
{\bfs K}_3&=&{T_1\over T_{13}p^+_3}{\bfs K}-{\delta(T_1{\bfs k}_1
+T_2{\bfs k}_2+T_3{\bfs k}_0)\over T_{13}(\delta+T_{13})}
\eea
We have for the ${\bfs q}$ integration
\bea
\int {d^2 q}\ e^{-(T_{13}+\delta){\bfs q}^2}
={\pi\over T_{13}+\delta},\qquad \int {d^2 q}\ q^\wedge q^\vee
e^{-(T_{13}+\delta){\bfs q}^2}
={\pi\over2(T_{13}+\delta)^2}\\
\int {d^2 q}\ q^\wedge e^{-(T_{13}+\delta){\bfs q}^2}=
\int {d^2 q}\ q^\vee e^{-(T_{13}+\delta){\bfs q}^2}
=\int {d^2 q}\ q^\wedge q^\wedge e^{-(T_{13}+\delta){\bfs q}^2}
=\int {d^2 q}\ q^\wedge q^\wedge q^\vee e^{-(T_{13}+\delta){\bfs q}^2}=0
\eea
From which we obtain
\bea
\Gamma^{\wedge\wedge\vee}&=&-{g^3\over4\pi^2}\sum_{q^+}\int dT_1dT_2dT_3
\delta(T_{13}q^+-T_{12}p^+_1-T_2p^+_2)\nonumber\\
&&\exp\left\{-{T_1T_3(k_1-k_0)^2+T_1T_2(k_1-k_2)^2
+T_2T_3(k_2-k_0)^2\over T_1+T_2+T_3}
-{\delta({\boldsymbol k}_0T_3+{\boldsymbol k}_1T_1
+{\boldsymbol k}_2 T_2)^2\over
T_{13}(T_{13}+\delta)}\right\}\nonumber\\
&&\bigg\{\left[{2K_1^\wedge K_2^\wedge K_3^\vee\over T_{13}+\delta}
+{K_1^\wedge+K_2^\wedge\over (T_{13}+\delta)^2}\right]A_3
+\left[{2K_1^\wedge K_2^\vee K_3^\wedge\over T_{13}+\delta}+
{K_1^\wedge + K_3^\wedge\over(T_{13}+\delta)^2}
\right]A_2\nonumber\\&&\qquad\qquad
+\left[{2K_1^\vee K_2^\wedge
K_3^\wedge\over T_{13}+\delta}+{K_2^\wedge
+K_3^\wedge\over (T_{13}+\delta)^2}\right]A_1 \bigg\}
\eea
Now consider the $\delta\to0$ limit. The last term in the exponent is
uniformly $O(\delta)$ and can be dropped since the divergences
at small $T_k$ are at worst logarithmic. The second terms in the
expressions for the $K_i$ are $O(\delta)$ for finite $T_i$, but $O(1)$
when all the $T_i$ are $O(\delta)$. That region of integration is
negligible for the $K_1K_2K_3$ term, but not for the terms linear
in the $K_i$. Thus it is legitimate to make the substitutions
\bea
2K_1^\wedge K_2^\vee K_3^\wedge&\to&{T_1T_2T_3\over 
T_{13}^3p_1^+p_2^+p_3^+}2K^\wedge K^\vee K^\wedge={T_1T_2T_3\over 
T_{13}^3p_1^+p_2^+p_3^+}{\bfs K}^2K^\wedge\\
K_i+K_{i+1}&\to&{T_{i+1}p_{i+1}^++T_{i+2}p_i^+\over
p_i^+p_{i+1}^+T_{13}}K^\wedge-2\delta{(T_1{\bfs k}_1
+T_2{\bfs k}_2+T_3{\bfs k}_0)\over T_{13}(\delta+T_{13})}
\label{klinear}\eea
The contribution of the second term on the
right of (\ref{klinear}) to the integration comes solely from the
region of all $T_i=O(\delta)$ so the exponential factor can be replaced by
unity and the integrals to be done can be simplified to
\bea
I^i&=&\int dT_1dT_2dT_3\delta(T_{13}q^+-T_{12}p^+_1-T_2p^+_2)
{\delta T_i\over T_{13}(\delta+T_{13})^3}
\eea
It is most convenient to use the delta function to eliminate
$T_3$ in favor of $T_{1,2}$ when $q^+<p_1^+$ but
$T_2$ in favor of $T_{1,3}$ when $q^+>p_1^+$. Then we find
\bea
(I_<^1,I_<^2)&=&{q^{+2}\over4 p_1^{+2}(p_1^++p_2^+)^2}(p_1^++p_2^+,p_1^+),
\qquad I_<^3={q^{+}\over4 p_1^{+}(p_1^++p_2^+)}\left[
2-{q^+\over p_1^{+} }-{q^+\over p_1^++p_2^+ }\right]\\
(I_>^1,I_>^3)&=&{(p_1^++p_2^+-q^{+})^2\over4 p_2^{+2}(p_1^++p_2^+)^2}(p_1^++p_2^+,p_2^+),
\qquad I_>^2={p_1^++p_2^+ -q^{+}\over4 p_2^{+}(p_1^++p_2^+)}\left[
{q^+-p_1^+\over p_2^{+} }+{q^+\over p_1^++p_2^+ }\right]
\eea
Note that the second line can be obtained from the first line by the
substitutions $q^+\to p_1^+ + p_2^+ - q^+$ and
$I^2, p_1^+\leftrightarrow I^3, p_2^+$.
The complete vertex should be proportional to $K^\wedge$, a property
not shared by the contribution of the second term. But we have not
yet included the swordfish graphs, which we turn to in the
next section. 

We close this section by giving the $\delta\to0$
limit of the triangle graphs with the contributions of 
the second terms of (\ref{klinear}) omitted.
\bea
\Gamma_{\triangle-}^{\wedge\wedge\vee}&=&-{g^3\over4\pi^2}
{p_3^+\over p_1^+p_2^+}K^\wedge
\sum_{q^+}\int dT_1dT_2dT_3
\delta(T_{13}q^+-T_{12}p^+_1-T_2p^+_2)\nonumber\\
&&\exp\left\{-{T_1T_3(k_1-k_0)^2+T_1T_2(k_1-k_2)^2
+T_2T_3(k_2-k_0)^2\over T_1+T_2+T_3}
\right\}\nonumber\\
&&\bigg\{\left[{{\bfs K}^2T_1T_2T_3\over T_{13}^4}
+{p_3^+(T_2p_2^++T_3p_1^+)\over T_{13}(T_{13}+\delta)^2}\right]
{A_3\over p_3^{+2}}
%\nonumber\\&&\qquad\qquad
+\left[{{\bfs K}^2T_1T_2T_3\over T_{13}^4}+
{p_2^+(T_2p_3^++T_1p_1^+)\over T_{13}(T_{13}+\delta)^2}
\right]{A_2\over p_3^{+2}}\nonumber\\&&\qquad\qquad
+\left[{{\bfs K}^2T_1T_2T_3\over T_{13}^4}+
{p_1^+(T_3p_3^++T_1p_2^+)\over T_{13}(T_{13}+\delta)^2}
\right]{A_1\over p_3^{+2}}\bigg\}
\eea
We simplify this expression by changing variables to $T=T_{13},
x=T_1/T_{13}, y=T_2/T_{13}$ and evaluating the integral over
$T$. Define
\bea
H(x,y)&=&x(1-x-y)p_1^2+xyp_2^2+y(1-x-y)p_3^2
\eea
and we obtain
\bea
\Gamma_{\triangle-}^{\wedge\wedge\vee}&=&-{g^3\over4\pi^2}
{p_3^+\over p_1^+p_2^+}K^\wedge
\sum_{q^+}\int_{x+y\leq1} dx dy
\delta(q^+-(x+y)p^+_1-yp^+_2)\nonumber\\
&&\bigg\{\left[{{\bfs K}^2xy(1-x-y)\over H}
-{p_3^+(yp_2^++(1-x-y)p_1^+)}\ln(\delta He^{\gamma+1})\right]
{A_3\over p_3^{+2}}
\nonumber\\&&
+\left[{{\bfs K}^2xy(1-x-y)\over H}-
{p_2^+(yp_3^++xp_1^+)}\ln(\delta He^{\gamma+1})
\right] {A_2\over p_3^{+2}}\nonumber\\&&
+\left[{{\bfs K}^2xy(1-x-y)\over H}-
{p_1^+((1-x-y)p_3^++xp_2^+)}\ln(\delta He^{\gamma+1})\right]
{A_1\over p_3^{+2}} \bigg\}
\label{3gtriangle}
\eea 
where we have used the $\delta\to0$
behavior of the integral
\bea
\int_0^\infty {T dT\over(T+\delta)^2}e^{-HT}&=&I(H\delta)
+H\delta I^\prime(H\delta)\\
&\sim& -\ln(\delta H)-1-\gamma= -\ln(\delta He^{\gamma+1}).
\eea
Finally, it is convenient to use an integration by parts to
convert the $\ln H$ terms to $H^{-1}dH/dx$, which enables an
explicit isolation of the divergent parts of the integral.
This is done by writing the coefficients of the $\ln$'s,
after using the delta function constraint to eliminate $y$
in favor of $x$, as derivatives with respect to $x$:
\bea
yp_2^+p_3^++(1-x-y)p_1^+p_3^+&=&-{d\over dx}\left[x\left(p_2^+(q^++p_1^+)
-p_1^+(q^+-p_1^+)\right)-x^2p_1^+p_2^+\right]\\
yp_2^+p_3^++xp_1^+p_2+&=&-{d\over dx}\left[xp_2^+(q^+-xp_1^+)\right]\\
(1-x-y)p_1^+p_3^++xp_1^+p_2^+&=&-{d\over dx}\left[xp_1^+
(p_1^++p_2^+-q^+-xp_2^+)\right]
\eea
The delta function for general $q^+\neq p_1^+$ limits the
range of $x$ to $0<x<q^+/p_1^+$ when $q^+<p_1^+$ and to
$0<x<(p_1^++p_2^+-q^+)/p_2^+$ when  $q^+>p_1^+$. The surface
terms from the integration by parts only contribute at the
upper limit. They are
\bea
&&\hskip-16pt{\rm Surface~Terms}^{\wedge\wedge\vee}\nonumber\\
&&\hskip-14pt=- {g^3\over4\pi^2}
{p_3^+\over p_1^+p_2^+}K^\wedge{1\over p_1^++p_2^+}\bigg\{
\sum_{q^+<p_1^+}\ln(\delta e^{\gamma+1}H_<)
\left[{q^{+}(p^+_1+p^+_2-q^+)A_3\over p_3^{+2}}
+{(p_1^++p_2^+)q^+(p_1^+-q^+)A_1
\over p_1^{+}p_3^{+2}} \right]
\nonumber\\
&&\hskip-14pt
+\sum_{q^+>p_1^+}\ln(\delta  e^{\gamma+1}H_>)
\left[{q^{+}(p^+_1+p^+_2-q^+)A_3\over p_3^{+2}}+
{(p_1^++p_2^+)(q^+-p_1^+)(p^+_1+p^+_2-q^+)A_2\over p_2^+p_3^{+2}
}\right]\bigg\}\\
&&\hskip-14pt=
-{g^3\over4\pi^2}
{p_3^+\over p_1^+p_2^+}K^\wedge{1\over p_1^++p_2^+}\bigg\{
\sum_{q^+<p_1^+}\ln(\delta e^{\gamma+1}H_<)
\left[{q^{+3}\over p_3^{+2}(p^+_1+p^+_2-q^+)}
+{(p^+_1+p^+_2-q^+)^3\over p_3^{+2}q^+}+{p_1^{+}(p_1^++p_2^+)
\over(p_1^+-q^+)q^{+}} \right]
\nonumber\\
&&\hskip-14pt
+\sum_{q^+>p_1^+}\ln(\delta  e^{\gamma+1}H_>)
\left[{q^{+3}\over p_3^{+2}(p^+_1+p^+_2-q^+)}
+{(p^+_1+p^+_2-q^+)^3\over p_3^{+2}q^+}+
{p_2^{+}(p_1^++p_2^+)\over(q^+-p_1^+)(p^+_1+p^+_2-q^+)}\right]\bigg\}\\
&&\hskip-14pt=
-{g^3\over4\pi^2}
{p_3^+\over p_1^+p_2^+}K^\wedge\bigg\{
\sum_{q^+<p_1^+}\ln(\delta e^{\gamma+1}H_<)
\left[{1\over p^+_1+p^+_2-q^+}+{2\over q^+}+{1\over p_1^+-q^+}\right]
\nonumber\\&&\hskip-14pt
+\sum_{q^+>p_1^+}\ln(\delta  e^{\gamma+1}H_>)
\left[{2\over p^+_1+p^+_2-q^+}+{1\over q^+}+{1\over q^+- p_1^+}\right]
\nonumber\\&&\hskip-14pt
-\sum_{q^+<p_1^+}\ln(\delta e^{\gamma+1}H_<)
\left[{2q^{+2}-2q^+(p_1^++p_2^+)+4(p_1^++p_2^+)^2
\over(p_1^++p_2^+)^3}\right]\nonumber\\
&&\hskip-14pt-\sum_{q^+>p_1^+}\ln(\delta  e^{\gamma+1}H_>)
\left[{2q^{+2}-2q^+(p_1^++p_2^+)+4(p_1^++p_2^+)^2
\over(p_1^++p_2^+)^3}\right]\bigg\}\\
&&\hskip-14pt=-{g^3\over4\pi^2}
{p_3^+\over p_1^+p_2^+}K^\wedge\bigg\{{\cal A}(p_1^2,p_1^+)
+{\cal A}(p_2^2,p_2^+)+{\cal A}(p_3^2,-p_3^+)\nonumber\\
&&\hskip-14pt+\sum_{q^+<p_1^+}
\left[{1\over p^+_1+p^+_2-q^+}+{2\over q^+}+{1\over p_1^+-q^+}\right]
+\sum_{q^+>p_1^+}
\left[{2\over p^+_1+p^+_2-q^+}+{1\over q^+}+{1\over q^+- p_1^+}\right]
\nonumber\\
&&\hskip-14pt+\sum_{q^+<p_1^+}
\left[{1\over p^+_1+p^+_2-q^+}+{1\over q^+}\right]\ln{p_1^2(p_1^{+}+p_2^+)^2
(p_1^{+}-q^+)\over(p_1+p_2)^2p_1^{+2}(p_1^{+}+p_2^+-q^+)}
\nonumber\\
&&\hskip-14pt+\sum_{q^+>p_1^+}
\left[{1\over p^+_1+p^+_2-q^+}+{1\over q^+}\right]\ln{p_2^2(p_1^{+}+p_2^+)^2
(q^+-p_1^{+})\over(p_1+p_2)^2p_2^{+2}q^+}
\nonumber\\
&&\hskip-14pt-\sum_{q^+<p_1^+}\ln(\delta e^{\gamma+1}H_<)
\left[{2q^{+2}-2q^+(p_1^++p_2^+)+4(p_1^++p_2^+)^2
\over(p_1^++p_2^+)^3}\right]\nonumber\\
&&\hskip-14pt-\sum_{q^+>p_1^+}\ln(\delta  e^{\gamma+1}H_>)
\left[{2q^{+2}-2q^+(p_1^++p_2^+)+4(p_1^++p_2^+)^2
\over(p_1^++p_2^+)^3}\right]\bigg\}\\
&&\hskip-14pt\sim-{g^3\over4\pi^2}
{p_3^+\over p_1^+p_2^+}K^\wedge\bigg\{{\cal A}(p_1^2,p_1^+)
+{\cal A}(p_2^2,p_2^+)+{\cal A}(p_3^2,-p_3^+)\nonumber\\
&&\hskip-14pt+\sum_{q^+}
\left[{2\over p^+_1+p^+_2-q^+}+{2\over q^+}\right]+
\sum_{q^+\neq p_1^+}{1\over |p_1^+-q^+|}
-\ln{(p_1^++p_2^+)^2\over p_1^+p_2^+}
\nonumber\\
&&\hskip-14pt+\sum_{q^+\neq p_1^+}
\left[{1\over q^+}\ln{p_1^2(p_1^{+}+p_2^+)
\over(p_1+p_2)^2p_1^{+}}+{1\over p^+_1+p^+_2-q^+}\ln{p_2^2(p_1^{+}+p_2^+)
\over(p_1+p_2)^2p_2^{+}}\right]
\nonumber\\
&&\hskip-14pt+\int_0^{p_1^+}dq^+
\left[{1\over p^+_1+p^+_2-q^+}+{1\over q^+}\right]
\ln{(p_1^{+}+p_2^+)(p_1^{+}-q^+)
\over p_1^{+}(p_1^{+}+p_2^+-q^+)}
\nonumber\\\hskip-14pt&&
+\int_{p_1^+}^{p_1^++p_2^+}dq^+
\left[{1\over p^+_1+p^+_2-q^+}+{1\over q^+}\right]\ln{(p_1^{+}+p_2^+)
(q^+-p_1^{+})\over p_2^{+}q^+}
\nonumber\\
&&\hskip-14pt-\int_0^{p_1^+}dq^+\ln(\delta e^{\gamma+1}H_<)
\left[{2q^{+2}-2q^+(p_1^++p_2^+)+4(p_1^++p_2^+)^2
\over(p_1^++p_2^+)^3}\right]\nonumber\\
&&\hskip-14pt-\int_{p_1^+}^{p_1^++p_2^+}dq^+\ln(\delta  e^{\gamma+1}H_>)
\left[{2q^{+2}-2q^+(p_1^++p_2^+)+4(p_1^++p_2^+)^2
\over(p_1^++p_2^+)^3}\right]\bigg\}
\label{surfaceterms}
\eea
where
\bea
H_<=H\left({q^+\over p_1^+}\right)
={q^+\over p_1^+}\left(1-{q^+\over p_1^+}\right)p_1^2,
\qquad H_>=H\left({p_1^++p_2^+-q^+\over p_2^+}\right)
={p_1^++p_2^+-q^+\over p_2^+}
\left({q^+-p_1^+\over p_2^+}\right)p_2^2
\eea
It is also useful to introduce
\bea
H_0\equiv H(0)={q^+\over p_1^++p_2^+}
\left(1-{q^+\over p_1^++p_2^+}\right)p_3^2
\eea
Then the last two lines can be written more compactly
as
\bea
&&-\int_0^{p_1^++p_2^+}dq^+\ln(\delta e^{\gamma+1}H_0)
\left[{2q^{+2}-2q^+(p_1^++p_2^+)+4(p_1^++p_2^+)^2
\over(p_1^++p_2^+)^3}\right]\nonumber\\
&&-\int_0^{p_1^+}dq^+\ln(H_</H_0)
\left[{2q^{+2}-2q^+(p_1^++p_2^+)+4(p_1^++p_2^+)^2
\over(p_1^++p_2^+)^3}\right]\nonumber\\
&&-\int_{p_1^+}^{p_1^++p_2^+}dq^+\ln(H_>/H_0)
\left[{2q^{+2}-2q^+(p_1^++p_2^+)+4(p_1^++p_2^+)^2
\over(p_1^++p_2^+)^3}\right]\nonumber\\
&=&-\int_0^1du\ln(\delta e^{\gamma+1}u(1-u)p_3^2)
\left[{2u^{2}-2u+4}\right]\nonumber\\
&&-\int_0^{p_1^++p_2^+}dq^+\ln\left({H(x_{max})\over H_0}\right)
\left[{2q^{+2}-2q^+(p_1^++p_2^+)+4(p_1^++p_2^+)^2
\over(p_1^++p_2^+)^3}\right]
\eea

The ultraviolet ($\delta\to0$) divergence of the triangle graph
is completely contained in these surface terms. The coefficient
of $\ln\delta$ involves the sums
\bea
\sum_{q^+<p_1^+}
\left[{1\over p^+_1+p^+_2-q^+}+{2\over q^+}+{1\over p_1^+-q^+}
-{2q^{+2}-2q^+(p_1^++p_2^+)+4(p_1^++p_2^+)^2
\over(p_1^++p_2^+)^3} \right]\hskip1.2in&& 
\nonumber\\
+\sum_{q^+>p_1^+}
\left[{2\over p^+_1+p^+_2-q^+}+{1\over q^+}+{1\over q^+- p_1^+}
-{2q^{+2}-2q^+(p_1^++p_2^+)+4(p_1^++p_2^+)^2
\over(p_1^++p_2^+)^3}\right]\to\hskip1in&&\nonumber\\
-{11\over3}+\sum_{q^+<p_1^+}\left[{1\over q^+}+{1\over p_1^+-q^+}\right]
+\sum_{q^+\neq p_1^+}\left[{1\over q^+}+{1\over p_1^++p_2^+-q^+}\right]
+\sum_{q^+>p_1^+}\left[{1\over q^+-p_1^+}+{1\over p_2^+-(q^+-p_1^+)}\right]
\hskip0in&&\nonumber\\
 \to 2\ln[M_1M_2(M_1+M_2)]+6\gamma-{11\over3}\hskip0in&&
\label{logdivver}
\eea 
where $M_i=p_i^+/m$ is a large positive integer.
In this evaluation we replace the sums by integrals 
for the non-singular terms. The sums are kept discrete for the 
singular terms with the interpretations
\bea
\sum_{q^+<p_1^+}{1\over q^+}&=&\sum_{q^+<p_1^+}{1\over p_1^+-q^+}
=\sum_{l=1}^{M_1-1} {1\over l}=\psi(M_1)+\gamma\sim \ln M_1+\gamma\\
\sum_{q^+>p_1^+}{1\over p_1^++p_2^+-q^+}&=&\sum_{q^+>p_1^+}{1\over q^+-p_1^+}
=\sum_{l=1}^{M_2-1} {1\over l}=\psi(M_2)+\gamma\sim \ln M_2+\gamma
\eea
where $\psi(z)=\Gamma^\prime(z)/\Gamma(z)$ is the digamma function and
$\gamma$ is Euler's constant.

But notice that the three discrete sums as grouped in the
third line of (\ref{logdivver}) are precisely those that occur in the
coefficient of the log divergences of the self energies for the
three legs coming into the vertex function. Furthermore the coefficients
of these sums are half those in the self energies and the sign is opposite.
Thus all these discrete divergent sums cancel up to half of the
ones on external legs. it is also noteworthy that the summands
actually cancel locally, i.e. independently for each $q^+$,
as already pointed out in \cite{gudmundssontt}. Since the
worldsheet organizes loops according to their location $\sigma$,
it is very satisfactory that a loop at fixed $\sigma$ will 
not have spurious $p^+$ divergences provided it is summed over all times.

After integration by parts the  amplitude reads:
\bea
\Gamma_{\triangle-}^{\wedge\wedge\vee}&=&
{\rm Surface~Terms}-{g^3\over4\pi^2}
{p_3^+\over p_1^+p_2^+}K^\wedge
\sum_{q^+}\int_{x+y\leq1} dx dy
\delta(q^+-(x+y)p^+_1-yp^+_2)\nonumber\\
&&\hskip-.5in{1\over H}\bigg\{\left[{{\bfs K}^2xy(1-x-y)}
-\left[xp_2^+q^+
+xp_1^+(p_1^++p_2^+-q^+)-x^2p_1^+p_2^+\right]{dH\over dx}\right]
{A_3\over p_3^{+2}}
\nonumber\\&&
+\left[{{\bfs K}^2xy(1-x-y)}-xp_1^+
(p_1^++p_2^+-q^+-xp_2^+){dH\over dx}
\right]{A_1\over p_3^{+2}}\nonumber\\&&
+\left[{{\bfs K}^2xy(1-x-y)}-xp_2^+(q^+-xp_1^+){dH\over dx}
\right]{A_2\over p_3^{+2}}\bigg\}\\
&=&
{\rm Surface~Terms}-{g^3\over4\pi^2}
{p_3^+\over p_1^+p_2^+}K^\wedge
\sum_{q^+}\int_0^{x_{max}} {dx\over p_1^++p_2^+}\nonumber\\
&&{1\over H}\bigg\{
-x\left[{p_2^+(p^+_1+p^+_2-q^+)^2\over q^+p_3^{+2}}
+{p_1^+q^{+2}\over (p^+_1+p^+_2-q^+)p_3^{+2}}\right]{dH\over dx}
\nonumber\\&&
+\left[{{\bfs K}^2xy(1-x-y)}-xp_1^+
(p_1^++p_2^+-q^+-xp_2^+){dH\over dx}
\right]{1\over q^{+2}}\left({p^{+2}_1\over (q^+-p^+_1)^{2}}
+{(p^+_1+p^+_2-q^+)^2\over p_3^{+2}}\right)\nonumber\\&&
+\left[{{\bfs K}^2xy(1-x-y)}-xp_2^+(q^+-xp_1^+){dH\over dx}
\right]{1\over(p^+_1+p^+_2-q^+)^2}\left({p^{+2}_2\over (q^+-p^+_1)^{2}}
+{q^{+2}\over p_3^{+2}}\right)\bigg\}\\
&=&
{\rm Surface~Terms}-{g^3\over4\pi^2}
{p_3^+\over p_1^+p_2^+}K^\wedge
\sum_{q^+}\int_0^{x_{max}} {dx\over p_1^++p_2^+}\nonumber\\
&&\bigg\{
-\left[{p_2^+(p^+_1+p^+_2-q^+)^2\over q^+p_3^{+2}}
+{p_1^+q^{+2}\over (p^+_1+p^+_2-q^+)p_3^{+2}}\right]I_1
%\nonumber\\&&
+{1\over q^{+2}}\left({p^{+2}_1\over (q^+-p^+_1)^{2}}
+{(p^+_1+p^+_2-q^+)^2\over p_3^{+2}}\right)I_3\nonumber\\&&
+{1\over(p^+_1+p^+_2-q^+)^2}\left({p^{+2}_2\over (q^+-p^+_1)^{2}}
+{q^{+2}\over p_3^{+2}}\right)I_2\bigg\}
\label{gammatriangle-}
\eea 
Here $x_{max}=q^+/p_1^+$ ($x_{max}=(p_1^++p_2^+-q^+)/p_2^+$) if
$q^+<p_1^+$ ($q^+>p_1^+$) respectively.
Evidently the continuum limit of the $q^+$ sums involves divergences 
due to the singularities in the $A_i$ when $q^+\sim 0, p_1^++p_2^+,
p_1^+$. Although the divergences seem to be linear, cancellations
soften those near $0, p_1^++p_2^+$. The divergence near
$p_1^+$ however is not softened in the triangle graph itself,
but we shall see that the swordfish diagrams cancel
the most divergent part leaving it logarithmic as well.

To show this we first note, using the delta function to eliminate $y$,
\bea
H&\to& x{p_1^++p_2^+ -q^+-xp_2^+\over p_1^++p_2^+}p_1^2
+x{q^+-xp_1^+\over p_1^++p_2^+}p_2^2
+{q^+-xp_1^+\over p_1^++p_2^+}{p_1^++p_2^+ -q^+-xp_2^+\over p_1^++p_2^+}
p_3^2\\
{dH\over dx}&\to&{p_1^++p_2^+ -q^+-2xp_2^+\over p_1^++p_2^+}p_1^2
+{q^+-2xp_1^+\over p_1^++p_2^+}p_2^2
-{p_1^+(p_1^++p_2^+ -q^+)+q^+p_2^+
-2xp_1^+p_2^+\over (p_1^++p_2^+)^2}
p_3^2
\eea
Then with a little rearrangement we find
\bea
I_1&\equiv& {x\over H}{dH\over dx}\nonumber\\
&=&2+{x\over H}\left[
{p_1^+(p_1^++p_2^+-q^+)\over p_1^++p_2^+}\left({p_3^2\over p_1^++p_2^+}
-{p_1^2\over p_1^+}\right)+{p_2^+q^+\over p_1^++p_2^+}
\left({p_3^2\over p_1^++p_2^+}-{p_2^2\over p_2^+}\right)
\right]\nonumber\\
&&\hskip3in-{2\over H}{q^+(p_1^++p_2^+-q^+)\over (p_1^++p_2^+)^2}p_3^2\\
I_2&\equiv&{1\over H}\left\{{\bfs K}^2xy(1-x-y)
-\left[xp_2^+(q^+-xp_1^+)\right]{dH\over dx}\right\}\nonumber\\
&=&\hskip0in
xp_1^+p_2^+ + x{q^+-p_1^+\over H}\left[xp_2^+p_1^2+(xp_1^+-q^+)p_2^2
-{p_2^+(xp_1^+-q^+)\over p_1^++p_2^+}p_3^2\right]\nonumber\\
&=&\hskip0in
xp_1^+p_2^+-(q^+-p_1^+)(p_1^++p_2^+) 
+ {(q^+-p_1^+)(p_1^++p_2^+-q^+)\over H}
\left[xp_1^2+{q^+-xp_1^+\over p_1^++p_2^+}p_3^2\right]\\
&\equiv&xp_1^+p_2^++{\hat I}_2\\
I_3&\equiv&{1\over H}\left\{{\bfs K}^2xy(1-x-y)
-\left[xp_1^+(p_1^++p_2^+-q^+-xp_2^+)\right]
{dH\over dx}\right\}\nonumber\\
&=&\hskip0in
xp_1^+p_2^+ + x{p_1^+-q^+\over H}\left[xp_1^+p_2^2+(xp_2^+-p_1^+
-p_2^++q^+)p_1^2
-{p_1^+(xp_2^+-p_1^+
-p_2^++q^+)\over p_1^++p_2^+}p_3^2\right]
\nonumber\\
&=&\hskip0in
xp_1^+p_2^++(q^+-p_1^+)(p_1^++p_2^+) 
+ {q^+(p_1^+-q^+)\over H}\left[xp_2^2
+{p_1^+
+p_2^+-xp_2^+-q^+\over p_1^++p_2^+}p_3^2\right]\\
&\equiv&xp_1^+p_2^++{\hat I}_3
\eea
We see that the second and third terms in the 
final expressions for $I_2,I_3$ 
supply a zero at $q^+=p_1^+$, softening the divergence near
$q^+\sim p_1^+$ to a logarithmic one. The linear divergence
comes entirely from the first terms.

The integrals of $I_1,I_2,I_3$ over $x$ are elementary,
and the ones involving
$H^{-1}$ are conveniently written
in terms of the roots $r_+,r_-$ of the quadratic polynomial
$H(x)=A(x-r_+)(x-r_-)$:
\bea
r_\pm&=& {-B\pm\sqrt{B^2-4AC}\over2A}\\
A&=&{p_1^+p_2^+\over p_1^++p_2^+}\left[{p_3^2\over p_1^++p_2^+}
-{p_1^2\over p_1^+}-{p_2^2\over p_2^+}\right]\\
B&=&{ p_1^++p_2^+-q^+\over  p_1^++p_2^+}p_1^2+{q^+\over  p_1^++p_2^+}p_2^2
-{q^+p_2^++(p_1^++p_2^+-q^+)p_1^+\over (p_1^++p_2^+)^2}p_3^2\\
C&=&{q^+(p_1^++p_2^+-q^+)\over (p_1^++p_2^+)^2}p_3^2\\
\int_0^{x_{max}}{dx\over H}&=&
{1\over A(r_+-r_-)}\left[\ln{r_+-x_{max}\over r_+}
-\ln{r_--x_{max}\over r_-}\right]\\
\int_0^{x_{max}}{xdx\over H}&=&
{r_+\over A(r_+-r_-)}\ln{r_+-x_{max}\over r_+}
-{r_-\over A(r_+-r_-)}\ln{r_--x_{max}\over r_-}
\eea
We now consider in turn the behavior of these expressions near
the singular values of $q^+=0,p_1^+,p_1^++p_2^+$. First
consider $q^+\sim 0$. Then $x_{max}=q^+/p_1^+$ and we find
\bea
\int_0^{x_{max}}{dx\over H}&\sim&
{p_1^++p_2^+\over (p_1^++p_2^+)p_1^2-p_1^+p_3^2}\ln{(p_1^++p_2^+)p_1^2\over 
 p_1^+p_3^2}\hskip1.5in {\rm for}~q^+\sim0\\
\int_0^{x_{max}}{xdx\over H}&\sim&{q^+(p_1^++p_2^+)/p_1^+\over 
(p_1^++p_2^+)p_1^2-p_1^+p_3^2}
\left[{p_1^+p_3^2\over p_1^+p_3^2-(p_1^++p_2^+)p_1^2}
\ln{(p_1^++p_2^+)p_1^2\over  p_1^+p_3^2}+1\right]
\qquad{\rm for}~q^+\sim0
\eea
Next we consider $q^+\sim p_1^++p_2^+$, for which $x_{max}=(p_1^++p_2^+
-q^+)/p_2^+$. Then
\bea
\int_0^{x_{max}}{dx\over H}&\sim&
{p_1^++p_2^+\over (p_1^++p_2^+)p_2^2-p_2^+p_3^2}\ln{(p_1^++p_2^+)p_2^2\over 
 p_2^+p_3^2}\hskip1.3in{\rm for}~q^+\sim p_1^++p_2^+\\
\int_0^{x_{max}}{xdx\over H}&\sim&{(p_1^++p_2^+-q^+)(p_1^++p_2^+)\over 
p_2^+\left[(p_1^++p_2^+)p_2^2-p_2^+p_3^2\right]}
\left[{p_2^+p_3^2\over p_2^+p_3^2-(p_1^++p_2^+)p_2^2}
\ln{(p_1^++p_2^+)p_2^2\over  p_2^+p_3^2}+1\right]\\
&&\hskip3.5in{\rm for}~q^+\sim p_1^++p_2^+\nonumber
\eea
Finally, for $q^+\sim p_1^+$ we have to treat separately the cases $q^+<p_1^+$
(when $x_{max}=q^+/p_1^+$) and $q^+>p_1^+$ 
(when $x_{max}=(p_1^++p_2^+-q^+)/p_2^+$). Actually, the singular factor
multiplying these integrals is just $(q^+-p_1^+)^{-1}$. Since the
$q^+$ sum is symmetric about $q^+=p_1^+$, a divergence will occur
only because of a discontinuity in the summand due to
the different behavior of $x_{max}$ on either side:
\bea
x_{max}=\cases{q^+/p_1^+\sim 1+\delta q^+/p_1^+ &for $q^+<p_1^+$\cr
\phantom{a}&\cr
(p_1^++p_2^+-q^+)/p_2^+\sim 1-\delta q^+/p_2^+&for $q^+>p_1^+$\cr}
\eea
where $\delta q^+=q^+-p_1^+\ll p_1^+$. Then we find
\bea
\int_0^{x_{max}}{dx\over H}&\sim&
-{1\over D}\left[\ln{|\delta q^+|\over D}-\ln{(p_1^++p_2^+)^2D
\over p_1^+p_2^+p_3^2} 
\right]-{1\over D}\cases{\ln{p_1^2\over p_1^+}&for $q^+<p_1^+$\cr
\ln{p_2^2\over p_2^+}&for $q^+>p_1^+$\cr}
\hskip.4in{\rm for}~q^+\sim p_1^+\\
\int_0^{x_{max}}{xdx\over H}&\sim&-{1\over D}\left[\ln{|\delta q^+|\over D}
-{A+D\over A}\ln{(p_1^++p_2^+)^2D\over p_1^+p_2^+p_3^2} 
\right]-{1\over D}
\cases{\ln{p_1^2\over p_1^+}&for $q^+<p_1^+$\cr
\ln{p_2^2\over p_2^+}&for $q^+>p_1^+$\cr}\hskip.1in{\rm for}~q^+\sim p_1^+\\
D&\equiv&{p_1^+p_2^+\over p_1^++p_2^+}\left[{p_1^2\over p_1^+}
+{p_2^2\over p_2^+}\right]
\eea
We see that the discontinuity (value for
$q_+<p_1^+-$ value for $q^+>p_1^+$) of either of
the right sides about the point
$q^+=p_1^+$ is simply
\bea
-{1\over D}\ln{p_1^2p_2^+\over p_2^2p_1^+}=-{p_1^++p_2^+\over
p_1^+p_2^2+p_2^+p_1^2}\ln{p_1^2p_2^+\over p_2^2p_1^+}
\eea

Now we combine these results to extract the remaining
divergences in $\Gamma_{\triangle-}^{\wedge\wedge\vee}$.
First of all we find that
\bea
\int dx I_1\sim\cases{{q^+\over p_1^+}\left[1+{p_1^+p_3^2
\over p_1^+p_3^2-(p_1^++p_2^+)p_1^2}\ln{(p_1^++p_2^+)p_1^2
\over p_1^+p_3^2}\right]&for $q^+\sim0$\cr
{p_1^++p_2^+-q^+\over p_2^+}\left[1+{p_2^+p_3^2
\over p_2^+p_3^2-(p_1^++p_2^+)p_2^2}\ln{(p_1^++p_2^+)p_2^2
\over p_2^+p_3^2}\right]&for 
$q^+\sim p_1^++p_2^+$\cr}
\eea
The singular factor multiplying this integral has only
simple poles at $q^+=0$ and $q^+=p_1^++p_2^+$ which are killed
by the above behavior, so the $I_1$ term in $\Gamma$ is
finite in the continuum limit. 

The $I_3$ term and the $I_2$ term involve
singular factors $q^{+-2}, (q^+-p_1^+)^{-2}$ 
and $(p_1^++p_2^+-q^+)^{-2}, (q^+-p_1^+)^{-2}$ respectively. 
We first examine the singular behavior for $q^+\sim0$ which is found
only in the $I_3$ term. We find
\bea
\int dx I_3\sim -q^+(p_1^++p_2^+)+{q^+p_1^+}{p_3^2(p_1^++p_2^+)
\over (p_1^++p_2^+)p_1^2-p_1^+p_3^2}\ln{(p_1^++p_2^+)p_1^2\over 
 p_1^+p_3^2}\hskip.5in {\rm for}~q^+\sim0
\eea
So the singular contribution of the triangle graph  at $q^+=0$ reads
\bea
-{g^3\over4\pi^2}
{p_3^+\over p_1^+p_2^+}K^\wedge
\sum_{q^+}{2\over q^+}\left[ -1+{p_1^+p_3^2
\over (p_1^++p_2^+)p_1^2-p_1^+p_3^2}\ln{(p_1^++p_2^+)p_1^2\over 
 p_1^+p_3^2}\right]
\label{divzero}
\eea
Very similarly, the singular behavior of the triangle graph near
$q^+=p_1^++p_2^+$, which comes from the $I_2$ term, reads
\bea
-{g^3\over4\pi^2}
{p_3^+\over p_1^+p_2^+}K^\wedge
\sum_{q^+}{2\over p_1^++p_2^+-q^+}\left[ -1+{p_2^+p_3^2
\over (p_1^++p_2^+)p_2^2-p_2^+p_3^2}\ln{(p_1^++p_2^+)p_2^2\over 
 p_2^+p_3^2}\right]
\label{divonetwo}
\eea

Finally we separate the singular contributions near $q^+=p_1^+$.
These are found in both the $I_2$ and $I_3$ terms. Unlike the
previous contributions, there are both linear and logarithmic divergence 
near $q^+=p_1^+$ in the triangle graph. Fortunately, the linear divergence 
comes only from the first terms of $I_2$ and $I_3$, and those terms
did not contribute to the divergences near $q^+=0,p_1^++p_2^+$, so it
is convenient to evaluate their contribution completely (i.e. not
just the singular parts.
\bea
&&{\rm First~Term}=-{g^3\over4\pi^2}{p_3^+\over p_1^++p_2^+}K^\wedge
\sum_{q^+}\int_0^{x_{max}} dx x {A_1+A_2\over p_3^{+2}}\nonumber\\
&=&
{g^3\over8\pi^2}K^\wedge
\sum_{q^+}{x_{max}^2}
\bigg[{1\over q^{+2}}\left({p^{+2}_1\over (q^+-p^+_1)^{2}}
+{(p^+_1+p^+_2-q^+)^2\over p_3^{+2}}\right)
+{1\over(p^+_1+p^+_2-q^+)^2}\left({p^{+2}_2\over (q^+-p^+_1)^{2}}
+{q^{+2}\over p_3^{+2}}\right)\bigg]\nonumber\\
&=&
{g^3\over8\pi^2}K^\wedge
\sum_{q^+<p_1^+}{1\over p_1^{+2}}
\bigg[{p^{+2}_1\over (q^+-p^+_1)^{2}}
+{(p^+_1+p^+_2-q^+)^2\over p_3^{+2}}
+{q^{+2}\over(p^+_1+p^+_2-q^+)^2}\left({p^{+2}_2\over (q^+-p^+_1)^{2}}
+{q^{+2}\over p_3^{+2}}\right)\bigg]\nonumber\\
&&+{g^3\over8\pi^2}K^\wedge
\sum_{q^+>p_1^+}{1\over p_2^{+2}}
\bigg[{(p^+_1+p^+_2-q^+)^2\over q^{+2}}\left({p^{+2}_1\over (q^+-p^+_1)^{2}}
+{(p^+_1+p^+_2-q^+)^2\over p_3^{+2}}\right)
+{p^{+2}_2\over (q^+-p^+_1)^{2}}
+{q^{+2}\over p_3^{+2}}\bigg]\nonumber\\
&=&{g^3\over8\pi^2}K^\wedge
\bigg[\sum_{q^+\neq p_1^+}{2\over(q^+-p_1^{+})^2}
+(p_1^++p_2^+)\sum_{q^+<p_1^+}
{q^{+}p_2^{+}+p_1^+(p_1^{+}+p_2^{+}-q^+)
\over (q^+-p_1^{+})p_1^{+2}(p_1^{+}+p_2^{+}-q^+)^2}\nonumber\\&&
+\sum_{q^+<p_1^+}{(p^+_1+p^+_2-q^+)^4+q^{+4}
\over p_1^{+2}p_3^{+2}(p^+_1+p^+_2-q^+)^2}
+(p_1^++p_2^+)\sum_{q^+>p_1^+}
{q^{+}p_2^{+}+p_1^+(p_1^{+}+p_2^{+}-q^+)\over(p_1^{+}-q^+)q^{+2}p_2^{+2}}
\nonumber\\&&
+\sum_{q^+>p_1^+}{(p^+_1+p^+_2-q^+)^4+q^{+4}
\over p_2^{+2}p_3^{+2}q^{+2}}\bigg]\nonumber\\
&=&{g^3\over8\pi^2}K^\wedge
\bigg[\sum_{q^+\neq p_1^+}\left\{{2\over(q^+-p_1^{+})^2}
-{p_1^++p_2^+\over p_1^+p_2^+}{2\over|q^+-p_1^+|}\right\}\nonumber\\
&&+\sum_{q^+<p_1^+}\left\{
{2(p_1^++p_2^+)\over p_1^+p_2^+(p_1^++p_2^+-q^+)}
+{(p^+_1+p^+_2-q^+)^4+q^{+4}+p_3^{+4}
\over p_1^{+2}p_3^{+2}(p^+_1+p^+_2-q^+)^2}\right\}
\nonumber\\&&
+\sum_{q^+>p_1^+}\left\{{2(p_1^++p_2^+)\over p_1^+p_2^+q^+}
+{(p^+_1+p^+_2-q^+)^4+q^{+4}+p_3^{+4}
\over p_2^{+2}p_3^{+2}q^{+2}}\right\}
\bigg]\\
&\to&{g^3\over8\pi^2}K^\wedge
\bigg[\sum_{q^+\neq p_1^+}\left\{{2\over(q^+-p_1^{+})^2}
-{p_1^++p_2^+\over p_1^+p_2^+}{2\over|q^+-p_1^+|}\right\}
+2{p_1^++p_2^+\over p_1^+p_2^+}\left\{\ln{(p_1^++p_2^+)^2\over p_1^+p_2^+}
+4\right\}\nonumber\\
&&+{2\over3}{1\over p_1^++p_2^+}-4{p_1^++p_2^+\over p_1^+p_2^+}
\left({p_1^+\over p_2^+}\ln{p_1^++p_2^+\over p_2^+}
+{p_2^+\over p_1^+}\ln{p_1^++p_2^+\over p_1^+}\right) \bigg]
\eea
Where we have evaluated the continuum limit of the
convergent terms.
The first term in square brackets is canceled by a corresponding
term from the swordfish diagrams (see 
Eq.~(\ref{swordfish3},\ref{swordfish2},\ref{swordfish1})
in the following section). Borrowing from (\ref{swordfish3})
we find
\bea
{\rm First~Term}+\Gamma^{\wedge\wedge\vee}_{\rm SF+}&\to&
{g^3\over8\pi^2}K^\wedge{p_1^++p_2^+\over p_1^+p_2^+}
\left[-\sum_{q^+\neq p_1^+}{2\over|q^+-p_1^+|}
+{50\over3}
-4\left({p_2^+\over p_1^{+}}+{p_1^+\over p_2^{+}}-{1\over2}\right)
\ln{(p_1^++p_2^+)^2\over p_1^+p_2^+}\right]
\nonumber\\
&&+{g^3\over12\pi^2}
{K^\wedge\over p_1^++p_2^+}-{g^3\over12\pi^2}(k_1+k_2
+k_0)^\wedge
\label{divonea}
\eea

The rest of $I_2$ and $I_3$ contribute a logarithmic divergence
near $q^+=p_1^+$. As already mentioned, since the $q^+$ sum
is symmetric about $q^+=p_1^+$, the singular behavior depends on the
discontinuity:
\bea
{1\over2}{\rm Disc}\int dx {\hat I_2}&\sim&-{1\over2}
(q^+-p_1^+){p_2^+(p_1^++p_2^+)p_1^2
\over p_1^+p_2^2+p_2^+p_1^2}\ln{p_2^+p_1^2\over p_1^+p_2^2}\\
{1\over2}{\rm Disc}\int dx {\hat I_3}&\sim&+{1\over2}
(q^+-p_1^+){p_1^+(p_1^++p_2^+)p_2^2
\over p_1^+p_2^2+p_2^+p_1^2}\ln{p_2^+p_1^2\over p_1^+p_2^2}
\eea
These two contributions to the divergence of the vertex function 
combine to
\bea 
-{g^3\over4\pi^2}
{p_3^+\over p_1^+p_2^+}K^\wedge
\sum_{q^+}\int_0^{x_{max}} {dx\over p_1^++p_2^+}{{\hat I}_2+{\hat I}_3
\over (q^+-p^+_1)^{2}}\sim-{g^3\over8\pi^2}
{p_3^+\over p_1^+p_2^+}K^\wedge
\sum_{q^+\neq p_1^+}{1\over |q^+-p^+_1|}{p_2^+p_1^2-p_1^+p_2^2
\over p_1^+p_2^2+p_2^+p_1^2}\ln{p_2^+p_1^2\over p_1^+p_2^2}
\label{divoneb}
\eea
Let us now combine (\ref{divzero}), (\ref{divonetwo}), (\ref{divonea}),
and (\ref{divoneb}),
with the surface terms (\ref{surfaceterms}):
\bea
&&\hskip-.3in-{g^3\over4\pi^2}
{p_3^+\over p_1^+p_2^+}K^\wedge\bigg\{{\cal A}(p_1^2,p_1^+)
+{\cal A}(p_2^2,p_2^+)+{\cal A}(p_3^2,-p_3^+)\nonumber\\
&&\hskip-.25in+\sum_{q^+}\left\{{1\over q^+}\left[{(p_1^++p_2^+)p_1^2+p_1^+p_3^2
\over (p_1^++p_2^+)p_1^2-p_1^+p_3^2}\ln{(p_1^++p_2^+)p_1^2\over 
 p_1^+p_3^2}\right]+{1\over p_1^++p_2^+-q^+}\left[
{(p_1^++p_2^+)p_2^2+p_2^+p_3^2
\over (p_1^++p_2^+)p_2^2-p_2^+p_3^2}\ln{(p_1^++p_2^+)p_2^2\over 
 p_2^+p_3^2}\right]\right\}\nonumber\\
&&+{1\over2}\sum_{q^+\neq p_1^+}{1\over |q^+-p^+_1|}{p_2^+p_1^2-p_1^+p_2^2
\over p_1^+p_2^2+p_2^+p_1^2}\ln{p_2^+p_1^2\over p_1^+p_2^2}
+{25\over3}
-2\left({p_2^+\over p_1^{+}}+{p_1^+\over p_2^{+}}\right)
\ln{(p_1^++p_2^+)^2\over p_1^+p_2^+}
\nonumber\\
&&+\int_0^{p_1^+}dq^+
\left[{1\over p^+_1+p^+_2-q^+}+{1\over q^+}\right]
\ln{(p_1^{+}+p_2^+)(p_1^{+}-q^+)
\over p_1^{+}(p_1^{+}+p_2^+-q^+)}
\nonumber\\&&
+\int_{p_1^+}^{p_1^++p_2^+}dq^+
\left[{1\over p^+_1+p^+_2-q^+}+{1\over q^+}\right]\ln{(p_1^{+}+p_2^+)
(q^+-p_1^{+})\over p_2^{+}q^+}
\nonumber\\
&&-\int_0^{p_1^+}dq^+\ln(\delta e^{\gamma+1}H_<)
\left[{2q^{+2}-2q^+(p_1^++p_2^+)+4(p_1^++p_2^+)^2
\over(p_1^++p_2^+)^3}\right]\nonumber\\
&&-\int_{p_1^+}^{p_1^++p_2^+}dq^+\ln(\delta  e^{\gamma+1}H_>)
\left[{2q^{+2}-2q^+(p_1^++p_2^+)+4(p_1^++p_2^+)^2
\over(p_1^++p_2^+)^3}\right]\bigg\}\nonumber\\&&
+{g^3\over12\pi^2}
{K^\wedge\over p_1^++p_2^+}-{g^3\over12\pi^2}(k_1+k_2
+k_0)^\wedge
\eea

\subsection{Swordfish Graphs}
We shall see that the swordfish graphs are nominally $O(\delta)$
and the integrals that define them give $O(1)$ only in the region
where all $T_i=O(\delta)$. Consequently, they are linear polynomials in
the transverse momenta. Specifically, there are three distinct graphs,
labeled according to the external leg attached to the cubic vertex:
\bea
\Gamma_{SF3}^{\wedge\wedge\vee}&=&-{g^3\over8\pi^2}\sum_{q^+}
{p_1^++p_2^+\over q^+(p_1^++p_2^+-q^+)}\left[
1-{(p_1^++q^+)(q^+-p_1^+-2p_2^+)\over(p_1^+-q^+)^2}\right]
(x_3k_2+(1-x_3)k_0)^\wedge\nonumber\\&&
\int dT{\delta\over (T+\delta)^2}
\exp\left\{-x_3(1-x_3)T(k_2-k_0)^2
-{\delta T\over T+\delta}(x_3{\bfs k}_2+(1-x_3){\bfs k}_0)^2
\right\}\\
&\to&-{g^3\over8\pi^2}\sum_{q^+}
{p_1^++p_2^+\over q^+(p_1^++p_2^+-q^+)}\left[
1-{(p_1^++q^+)(q^+-p_1^+-2p_2^+)\over(p_1^+-q^+)^2}\right]
(x_3k_2+(1-x_3)k_0)^\wedge\nonumber\\
&=&-{g^3\over8\pi^2}{1\over p_1^++p_2^+}\sum_{q^+}
{1\over x_3(1-x_3)}\left[
1-{(x_3+\eta)(x_3+\eta-2)\over(\eta-x_3)^2}\right]
(x_3k_2+(1-x_3)k_0)^\wedge
\eea
Here $x_3=q^+/(p_1^++p_2^+)$ and $\eta=p_1^+/(p_1^++p_2^+)$.
\bea
\Gamma_{SF1}^{\wedge\wedge\vee}&=&-{g^3\over8\pi^2}\sum_{q^+<p_1^+}
\bigg\{{q^+\over p_1^+(p_1^+-q^+)}\left[
-2-{(p_1^++p_2^++q^+)(q^+-p_1^++p_2^+)\over(p_1^++p_2^+-q^+)^2}
-{(2q^+-p_1^+)(p_1^++2p_2^+)\over p_1^{+2}}\right]\nonumber\\&&
+{p_1^+-q^+\over p_1^+q^+}\left[
1-{(2q^+-p_1^+)(p_1^++2p_2^+)\over p_1^{+2}}\right]\bigg\}
(x_1k_1+(1-x_1)k_0)^\wedge\nonumber\\&&
\int dT{\delta\over (T+\delta)^2}\exp\left\{-x_1(1-x_1)T(k_1-k_0)^2
-{\delta T\over T+\delta}(x_1{\bfs k}_1+(1-x_1){\bfs k}_0)^2
\right\}\\
&\to&-{g^3\over8\pi^2}\sum_{q^+<p_1^+}
\bigg\{{q^+\over p_1^+(p_1^+-q^+)}\left[
-3-{(p_1^++p_2^++q^+)(q^+-p_1^++p_2^+)\over(p_1^++p_2^+-q^+)^2}\right]
\nonumber\\&&
+\left[{q^+\over p_1^+(p_1^+-q^+)}+{p_1^+-q^+\over p_1^+q^+}\right]\left[
1-{(2q^+-p_1^+)(p_1^++2p_2^+)\over p_1^{+2}}\right]\bigg\}
(x_1k_1+(1-x_1)k_0)^\wedge\nonumber\\
&=&-{g^3\over8\pi^2}{1\over p_1^+}\sum_{q^+<p_1^+}
\bigg\{{x_1\over 1-x_1}\left[
-3-{(x_1+\eta^{-1})(x_1+\eta^{-1}-2)\over(\eta^{-1}-x_1)^2}\right]
\nonumber\\&&
+\left[{x_1\over 1-x_1}+{1-x_1\over x_1}\right]\left[
1-{(2x_1-1)(2\eta^{-1}-1)}\right]\bigg\}
(x_1k_1+(1-x_1)k_0)^\wedge
\eea
This time $x_1=q^+/p_1^+$.
\bea
\Gamma_{SF2}^{\wedge\wedge\vee}&=&-{g^3\over8\pi^2}\sum_{q^+>p_1^+}
\bigg\{{p_1^++p_2^+-q^+\over p_2^+(q^+-p_1^+)}\bigg[
-2-{(q^+-2p_1^+-2p_2^+)(q^+-2p_1^+)\over q^{+2}}\nonumber\\
&&\hskip3in-{(2p_1^++p_2^+-2q^+)(2p_1^++p_2^+)\over p_2^{+2}}\bigg]
\nonumber\\&&\hskip.75in
+{q^+-p_1^+\over p_2^+(p_1^++p_2^+-q^+)}\left[
1-{(2p_1^++p_2^+-2q^+)(2p_1^++p_2^+)\over p_2^{+2}}\right]\bigg\}
(x_2k_1+(1-x_2)k_2)^\wedge\nonumber\\&&
\int dT{\delta\over (T+\delta)^2}\exp\left\{-x_2(1-x_2)T(k_1-k_2)^2
-{\delta T\over T+\delta}(x_2{\bfs k}_1+(1-x_2){\bfs k}_2)^2
\right\}\\
&\to&-{g^3\over8\pi^2}\sum_{q^+>p_1^+}
\bigg\{{p_1^++p_2^+-q^+\over p_2^+(q^+-p_1^+)}\left[
-3-{(q^+-2p_1^+-2p_2^+)(q^+-2p_1^+)\over q^{+2}}\right]\nonumber\\&&
\hskip-.25in+\left[{p_1^++p_2^+-q^+\over p_2^+(q^+-p_1^+)}+{q^+-p_1^+\over p_2^+(p_1^++p_2^+-q^+)}\right]\left[
1-{(2p_1^++p_2^+-2q^+)(2p_1^++p_2^+)\over p_2^{+2}}\right]\bigg\}
(x_2k_1+(1-x_2)k_2)^\wedge\nonumber\\
&=&-{g^3\over8\pi^2}{1\over p_2^+}\sum_{q^+>p_1^+}
\bigg\{{x_2\over1-x_2}\left[
-3-{(x_2+(1-\eta)^{-1})(x_2+(1-\eta)^{-1}-2)\over 
(x_2-(1-\eta)^{-1})^{2}}\right]\nonumber\\&&
\hskip.25in+\left[{x_2\over1-x_2}+{1-x_2\over x_2}\right]\left[
1-{(2x_2-1)(2(1-\eta)^{-1}-1)}\right]\bigg\}
(x_2k_1+(1-x_2)k_2)^\wedge
\eea
Finally $x_2=(p_1^++p_2^+-q^+)/p_2^+$.
The arrows indicate the $\delta\to0$ limit of each result, which as
promised is seen to be linear in the transverse momenta,
though not simply proportional to $K^\wedge$. The dependence
on longitudinal momenta is far from simple. However these complicated
expressions combine nicely with the contributions of the 
the second terms of (\ref{klinear}) discussed at the end of the
previous subsection. Indeed collecting together all of the
singular terms from those and the swordfish diagrams shows
that they are all proportional to $K^\wedge$:
\bea
{\rm Singular}(\wedge\wedge\vee)&=&
{g^3\over8\pi^2}K^\wedge\bigg\{\sum_{q^+<p_1^+}\left[{2(p_1^++p_2^+)^2
\over p_1^{+2}(p_1^++p_2^+-q^+)^2}-{4(p_1^++p_2^+)
\over p_1^{+2}(p_1^++p_2^+-q^+)}-{2\over(q^+-p_1^+)^2}\right]\nonumber\\
&&\hskip1in+\sum_{q^+>p_1^+}\left[{2(p_1^++p_2^+)^2
\over p_2^{+2}q^{+2}}-{4(p_1^++p_2^+)
\over p_{2}^{+2}q^+}-{2\over(q^+-p_1^+)^2}\right]\bigg\}
\eea
The corresponding singular contributions for the other
spin configurations are
\bea
{\rm Singular}(\wedge\vee\wedge)&=&
{g^3\over8\pi^2}K^\wedge\bigg\{\sum_{q^+<p_1^+}\left[{2p_2^{+2}
\over p_1^{+2}(p_1^++p_2^+-q^+)^2}-{4p_2^+\quad??
\over p_1^{+2}(p_1^++p_2^+-q^+)}-{2p_2^{+2}\over p_3^{+2}
(q^+-p_1^+)^2}\right]\nonumber\\
&&-\sum_{q^+\neq p_1^+}{4p_2^+\over p_3^{+2}(p_1^+-q^+)}
+\sum_{q^+>p_1^+}\left[{2\over q^{+2}}
-{2p_2^{+2}\over p_3^{+2}((q^+-p_1^+)^2}\right]\bigg\}\\
{\rm Singular}(\vee\wedge\wedge)&=&
{g^3\over8\pi^2}K^\wedge\bigg\{\sum_{q^+<p_1^+}\left[{2
\over (p_1^++p_2^+-q^+)^2}-{2p_1^{+2}\over p_3^{+2}
(q^+-p_1^+)^2}\right]\nonumber\\
&&-\sum_{q^+\neq p_1^+}{4p_1^+\over p_3^{+2}(q^+-p_1^+)}
+\sum_{q^+>p_1^+}\left[{2p_1^{+2}
\over p_2^{+2}q^{+2}}-{4p_1^+
\over p_2^{+2}q^+}-{2p_1^{+2}\over p_3^{+2}(q^+-p_1^+)^2}\right]\bigg\}
\eea
In addition to these there are polynomials in $q^+$ which depend
separately on the ${\bfs k}_1,{\bfs k}_2,{\bfs k}_0$:
\bea
&&\hskip-.9cm{g^3\over8\pi^2}\sum_{q^+<p_1^+}\bigg[
k^{\prime\wedge}\left({2q^{+2}\over p_1^+(p_1^++p_2^+)^2}+{4\over p_1^+ }
\right)+k^\wedge\left({q^{+2}\over p_1^{+2}}
\left({2\over p_1^++p_2^+ }-{4(p_1^++2p_2^+)\over p_1^{+2}}\right)
-{4(p_1^++p_2^+)\over p_1^{+2} }+{4q^+p_2^+\over p_1^{+3} }\right)
\nonumber\\
&&+k^{\prime\prime\wedge}\left({q^{+2}\over p_1^{+2}}
\left({4(p_1^++2p_2^+)\over p_1^{+2}}-{2(2p_1^++p_2^+)\over (p_1^++p_2^+)^2
}\right)+{4q^{+}\over p_1^{+}}\left({1\over p_1^++p_2^+ }-{(p_1^++3p_2^+)
\over p_1^{+2}}\right)+{8p_2^+\over p_1^{+2}}\right)\bigg]
\eea
plus a sum over $q^+>p_1^+$ whose summand is obtained from the
above by the substitutions $q^+\to p_1^++p_2^+-q^+$,
${\bfs k}_2, p_1^+\leftrightarrow{\bfs k}_0, p_2^+$.
Since these summands are nonsingular it is safe to replace
the sums by integrals and perform them, after which spectacular simplification
takes place: 
\bea
{\rm Polynomial~Contribution}({\wedge\wedge\vee})\to
{g^3\over8\pi^2}\left[{14\over3}{p_1^++p_2^+\over p_1^+p_2^+}K^\wedge
-{2\over3}(k_1+k_2+k_0)^\wedge\right]
\eea
The coefficients break down as $14/3=26/3 - 4$, $-2/3=-4/3 +2/3$, with
the first terms  coming
from the swordfish graphs while the second ones  come from
the triangle graphs.
The only thing that changes in the analogous contribution for the other
spin configurations is the coefficient of $K^\wedge$, which
just matches the tree coefficient:
\bea
{\rm Polynomial~Contribution}({\wedge\vee\wedge})\to
{g^3\over8\pi^2}\left[{14\over3}{p_2^+\over p_1^+(p_1^++p_2^+)}K^\wedge
-{2\over3}(k_1+k_2+k_0)^\wedge\right]\\
{\rm Polynomial~Contribution}({\vee\wedge\wedge})\to
{g^3\over8\pi^2}\left[{14\over3}{p_1^+\over p_2^+(p_1^++p_2^+)}K^\wedge
-{2\over3}(k_1+k_2+k_0)^\wedge\right]
\eea
The second term in the square brackets of each of these
contributions is the only regularization 
artifact that will require a new counter-term, beyond the
usual coupling, self-energy, and wave function renormalization. 
It is spin independent and can be given
a local worldsheet interpretation if we simply rewrite it
in the form 
\bea
-{g^3\over12\pi^2}\left[(k_2^{\wedge}-k_1^\wedge)
+(k_0^{\wedge}-k_1^\wedge)+3k_1^\wedge\right]
\eea
The first two terms can be produced by appropriate insertions
of $\partial q^\wedge/\partial\sigma$ near the interaction point
on the worldsheet,
and the last term is already local since $k$ is the value
of $q^\wedge$ at the interaction point.

In summary the contribution of the swordfish diagrams
combined with the delta terms from the triangle diagrams
is given by
\bea
\Gamma^{\wedge\wedge\vee}_{\rm SF+}
&\to&{g^3\over8\pi^2}K^\wedge{p_1^++p_2^+\over p_1^+p_2^+}
\left[{26\over3}
-{4p_2^+\over p_1^{+}}\ln{p_1^++p_2^+\over p_2^+}
-{4p_1^+
\over p_2^{+}}\ln{p_1^++p_2^+\over p_1^+}\right]\nonumber\\&&\hskip1in
-{g^3\over8\pi^2}K^\wedge\sum_{q^+\neq p_1^+}{2\over(q^+-p_1^+)^2}
-{g^3\over12\pi^2}(k_1+k_2+k_0)^\wedge\label{swordfish3}\\
\Gamma^{\wedge\vee\wedge}_{\rm SF+}&\to&{g^3\over8\pi^2}K^\wedge
{p_2^+\over p_1^+(p_1^++p_2^+)}
\left[{26\over3}
-{4(p_1^++p_2^+)\over p_1^{+}}\ln{p_1^++p_2^+\over p_2^+}
-{4p_1^+
\over p_1^{+}+p_2^+}\ln{p_1^+\over p_2^+}\right]
\nonumber\\&&\hskip1in
-{g^3\over8\pi^2}K^\wedge
\sum_{q^+\neq p_1^+}{2p_2^{+2}\over p_3^{+2}(q^+-p_1^+)^2}
-{g^3\over12\pi^2}(k_1+k_2+k_0)^\wedge\label{swordfish2}\\
\Gamma^{\vee\wedge\wedge}_{\rm SF+}&\to&{g^3\over8\pi^2}K^\wedge
{p_1^+\over p_2^+(p_1^++p_2^+)}
\left[{26\over3}
-{4p_2^+
\over p_1^{+}+p_2^+}\ln{p_2^+\over p_1^+}
-{4(p_1^++p_2^+)
\over p_2^{+}}\ln{p_1^++p_2^+\over p_1^+}\right]
\nonumber\\&&\hskip1in
-{g^3\over8\pi^2}K^\wedge
\sum_{q^+\neq p_1^+}{2p_1^{+2}\over p_3^{+2}(q^+-p_1^+)^2}
-{g^3\over12\pi^2}(k_1+k_2+k_0)^\wedge
\label{swordfish1}
\eea
The arrows signify that the sums over discretized $q^+$ have been
replaced by integrals and performed wherever possible. The only
term where this is not possible is shown as a discretized sum.
As mentioned in the previous section, this term cancels a 
corresponding term in the triangle diagram calculation.

\subsection{Renormalization at One Loop}
When we studied wave function renormalization, we found that the
log divergence had a divergent $p^+$ dependent coefficient. But
then we found that this $p^+$ dependence was exactly canceled by
corresponding contributions from the triangle vertex corrections.
Thus we can drop the $p^+$ dependence and use the wave function 
renormalization constant (\ref{zee3}):
\bea
Z_3=1-{g^2N_c\over4\pi^2}\left({11\over6}\right)\ln(\mu^2\delta).
\eea 
We can similarly drop the $p^+$ dependent part of the log
divergence in the vertex renormalization and use the vertex
renormalization constant
\bea
{1\over Z_1}=1+{g^2N_c\over8\pi^2}\left({11\over3}\right)\ln(\mu^2\delta)
\eea

We can now write the relation between renormalized and bare
coupling
\bea
{g}_R&=&g{Z^{3/2}\over Z_1}\nonumber\\
&=&{g}\left(1+{g^2N_c\over8\pi^2}\left({11\over3}
-{3\over2}{11\over3}\right)\ln(\mu^2\delta)\right)\nonumber\\
&=&{g}\left(1-{11\over3}{g^2N_c\over16\pi^2}\ln(\mu^2\delta)\right)
\eea
and for the Callan-Symanzik beta function
\bea
\beta(g)\equiv\mu{dg_R\over d\mu}=-{11g^3N_c\over24\pi^2}+O(g^5). 
\label{largenbeta}
\eea
This is the known result for the beta function
for {\it planar} Yang-Mills field theory.
\subsection{Three gluon vertex contribution to scattering of Glue by Glue}
The one-loop three gluon vertex contribution to the four gluon
scattering amplitude requires putting two of the three gluons on shell,
and there are three distinct cases. First we put $p_1^2=p_3^2=0$,
so we have
\bea
H&\to& xyp_2^2,\qquad {\bfs K}^2\to p_1^{+}(p_1^++p_2^+)p_2^2\\
\Gamma_{\triangle-}^{\wedge\wedge\vee}&\to&-{g^3\over4\pi^2}
{p_3^+\over p_1^+p_2^+}K^\wedge
\sum_{q^+}\int_{x+y\leq1} dx dy
\delta(q^+-(x+y)p^+_1-yp^+_2)
\bigg\{\nonumber\\&&
p_1^+(p_1^++p_2^+)(1-x-y){A_1+A_2+A_3\over p_3^{+2}}\nonumber\\
&&-{p_3^+(yp_2^++(1-x-y)p_1^+)}\ln(\delta xyp_2^2e^{\gamma+1})
{A_3\over p_3^{+2}}
-{p_2^+(yp_3^++xp_1^+)}\ln(\delta xyp_2^2e^{\gamma+1})
{A_2\over p_3^{+2}}\nonumber\\&&
-{p_1^+((1-x-y)p_3^++xp_2^+)}\ln(\delta xyp_2^2e^{\gamma+1})
{A_1\over p_3^{+2}} \bigg\}\nonumber\\
&=&-{g^3\over4\pi^2}
{p_3^+\over p_1^+p_2^+}K^\wedge
\sum_{q^+}\int_0^{x_m} {dx\over p_1^++p_2^+} 
\bigg\{p_1^+(p_1^++p_2^+-q^+-xp_2^+)
{A_1+A_2+A_3\over p_3^{+2}}\\
&&\bigg[p_2^+q^+{A_2+A_3\over p_3^{+2}}
+p_1^+(p_1^++p_2^+-q^+){A_1+A_3\over p_3^{+2}}-2xp_1^+p_2^+
{A_1+A_2+A_3\over p_3^{+2}}\bigg]\ln(\delta xyp_2^2e^{\gamma+1})
\bigg\}\nonumber
\eea 
Formulas that enable the explicit evaluation of the $x,y$ integrals
are listed in an appendix.
Then 
\bea
\Gamma_{\triangle-}^{\wedge\wedge\vee}
&=&-{g^3\over4\pi^2}
{p_3^+\over p_1^+p_2^+}K^\wedge
\sum_{q^+<p_1^+} \bigg\{B_0
+B_1\left(\ln(\delta p_2^2e^{\gamma})+\ln{q^+\over p_1^+}\right)
\bigg\}\nonumber\\
&&-{g^3\over4\pi^2}
{p_3^+\over p_1^+p_2^+}K^\wedge
\sum_{q^+>p_1^+}\bigg\{B_0^\prime+B_1^\prime\left(\ln(\delta p_2^2e^{\gamma})
+\ln{p_1^++p_2^+-q^+\over p_2^+}\right)+B_2\ln{q^+-p_1^+\over p_2^+}
\bigg\}\nonumber\\
&&-{g^3\over4\pi^2}
{p_3^+\over p_1^+p_2^+}K^\wedge
\sum_{q^+\neq p_1^+} 
\bigg\{B_1\ln{q^+\over p_1^++p_2^+}\bigg\},
\quad{\rm for}~p_1^2=p_3^2=0
\eea
Where the $B_i$ are given by
\bea
B_0&=&{q^+\over p_1^++p_2^+}
\left(p_1^++p_2^+-q^+-{q^+\over2p_1^+}p_2^+\right){A_2\over p_3^{+2}}
-{q^{+2}p_2^+\over 2p_1^+(p_1^++p_2^+)}{A_3-A_1\over p_3^{+2}}\nonumber\\
B_1&=&{q^+(p_1^++p_2^+-q^+)\over p_1^++p_2^+}{A_3\over p_3^{+2}}
+{q^+(p_1^+-q^+)\over p_1^+}{A_1\over p_3^{+2}}\nonumber\\
B_0^\prime&=&{p_1^++p_2^+-q^+\over p_1^++p_2^+}
\left(q^+-{p_1^++p_2^+-q^+\over2p_2^+}p_1^+\right){A_1\over p_3^{+2}}
-{(p_1^++p_2^+-q^+)^2p_1^+\over 2p_2^+(p_1^++p_2^+)}
{A_3-A_2\over p_3^{+2}}\nonumber\\
B_1^\prime&=&{q^+(p_1^++p_2^+-q^+)\over p_1^++p_2^+}{A_3\over p_3^{+2}}
+{(p_1^++p_2^+-q^+)(q^+-p_1^+)\over p_2^+}{A_2\over p_3^{+2}}\nonumber\\
B_2&=&{q^+(q^+-p_1^+)\over p_1^+}{A_1\over p_3^{+2}}
+{(p_1^++p_2^+-q^+)(q^+-p_1^+)\over p_2^+}{A_2\over p_3^{+2}}
\eea
The various spin configurations are obtained by substituting the 
appropriate expressions for $A_1,A_2,A_3$:
\bea
B_0^{\wedge\wedge\vee}&=&{p_1^+p_2^+\over p_1^++p_2^+}
{1\over(q^+-p^+_1)^2}-{p^+_2\over2 p^+_1(p^+_1+p^+_2)^3}
{q^{+4}+(p_1^++p_2^+)^4+(p_1^++p_2^+-q^+)^4\over(p_1^++p_2^+-q^+)^2}\nonumber\\
B_0^{\prime\wedge\wedge\vee}&=&{p_1^+p_2^+\over p_1^++p_2^+}
{1\over(q^+-p^+_1)^2}-{p^+_1\over2 p^+_2(p^+_1+p^+_2)^3}
{q^{+4}+(p_1^++p_2^+)^4+(p_1^++p_2^+-q^+)^4\over q^{+2}}\nonumber\\
B_0^{\wedge\vee\wedge}&=&{p_2^{+2}\over(p_1^++p_2^+)^2}
\bigg[{p_1^+[p_2^++2(p_1^+-q^+)]\over(p_1^++p_2^+)(q^+-p^+_1)^2}
+p_1^+{(q^+-p_1^+)^2+2p_2^+(p_1^+-q^+)+3p_2^{+2}\over p_2^{+3}(p_1^++p_2^+)}
\nonumber\\
&&-{(p_1^++p_2^+)[(p_1^+-q^{+})^4+p_2^{+4}
+(p_1^++p_2^+-q^+)^4]\over2p_1^+p_2^{+3}(p_1^++p_2^+-q^+)^2}\bigg]\nonumber\\
B_0^{\prime\wedge\vee\wedge}&=&{p_2^{+2}\over(p_1^++p_2^+)^2}
\bigg[{p_1^+[p_2^++2(p_1^+-q^+)]\over(p_1^++p_2^+)(q^+-p^+_1)^2}
+p_1^+{(q^+-p_1^+)^2+2p_2^+(p_1^+-q^+)+3p_2^{+2}\over p_2^{+3}(p_1^++p_2^+)}
\nonumber\\
&&-{(p_1^++p_2^+)p_1^+\over p_2^+q^{+2}}\bigg]\nonumber\\
B_0^{\vee\wedge\wedge}&=&{p_1^{+2}\over(p_1^++p_2^+)^2}
\bigg[{p_2^+[p_1^++2(q^+-p_1^+)]\over(p_1^++p_2^+)(q^+-p^+_1)^2}
+p_2^+{(q^+-p_1^+)^2+2p_1^+(q^+-p_1^+)+3p_1^{+2}\over p_1^{+3}(p_1^++p_2^+)}
\nonumber\\
&&-{(p_1^++p_2^+)p_2^+\over p_1^+(p_1^++p_2^+-q^{+})^2}\bigg]\nonumber\\
B_0^{\prime\vee\wedge\wedge}&=&{p_1^{+2}\over(p_1^++p_2^+)^2}
\bigg[{p_2^+[p_1^++2(q^+-p_1^+)]\over(p_1^++p_2^+)(q^+-p^+_1)^2}
+p_2^+{(q^+-p_1^+)^2+2p_1^+(q^+-p_1^+)+3p_1^{+2}\over p_1^{+3}(p_1^++p_2^+)}
\nonumber\\
&&-{(p_1^++p_2^+)[(p_1^+-q^{+})^4+p_1^{+4}
+q^{+4}]\over2p_2^+p_1^{+3}q^{+2}}\bigg]
\eea
It is worthwhile to immediately integrate these rather unwieldy expressions
for $B_0, B_0^\prime$ and combine them with the corresponding
swordfish amplitude:
\bea
\Gamma^{\wedge\wedge\vee}_{\rm S.F.}-{g^3\over4\pi^2}
{p_3^+\over p_1^+p_2^+}K^\wedge
\left(\sum_{q^+<p_1^+}B_0+\sum_{q^+>p_1^+}B_0^\prime\right)
&=&{g^3\over12\pi^2}\left[-{p_3^+\over p_1^+p_2^+}K^\wedge
+{K^\wedge\over p_3^+}-(k_1+k_2+k_0)^\wedge\right]\nonumber\\
\Gamma^{\wedge\vee\wedge}_{\rm S.F.}-{g^3\over4\pi^2}
{p_3^+\over p_1^+p_2^+}K^\wedge
\left(\sum_{q^+<p_1^+}B_0+\sum_{q^+>p_1^+}B_0^\prime\right)
&=&{g^3\over12\pi^2}\left[-{p_2^+\over p_1^+p_3^+}K^\wedge
+{K^\wedge\over p_2^+}-(k_1+k_2+k_0)^\wedge\right]\nonumber\\
\Gamma^{\vee\wedge\wedge}_{\rm S.F.}-{g^3\over4\pi^2}
{p_3^+\over p_1^+p_2^+}K^\wedge
\left(\sum_{q^+<p_1^+}B_0+\sum_{q^+>p_1^+}B_0^\prime\right)
&=&{g^3\over12\pi^2}\left[-{p_1^+\over p_3^+p_2^+}K^\wedge
+{K^\wedge\over p_1^+}-(k_1+k_2+k_0)^\wedge\right]
\eea
Note here that one spin configuration can be obtained from 
another by suitably cycling the indices. As noted later $B_0, B_0^\prime$
enter the triangle amplitude in the same way for all choices of 
pairs of on-shell external lines. This will not be the case
for the other $B$'s.
\bea
B_1^{\wedge\wedge\vee}&=&{2\over q^+}
+{1\over p_1^++p_2^+-q^+}+{1\over p_1^+-q^+}+{2q^+(p_1^++p_2^+-q^+)
\over (p_1^++p_2^+)^3}-{4\over p_1^++p_2^+}\nonumber\\
B_1^{\wedge\vee\wedge}&=&{p_2^{+2}\over p_3^{+2}}
\left[{2\over q^+}
+{1\over p_1^++p_2^+-q^+}+{1\over p_1^+-q^+}\right]\nonumber\\
B_1^{\vee\wedge\wedge}&=&{p_1^{+2}\over p_3^{+2}}
\left[{2\over q^+}
+{1\over p_1^++p_2^+-q^+}+{1\over p_1^+-q^+}+{2q^+(p_1^+-q^+)
\over p_1^{+3}}-{4\over p_1^+}\right]\nonumber\\
B_1^{\prime\wedge\wedge\vee}&=&{1\over q^+}
+{2\over p_1^++p_2^+-q^+}+{1\over q^+-p_1^+}+{2q^+(p_1^++p_2^+-q^+)
\over (p_1^++p_2^+)^3}-{4\over p_1^++p_2^+}\nonumber\\
B_1^{\prime\wedge\vee\wedge}&=&{p_2^{+2}\over p_3^{+2}}
\left[{1\over q^+}
+{2\over p_1^++p_2^+-q^+}+{1\over p_1^+-q^+}+{2(q^+-p_1^+)(p_1^++p_2^+-q^+)
\over p_2^{+3}}-{4\over p_2^+}\right]\nonumber\\
B_1^{\prime\vee\wedge\wedge}&=&{p_1^{+2}\over p_3^{+2}}
\left[{1\over q^+}
+{2\over p_1^++p_2^+-q^+}+{1\over q^+-p_1^+}\right]\nonumber\\
B_2^{\wedge\wedge\vee}&=&{2\over q^+-p_1^+}
+{1\over p_1^++p_2^+-q^+}-{1\over q^+}\nonumber\\
B_2^{\wedge\vee\wedge}&=&{p_2^{+2}\over p_3^{+2}}
\left[{2\over q^+-p_1^+}
+{1\over p_1^++p_2^+-q^+}-{1\over q^+}+{2(q^+-p_1^+)(p_1^++p_2^+-q^+)
\over p_2^{+3}}-{4\over p_2^+}\right]\nonumber\\
B_2^{\vee\wedge\wedge}&=&{p_1^{+2}\over p_3^{+2}}
\left[{2\over q^+-p_1^+}
+{1\over p_1^++p_2^+-q^+}-{1\over q^+}+{2(q^+-p_1^+)q^+
\over p_1^{+3}}+{4\over p_1^+}\right]
\eea
%%%%%%

\bea
\Gamma_{\triangle}^{\wedge\wedge\vee}
&=&\Gamma_{\triangle-}^{\wedge\wedge\vee}
+\Gamma_{{\rm S.F.+}}^{\wedge\wedge\vee}\nonumber\\
&=&{g^3\over12\pi^2}\left(-{p_3^+\over p_1^+p_2^+}K^\wedge
-{K^\wedge\over p_1^++p_2^+}-(k_1+k_2+k_0)^\wedge\right)
\nonumber\\
&&-{g^3\over4\pi^2}{p_3^+\over p_1^+p_2^+}K^\wedge\bigg(
\sum_{q^+\neq p_1^+}
%\nonumber\\ &&
B_1\ln{q^+\over p_1^++p_2^+}+\sum_{q^+<p_1^+}  
B_1\left(\ln(\delta p_2^2e^{\gamma})+\ln{q^+\over p_1^+}\right)
\nonumber\\
&&\hskip0in+\sum_{q^+>p_1^+}\bigg\{
B_1^\prime\left(\ln(\delta p_2^2e^{\gamma})
+\ln{p_1^++p_2^+-q^+\over p_2^+}\right)
+B_2\ln{q^+-p_1^+\over p_2^+}\bigg\}\bigg),
\quad{\rm for}~p_1^2=p_3^2=0
\eea
The final simplification step is to explicitly evaluate the
$q^+$ summation for the polynomial parts of the $B$'s:
\bea
\Gamma_{\triangle}^{\wedge\wedge\vee}
&=&{g^3\over12\pi^2}\left[-{70\over3}{p_3^+\over p_1^+p_2^+}K^\wedge
-(k_1+k_2+k_0)^\wedge\right]
-{g^3\over4\pi^2}{p_3^+\over p_1^+p_2^+}K^\wedge\bigg(-{11\over3}
\ln(\delta p_2^2e^{\gamma})
\nonumber\\
&&+\sum_{q^+\neq p_1^+}
%\nonumber\\ &&
\left[{2\over q^+}+{1\over p_1^++p_2^+-q^+}+{1\over p_1^+-q^+}\right]
\ln{q^+\over p_1^++p_2^+}\nonumber\\
&&+\sum_{q^+<p_1^+}  
\bigg\{\bigg[{2\over q^+}+{1\over p_1^++p_2^+-q^+}+{1\over p_1^+-q^+}
\bigg]\left(\ln(\delta p_2^2e^{\gamma})+\ln{q^+\over p_1^+}\right)
\bigg\}\nonumber\\
&&+\sum_{q^+>p_1^+}\bigg\{
\bigg[{1\over q^+}+{2\over p_1^++p_2^+-q^+}+{1\over q^+-p_1^+}
\bigg]\left(\ln(\delta p_2^2e^{\gamma})
+\ln{p_1^++p_2^+-q^+\over p_2^+}\right)
\nonumber\\
&&\qquad\qquad
+\bigg[{2\over q^+-p_1^+}+{1\over p_1^++p_2^+-q^+}-{1\over q^+}
\bigg]
\ln{q^+-p_1^+\over p_2^+}\bigg\}\bigg),
\quad{\rm for}~p_1^2=p_3^2=0
\eea
For the other two spin configurations we merely quote the final answers:
\bea
\Gamma_{\triangle}^{\wedge\vee\wedge}
&=&{g^3\over12\pi^2}\left[-{70\over3}{p_2^+\over p_1^+p_3^+}K^\wedge
+{K^\wedge\over p_2^+}
-(k_1+k_2+k_0)^\wedge\right]
-{g^3\over4\pi^2}{p_2^+\over p_1^+p_3^+}K^\wedge\bigg(-{11\over3}
\ln(\delta p_2^2e^{\gamma})
\nonumber\\
&&+\sum_{q^+\neq p_1^+}
%\nonumber\\ &&
\left[{2\over q^+}+{1\over p_1^++p_2^+-q^+}+{1\over p_1^+-q^+}\right]
\ln{q^+\over p_1^++p_2^+}\nonumber\\
&&+\sum_{q^+<p_1^+}  
\bigg\{\bigg[{2\over q^+}+{1\over p_1^++p_2^+-q^+}+{1\over p_1^+-q^+}
\bigg]\left(\ln(\delta p_2^2e^{\gamma})+\ln{q^+\over p_1^+}\right)
\bigg\}\nonumber\\
&&+\sum_{q^+>p_1^+}\bigg\{
\bigg[{1\over q^+}+{2\over p_1^++p_2^+-q^+}+{1\over q^+-p_1^+}
\bigg]\left(\ln(\delta p_2^2e^{\gamma})
+\ln{p_1^++p_2^+-q^+\over p_2^+}\right)
\nonumber\\
&&\qquad\qquad
+\bigg[{2\over q^+-p_1^+}+{1\over p_1^++p_2^+-q^+}-{1\over q^+}
\bigg]
\ln{q^+-p_1^+\over p_2^+}\bigg\}\bigg)\nonumber\\
\Gamma_{\triangle}^{\vee\wedge\wedge}
&=&{g^3\over12\pi^2}\left[-{70\over3}{p_1^+\over p_2^+p_3^+}K^\wedge
-(k_1+k_2+k_0)^\wedge\right]
-{g^3\over4\pi^2}{p_1^+\over p_2^+p_3^+}K^\wedge\bigg(-{11\over3}
\ln(\delta p_2^2e^{\gamma})
\nonumber\\
&&+\sum_{q^+\neq p_1^+}
%\nonumber\\ &&
\left[{2\over q^+}+{1\over p_1^++p_2^+-q^+}+{1\over p_1^+-q^+}\right]
\ln{q^+\over p_1^++p_2^+}\nonumber\\
&&+\sum_{q^+<p_1^+}  
\bigg\{\bigg[{2\over q^+}+{1\over p_1^++p_2^+-q^+}+{1\over p_1^+-q^+}
\bigg]\left(\ln(\delta p_2^2e^{\gamma})+\ln{q^+\over p_1^+}\right)
\bigg\}\nonumber\\
&&+\sum_{q^+>p_1^+}\bigg\{
\bigg[{1\over q^+}+{2\over p_1^++p_2^+-q^+}+{1\over q^+-p_1^+}
\bigg]\left(\ln(\delta p_2^2e^{\gamma})
+\ln{p_1^++p_2^+-q^+\over p_2^+}\right)
\nonumber\\
&&\qquad\qquad
+\bigg[{2\over q^+-p_1^+}+{1\over p_1^++p_2^+-q^+}-{1\over q^+}
\bigg]
\ln{q^+-p_1^+\over p_2^+}\bigg\}\bigg),
\quad{\rm for}~p_1^2=p_3^2=0
\eea
We note that apart from suitable relabeling of indices in passing from 
one spin configuration to another, there is a breaking of the cyclic symmetry
through putting legs $1,3$ on shell. The term $K/p_i^+$ in the
square brackets on first line of each case is uncanceled when the
on-shell lines have like helicity and canceled otherwise.

Next we choose the on-shell pair $p_1^2=p_2^2=0$ and obtain
\bea
H&\to& y(1-x-y)p_3^2,\qquad {\bfs K}^2\to -p_1^{+}p_2^+p_3^2\\
\Gamma_{\triangle-}^{\wedge\wedge\vee}&\to&-{g^3\over4\pi^2}
{p_3^+\over p_1^+p_2^+}K^\wedge
\sum_{q^+}\int_{x+y\leq1} dx dy
\delta(q^+-(x+y)p^+_1-yp^+_2)
\bigg\{-p_1^+p_2^+x{A_1+A_2+A_3\over p_3^{+2}}\nonumber\\
&&-{p_3^+(yp_2^++(1-x-y)p_1^+)}\ln(\delta y(1-x-y)p_3^2e^{\gamma+1})
{A_3\over p_3^{+2}}\nonumber\\
&&
-{p_2^+(yp_3^++xp_1^+)}\ln(\delta y(1-x-y)p_3^2e^{\gamma+1})
{A_2\over p_3^{+2}}\nonumber\\&&
-{p_1^+((1-x-y)p_3^++xp_2^+)}\ln(\delta y(1-x-y)p_3^2e^{\gamma+1})
{A_1\over p_3^{+2}} \bigg\}\nonumber\\
&=&-{g^3\over4\pi^2}
{p_3^+\over p_1^+p_2^+}K^\wedge
\sum_{q^+}\int_0^{x_m} {dx\over p_1^++p_2^+} 
\bigg\{-p_1^+p_2^+x
{A_1+A_2+A_3\over p_3^{+2}}\nonumber\\
&&\hskip-.75in+\bigg[p_2^+q^+{A_2+A_3\over p_3^{+2}}
+p_1^+(p_1^++p_2^+-q^+){A_1+A_3\over p_3^{+2}}-2xp_1^+p_2^+
{A_1+A_2+A_3\over p_3^{+2}}\bigg]\ln(\delta y(1-x-y)p_3^2e^{\gamma+1})
\bigg\}
\eea
Integrating over $x,y$ then yields
\bea
\Gamma_{\triangle-}^{\wedge\wedge\vee}
&=&-{g^3\over4\pi^2}
{p_3^+\over p_1^+p_2^+}K^\wedge
\sum_{q^+<p_1^+} \bigg\{B_0+B_1\ln(\delta p_3^2e^{\gamma})
-B_2\ln{p_1^+-q^+\over p_1^+}\bigg\}\nonumber\\
&&-{g^3\over4\pi^2}
{p_3^+\over p_1^+p_2^+}K^\wedge
\sum_{q^+>p_1^+} \bigg\{B_0^\prime
+B_1^\prime\ln(\delta p_3^2e^{\gamma})
+B_2\ln{q^+-p_1^+\over p_2^+}\bigg\}\nonumber\\
&&-{g^3\over4\pi^2}
{p_3^+\over p_1^+p_2^+}K^\wedge
\sum_{q^+\neq p_1^+} 
\bigg\{B_1^\prime\ln{p_1^++p_2^+-q^+\over p_1^++p_2^+}
+B_1\ln{q^+\over p_1^++p_2^+}\bigg\},
\quad{\rm for}~p_1^2=p_2^2=0
\eea

Adding in the swordfish diagram and simplifying
\bea
\Gamma_{\triangle}^{\wedge\wedge\vee}
&=&\Gamma_{\triangle-}^{\wedge\wedge\vee}
+\Gamma_{{\rm S.F.+}}^{\wedge\wedge\vee}\nonumber\\
&=&-{g^3\over12\pi^2}(k_1+k_2+k_0)^\wedge
-{g^3\over4\pi^2}{p_3^+\over p_1^+p_2^+}K^\wedge\bigg({70\over9}
-{p_1^+p_2^+\over 3p_3^{+2}}-{11\over3}
\ln(\delta p_3^2e^{\gamma})
\nonumber\\
&&+\sum_{q^+<p_1^+}  
\bigg\{
\bigg[{2\over q^+}+{1\over p_1^++p_2^+-q^+}+{1\over p_1^+-q^+}
\bigg]\left(\ln(\delta p_3^2e^{\gamma})
+\ln{q^+\over p_1^++p_2^+}\right)\nonumber\\
&&+\bigg[{1\over q^+}+{2\over p_1^++p_2^+-q^+}+{1\over q^+-p_1^+}
\bigg]\ln{p_1^++p_2^+-q^+\over p_1^++p_2^+}\nonumber\\
&&+\bigg[{2\over p_1^+-q^+}-{1\over p_1^++p_2^+-q^+}+{1\over q^+}
\bigg]\ln{p_1^+-q^+\over p_1^+}\bigg\}\nonumber\\
&&+\sum_{q^+>p_1^+} \bigg\{
\bigg[{1\over q^+}+{2\over p_1^++p_2^+-q^+}+{1\over q^+-p_1^+}
\bigg]\left(\ln(\delta p_3^2e^{\gamma})
+\ln{p_1^++p_2^+-q^+\over p_1^++p_2^+}\right)
\nonumber\\
&&+\bigg[{2\over q^+}+{1\over p_1^++p_2^+-q^+}+{1\over p_1^+-q^+}
\bigg]\ln{q^+\over p_1^++p_2^+}
\nonumber\\ &&
+\bigg[{2\over q^+-p_1^+}+{1\over p_1^++p_2^+-q^+}-{1\over q^+}
\bigg]\ln{q^+-p_1^+\over p_2^+}\bigg\}\bigg), \qquad{\rm for}~p_1^2=p_2^2=0
\eea
The other spin configurations are
\bea
\Gamma_{\triangle}^{\wedge\vee\wedge}
&=&-{g^3\over12\pi^2}(k_1+k_2+k_0)^\wedge
-{g^3\over4\pi^2}{p_2^+\over p_1^+p_3^+}K^\wedge\bigg({70\over9}
-{11\over3}\ln(\delta p_3^2e^{\gamma})
\nonumber\\
&&+\sum_{q^+<p_1^+}  
\bigg\{
\bigg[{2\over q^+}+{1\over p_1^++p_2^+-q^+}+{1\over p_1^+-q^+}
\bigg]\left(\ln(\delta p_3^2e^{\gamma})
+\ln{q^+\over p_1^++p_2^+}\right)\nonumber\\
&&+\bigg[{1\over q^+}+{2\over p_1^++p_2^+-q^+}+{1\over q^+-p_1^+}
\bigg]\ln{p_1^++p_2^+-q^+\over p_1^++p_2^+}\nonumber\\
&&+\bigg[{2\over p_1^+-q^+}-{1\over p_1^++p_2^+-q^+}+{1\over q^+}
\bigg]\ln{p_1^+-q^+\over p_1^+}\bigg\}\nonumber\\
&&+\sum_{q^+>p_1^+} \bigg\{
\bigg[{1\over q^+}+{2\over p_1^++p_2^+-q^+}+{1\over q^+-p_1^+}
\bigg]\left(\ln(\delta p_3^2e^{\gamma})
+\ln{p_1^++p_2^+-q^+\over p_1^++p_2^+}\right)
\nonumber\\
&&+\bigg[{2\over q^+}+{1\over p_1^++p_2^+-q^+}+{1\over p_1^+-q^+}
\bigg]\ln{q^+\over p_1^++p_2^+}
\nonumber\\ &&
+\bigg[{2\over q^+-p_1^+}+{1\over p_1^++p_2^+-q^+}-{1\over q^+}
\bigg]\ln{q^+-p_1^+\over p_2^+}\bigg\}\bigg), \qquad{\rm for}~p_1^2=p_2^2=0\\
\Gamma_{\triangle}^{\vee\wedge\wedge}
&=&-{g^3\over12\pi^2}(k_1+k_2+k_0)^\wedge
-{g^3\over4\pi^2}{p_1^+\over p_2^+p_3^+}K^\wedge\bigg({70\over9}
-{11\over3}\ln(\delta p_3^2e^{\gamma})
\nonumber\\
&&+\sum_{q^+<p_1^+}  
\bigg\{
\bigg[{2\over q^+}+{1\over p_1^++p_2^+-q^+}+{1\over p_1^+-q^+}
\bigg]\left(\ln(\delta p_3^2e^{\gamma})
+\ln{q^+\over p_1^++p_2^+}\right)\nonumber\\
&&+\bigg[{1\over q^+}+{2\over p_1^++p_2^+-q^+}+{1\over q^+-p_1^+}
\bigg]\ln{p_1^++p_2^+-q^+\over p_1^++p_2^+}\nonumber\\
&&+\bigg[{2\over p_1^+-q^+}-{1\over p_1^++p_2^+-q^+}+{1\over q^+}
\bigg]\ln{p_1^+-q^+\over p_1^+}\bigg\}\nonumber\\
&&+\sum_{q^+>p_1^+} \bigg\{
\bigg[{1\over q^+}+{2\over p_1^++p_2^+-q^+}+{1\over q^+-p_1^+}
\bigg]\left(\ln(\delta p_3^2e^{\gamma})
+\ln{p_1^++p_2^+-q^+\over p_1^++p_2^+}\right)
\nonumber\\
&&+\bigg[{2\over q^+}+{1\over p_1^++p_2^+-q^+}+{1\over p_1^+-q^+}
\bigg]\ln{q^+\over p_1^++p_2^+}
\nonumber\\ &&
+\bigg[{2\over q^+-p_1^+}+{1\over p_1^++p_2^+-q^+}-{1\over q^+}
\bigg]\ln{q^+-p_1^+\over p_2^+}\bigg\}\bigg), \qquad{\rm for}~p_1^2=p_2^2=0
\eea
Finally, we choose $p_2^2=p_3^2=0$, 
\bea
H&\to& x(1-x-y)p_1^2,\qquad {\bfs K}^2\to p_2^{+}(p_1^++p_2^+)p_2^2\\
\Gamma_{\triangle-}^{\wedge\wedge\vee}&\to&-{g^3\over4\pi^2}
{p_3^+\over p_1^+p_2^+}K^\wedge
\sum_{q^+}\int_{x+y\leq1} dx dy
\delta(q^+-(x+y)p^+_1-yp^+_2)
\bigg\{p_2^+(p_1^++p_2^+)y{A_1+A_2+A_3\over p_3^{+2}}\nonumber\\
&&-{p_3^+(yp_2^++(1-x-y)p_1^+)}\ln(\delta x(1-x-y)p_1^2e^{\gamma+1})
{A_3\over p_3^{+2}}\nonumber\\
&&
-{p_2^+(yp_3^++xp_1^+)}\ln(\delta x(1-x-y)p_1^2e^{\gamma+1})
{A_2\over p_3^{+2}}\nonumber\\&&
-{p_1^+((1-x-y)p_3^++xp_2^+)}\ln(\delta x(1-x-y)p_1^2e^{\gamma+1})
{A_1\over p_3^{+2}} \bigg\}\nonumber\\
&=&-{g^3\over4\pi^2}
{p_3^+\over p_1^+p_2^+}K^\wedge
\sum_{q^+}\int_0^{x_m} {dx\over p_1^++p_2^+} 
\bigg\{p_2^+(q^+-xp_1^+)
{A_1+A_2+A_3\over p_3^{+2}}\nonumber\\
&&\hskip-.75in+\bigg[p_2^+q^+{A_2+A_3\over p_3^{+2}}
+p_1^+(p_1^++p_2^+-q^+){A_1+A_3\over p_3^{+2}}-2xp_1^+p_2^+
{A_1+A_2+A_3\over p_3^{+2}}\bigg]\ln(\delta x(1-x-y)p_1^2e^{\gamma+1})
\bigg\}
\eea 

\bea
\Gamma_{\triangle-}^{\wedge\wedge\vee}
&=&-{g^3\over4\pi^2}
{p_3^+\over p_1^+p_2^+}K^\wedge
\sum_{q^+<p_1^+} \bigg\{B_0+B_1\left(\ln(\delta p_1^2e^{\gamma})
+\ln{q^+\over p_1^+}\right)-B_2\ln{p_1^+-q^+\over p_1^+}
\bigg\}\nonumber\\
&&-{g^3\over4\pi^2}
{p_3^+\over p_1^+p_2^+}K^\wedge
\sum_{q^+>p_1^+} \bigg\{B_0^\prime+B_1^\prime\left(\ln(\delta p_1^2e^{\gamma})
+\ln{p_1^++p_2^+-q^+\over p_2^+}\right)\bigg\}\nonumber\\
&&-{g^3\over4\pi^2}
{p_3^+\over p_1^+p_2^+}K^\wedge
\sum_{q^+\neq p_1^+} \bigg\{
B_1^\prime\ln{p_1^++p_2^+-q^+\over p_1^++p_2^+}\bigg\},
\quad{\rm for}~p_2^2=p_3^2=0
\eea
Adding in the swordfish diagram and simplifying gives
\bea
\Gamma_{\triangle}^{\wedge\wedge\vee}
&=&\Gamma_{\triangle-}^{\wedge\wedge\vee}
+\Gamma_{{\rm S.F.+}}^{\wedge\wedge\vee}\nonumber\\
&=&-{g^3\over12\pi^2}(k_1+k_2+k_0)^\wedge
-{g^3\over4\pi^2}{p_3^+\over p_1^+p_2^+}K^\wedge\bigg({70\over9}-{11\over3}
\ln(\delta p_1^2e^{\gamma})
\nonumber\\
&&+\sum_{q^+<p_1^+}  
\bigg\{
\bigg[{2\over q^+}+{1\over p_1^++p_2^+-q^+}+{1\over p_1^+-q^+}
\bigg]\left(\ln(\delta p_1^2e^{\gamma})
+\ln{q^+\over p_1^+}\right)\nonumber\\
&&+\bigg[{2\over p_1^+-q^+}-{1\over p_1^++p_2^+-q^+}+{1\over q^+}
\bigg]\ln{p_1^+-q^+\over p_1^+}\bigg\}\nonumber\\
&&+\sum_{q^+>p_1^+} \bigg\{
\bigg[{1\over q^+}+{2\over p_1^++p_2^+-q^+}+{1\over q^+-p_1^+}
\bigg]\left(\ln(\delta p_1^2e^{\gamma})
+\ln{p_1^++p_2^+-q^+\over p_2^+}\right)
\nonumber\\
&&+\sum_{q^+\neq p_1^+}\bigg[{1\over q^+}+{2\over p_1^++p_2^+-q^+}
+{1\over q^+-p_1^+}
\bigg]\ln{p_1^++p_2^+-q^+\over p_1^++p_2^+}
\bigg\}\bigg), \qquad{\rm for}~p_2^2=p_3^2=0
\eea
The other two spin configurations are then:
\bea
\Gamma_{\triangle}^{\wedge\vee\wedge}
&=&-{g^3\over12\pi^2}
(k_1+k_2+k_0)^\wedge
-{g^3\over4\pi^2}{p_2^+\over p_1^+p_3^+}K^\wedge\bigg({70\over9}-{11\over3}
\ln(\delta p_1^2e^{\gamma})
\nonumber\\
&&+\sum_{q^+<p_1^+}  
\bigg\{
\bigg[{2\over q^+}+{1\over p_1^++p_2^+-q^+}+{1\over p_1^+-q^+}
\bigg]\left(\ln(\delta p_1^2e^{\gamma})
+\ln{q^+\over p_1^+}\right)\nonumber\\
&&+\bigg[{2\over p_1^+-q^+}-{1\over p_1^++p_2^+-q^+}+{1\over q^+}
\bigg]\ln{p_1^+-q^+\over p_1^+}\bigg\}\nonumber\\
&&+\sum_{q^+>p_1^+} \bigg\{
\bigg[{1\over q^+}+{2\over p_1^++p_2^+-q^+}+{1\over q^+-p_1^+}
\bigg]\left(\ln(\delta p_1^2e^{\gamma})
+\ln{p_1^++p_2^+-q^+\over p_2^+}\right)
\nonumber\\
&&+\sum_{q^+\neq p_1^+}\bigg[{1\over q^+}+{2\over p_1^++p_2^+-q^+}
+{1\over q^+-p_1^+}
\bigg]\ln{p_1^++p_2^+-q^+\over p_1^++p_2^+}
\bigg\}\bigg), \qquad{\rm for}~p_2^2=p_3^2=0\\
\Gamma_{\triangle}^{\vee\wedge\wedge}
&=&-{g^3\over12\pi^2}
(k_1+k_2+k_0)^\wedge
-{g^3\over4\pi^2}{p_1^+\over p_2^+p_3^+}K^\wedge\bigg({70\over9}
-{p_2^+p_3^+\over3 p_1^{+2}}-{11\over3}
\ln(\delta p_1^2e^{\gamma})
\nonumber\\
&&+\sum_{q^+<p_1^+}  
\bigg\{
\bigg[{2\over q^+}+{1\over p_1^++p_2^+-q^+}+{1\over p_1^+-q^+}
\bigg]\left(\ln(\delta p_1^2e^{\gamma})
+\ln{q^+\over p_1^+}\right)\nonumber\\
&&+\bigg[{2\over p_1^+-q^+}-{1\over p_1^++p_2^+-q^+}+{1\over q^+}
\bigg]\ln{p_1^+-q^+\over p_1^+}\bigg\}\nonumber\\
&&+\sum_{q^+>p_1^+} \bigg\{
\bigg[{1\over q^+}+{2\over p_1^++p_2^+-q^+}+{1\over q^+-p_1^+}
\bigg]\left(\ln(\delta p_1^2e^{\gamma})
+\ln{p_1^++p_2^+-q^+\over p_2^+}\right)
\nonumber\\
&&+\sum_{q^+\neq p_1^+}\bigg[{1\over q^+}+{2\over p_1^++p_2^+-q^+}
+{1\over q^+-p_1^+}
\bigg]\ln{p_1^++p_2^+-q^+\over p_1^++p_2^+}
\bigg\}, \qquad{\rm for}~p_2^2=p_3^2=0
\eea
All of the triangle amplitudes listed in this subsection
are appropriate to two incoming and one outgoing particle,
$p_1^+,p_2^+>0$. We get the case of two outgoing particles by
applying the dictionary (\ref{signflips}). In particular when
we assemble the triangle contributions to the scattering of glue
by glue, there are four contributing diagrams in which, respectively,
the gluon pairs $(1,2), (2,3), (3,4), (4,1)$ hook onto the
triangle sub-diagram. If $p_1^++p_4^+>0$ the first two have two incoming
gluons and the last two have two outgoing gluons. When $p_1^++p_4^+<0$
it is the first and last which have two incoming gluons.

As an example take the triangle diagram attached to gluons $(34)$,
with $p_3^+,p_4^+<0$.
Then, for example,
\bea
&&\hskip-.7in\Gamma^{\vee\wedge\wedge}(p_{12},p_3,p_4; k_0, k_2, k_3)=
\Gamma^{\wedge\wedge\vee}(-p_4,-p_3, -p_{12}; k_0, k_3, k_2)\nonumber\\
&&\hskip-.2in=-{g^3\over12\pi^2}(k_3+k_2+k_0)^\wedge
-{g^3\over4\pi^2}{p_{12}^+\over p_3^+p_4^+}K_{34}^\wedge\bigg({70\over9}
-{p_3^+p_4^+\over 3p_{12}^{+2}}-{11\over3}
\ln(\delta p_{12}^2e^{\gamma})
\nonumber\\
&&+\sum_{q^+<|p_4^+|}  
\bigg\{
\bigg[{2\over q^+}+{1\over p_1^++p_2^+-q^+}+{1\over |p_4^+|-q^+}
\bigg]\left(\ln(\delta p_{12}^2e^{\gamma})
+\ln{q^+\over p_1^++p_2^+}\right)\nonumber\\
&&+\bigg[{1\over q^+}+{2\over p_1^++p_2^+-q^+}+{1\over q^+-|p_4^+|}
\bigg]\ln{p_1^++p_2^+-q^+\over p_1^++p_2^+}\nonumber\\
&&+\bigg[{2\over |p_4^+|-q^+}-{1\over p_1^++p_2^+-q^+}+{1\over q^+}
\bigg]\ln{|p_4^+|-q^+\over |p_4^+|}\bigg\}\nonumber\\
&&+\sum_{q^+>|p_4^+|} \bigg\{
\bigg[{1\over q^+}+{2\over p_1^++p_2^+-q^+}+{1\over q^+-|p_4^+|}
\bigg]\left(\ln(\delta p_{12}^2e^{\gamma})
+\ln{p_1^++p_2^+-q^+\over p_1^++p_2^+}\right)
\nonumber\\
&&+\bigg[{2\over q^+}+{1\over p_1^++p_2^+-q^+}+{1\over |p_4^+|-q^+}
\bigg]\ln{q^+\over p_1^++p_2^+}
\nonumber\\ &&
+\bigg[{2\over q^+-|p_4^+|}+{1\over p_1^++p_2^+-q^+}-{1\over q^+}
\bigg]\ln{q^+-|p_4^+|\over |p_3^+|}\bigg\}\bigg), \qquad{\rm for}~p_3^2=p_4^2=0
\eea
In summary, all of the 2 like-helicity 
one loop cubic vertices can be put in the
form
\bea
\Gamma_{\rm 1~loop}=-{g^3\over12\pi^2}\sum_i k_i-{g^2\over8\pi^2}
\Gamma_{\rm tree}\left({70\over9}
-{11\over3}\ln(\delta p_o^2e^{\gamma})+S\right)
+\alpha{g^3\over12\pi^2}{K\over p_o^+}
\eea
where the vectors $k_i,K$ carry the polarization of the two like-
helicity gluons, $p_o$ is the four-momentum of the off-shell gluon,
and $\alpha=1$ when the on-shell gluons have like-helicity,
and $\alpha=0$ otherwise. Finally $S$
is an infrared sensitive term that depends on the location of
the off-shell gluon, but not on any of the gluon helicities. In the
case $p_1^+, p_2^+>0$:
\bea
S_1&=&\sum_{q^+<p_1^+}  
\bigg\{
\bigg[{2\over q^+}+{1\over p_1^++p_2^+-q^+}+{1\over p_1^+-q^+}
\bigg]\left(\ln(\delta p_1^2e^{\gamma})
+\ln{q^+\over p_1^+}\right)\nonumber\\
&&+\bigg[{2\over p_1^+-q^+}-{1\over p_1^++p_2^+-q^+}+{1\over q^+}
\bigg]\ln{p_1^+-q^+\over p_1^+}\bigg\}\nonumber\\
&&+\sum_{q^+>p_1^+} \bigg\{
\bigg[{1\over q^+}+{2\over p_1^++p_2^+-q^+}+{1\over q^+-p_1^+}
\bigg]\left(\ln(\delta p_1^2e^{\gamma})
+\ln{p_1^++p_2^+-q^+\over p_2^+}\right)
\nonumber\\
&&+\sum_{q^+\neq p_1^+}\bigg[{1\over q^+}+{2\over p_1^++p_2^+-q^+}
+{1\over q^+-p_1^+}
\bigg]\ln{p_1^++p_2^+-q^+\over p_1^++p_2^+}\\
S_2&=&\sum_{q^+\neq p_1^+}
%\nonumber\\ &&
\left[{2\over q^+}+{1\over p_1^++p_2^+-q^+}+{1\over p_1^+-q^+}\right]
\ln{q^+\over p_1^++p_2^+}\nonumber\\
&&+\sum_{q^+<p_1^+}  
\bigg\{\bigg[{2\over q^+}+{1\over p_1^++p_2^+-q^+}+{1\over p_1^+-q^+}
\bigg]\left(\ln(\delta p_2^2e^{\gamma})+\ln{q^+\over p_1^+}\right)
\bigg\}\nonumber\\
&&+\sum_{q^+>p_1^+}\bigg\{
\bigg[{1\over q^+}+{2\over p_1^++p_2^+-q^+}+{1\over q^+-p_1^+}
\bigg]\left(\ln(\delta p_2^2e^{\gamma})
+\ln{p_1^++p_2^+-q^+\over p_2^+}\right)
\nonumber\\
&&\qquad\qquad
+\bigg[{2\over q^+-p_1^+}+{1\over p_1^++p_2^+-q^+}-{1\over q^+}
\bigg]
\ln{q^+-p_1^+\over p_2^+}\bigg\}\\
S_3&=&\sum_{q^+<p_1^+}  
\bigg\{
\bigg[{2\over q^+}+{1\over p_1^++p_2^+-q^+}+{1\over p_1^+-q^+}
\bigg]\left(\ln(\delta p_3^2e^{\gamma})
+\ln{q^+\over p_1^++p_2^+}\right)\nonumber\\
&&+\bigg[{1\over q^+}+{2\over p_1^++p_2^+-q^+}+{1\over q^+-p_1^+}
\bigg]\ln{p_1^++p_2^+-q^+\over p_1^++p_2^+}\nonumber\\
&&+\bigg[{2\over p_1^+-q^+}-{1\over p_1^++p_2^+-q^+}+{1\over q^+}
\bigg]\ln{p_1^+-q^+\over p_1^+}\bigg\}\nonumber\\
&&+\sum_{q^+>p_1^+} \bigg\{
\bigg[{1\over q^+}+{2\over p_1^++p_2^+-q^+}+{1\over q^+-p_1^+}
\bigg]\left(\ln(\delta p_3^2e^{\gamma})
+\ln{p_1^++p_2^+-q^+\over p_1^++p_2^+}\right)
\nonumber\\
&&+\bigg[{2\over q^+}+{1\over p_1^++p_2^+-q^+}+{1\over p_1^+-q^+}
\bigg]\ln{q^+\over p_1^++p_2^+}
\nonumber\\ &&
+\bigg[{2\over q^+-p_1^+}+{1\over p_1^++p_2^+-q^+}-{1\over q^+}
\bigg]\ln{q^+-p_1^+\over p_2^+}\bigg\}
\eea
%%%%%%%%%%%%

\section{Four Point Vertex Function}
\subsection{Box Diagrams}
\begin{figure}[ht]
\psfrag{'q'}{$q$}
\psfrag{'k0'}{$k_0$}
\psfrag{'k1'}{$k_1$}
\psfrag{'k2'}{$k_2$}
\psfrag{'k3'}{$k_3$}
\begin{center}
\includegraphics[width=2in]{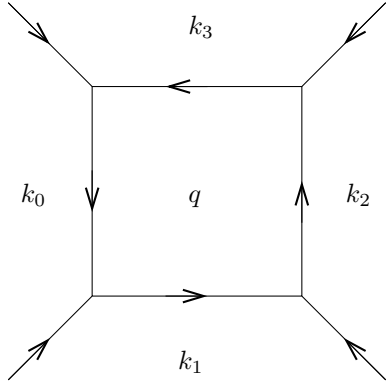}
\caption{A typical box diagram showing the assignment of dual momenta.}
\label{box}
\end{center}
\end{figure}

The simplest spin configuration is all like helicity, for definiteness
take $\wedge\wedge\wedge\wedge$. The box is the only
one-loop 1PIR diagram contributing to this process. Fig.~\ref{box}
shows one of the two diagrams for this
process and the assignment of dual momentum variables to it.
Because their are no divergences for this process, we can immediately
write down the Schwinger representation for it:
\bea
\Gamma^{\wedge\wedge\wedge\wedge}&=&16g^4\sum_{q^+}\int dT_1dT_2dT_3dT_4
\delta(T_{14}q^+-T_{24}p^+_1-T_{34}p^+_2-T_4p_3^+)
\int {d^2 q\over(2\pi)^3}\nonumber\\
&&\exp\bigg\{
-{\delta({\boldsymbol k}_0T_1
+{\boldsymbol k}_1T_2+{\boldsymbol k}_2T_3+{\boldsymbol k}_3T_4)^2\over
T_{14}(T_{14}+\delta)}-(T_{14}+\delta){\bfs q}^2\bigg\}\nonumber\\
&&\exp\left\{
-{T_1T_2p_1^2
+T_1T_3(p_1+p_2)^2+T_1T_4p_4^2
+T_2T_3p_2^2+T_2T_4(p_2+p_3)^2+T_3T_4p_3^2\over T_{14}}
\right\}\nonumber\\
&&(q+K_1)^\wedge(q+K_2)^\wedge(q+K_3)^\wedge(q+K_4)^\wedge\\
&\sim&{2g^4\over\pi^2}\int {dT_1dT_2dT_3dT_4\over T_{14}^6p_1^+p_2^+p_3^+p_4^+}
(T_3K_{12}^\wedge+T_4K_{41}^\wedge)(T_4K_{23}^\wedge+T_1K_{12}^\wedge)
(T_2K_{23}^\wedge+T_1K_{34}^\wedge)
(T_3K_{34}^\wedge+T_2K_{41}^\wedge)\nonumber\\
&&\exp\left\{
-{T_1T_2p_1^2
+T_1T_3(p_1+p_2)^2+T_1T_4p_4^2
+T_2T_3p_2^2+T_2T_4(p_2+p_3)^2+T_3T_4p_3^2\over T_{14}}
\right\}
\eea
Where we have taken $\delta\to0$ and $q^+$ continuous in the
last line. Actually, this box diagram is even finite on shell
so one can safely set $p_i^2=0$ from the beginning as well. 
For completeness we quote the $K_i$ with $\delta>0$
even though for this box it is safe to put $\delta=0$ from the
beginning:
\bea
K_1&=&{T_3K_{12}+T_4K_{41}\over T_{14}p_1^+}-{\delta\over T_{14}(T_{14}
+\delta)}(T_1k_0+T_2k_1+T_3k_2+T_4k_3)\\
K_2&=&{T_4K_{23}+T_1K_{12}\over T_{14}p_2^+}-{\delta\over T_{14}(T_{14}
+\delta)}(T_1k_0+T_2k_1+T_3k_2+T_4k_3)\\
K_3&=&{T_2K_{23}+T_1K_{34}\over T_{14}p_3^+}-{\delta\over T_{14}(T_{14}
+\delta)}(T_1k_0+T_2k_1+T_3k_2+T_4k_3)\\
K_4&=&{T_3K_{34}+T_2K_{41}\over T_{14}p_4^+}-{\delta\over T_{14}(T_{14}
+\delta)}(T_1k_0+T_2k_1+T_3k_2+T_4k_3)
\label{bare++++box}
\eea

The next simplest spin configuration is three like and one unlike helicity,
e.g. $\wedge\wedge\wedge\vee$. There are also other 1PIR diagrams
involving no more than one quartic vertex contributing to this
process. As in the case of all like helicity this process is
free of both IR and UV divergences. However, this will require cancellations
against the reducible diagrams involving triangle sub-graphs.
The box diagrams combine to:
\bea
\Gamma^{\wedge\wedge\wedge\vee}&=&8g^4\sum_{q^+}\int dT_1dT_2dT_3dT_4
\delta(T_{14}q^+-T_{24}p^+_1-T_{34}p^+_2-T_4p_3^+)
\int {d^2 q\over(2\pi)^3}\nonumber\\
&&\exp\bigg\{
-{\delta({\boldsymbol k}_0T_1
+{\boldsymbol k}_1T_2+{\boldsymbol k}_2T_3+{\boldsymbol k}_3T_4)^2\over
T_{14}(T_{14}+\delta)}-(T_{14}+\delta){\bfs q}^2\bigg\}\nonumber\\
&&\exp\left\{
-{T_1T_2p_1^2
+T_1T_3(p_1+p_2)^2+T_1T_4p_4^2
+T_2T_3p_2^2+T_2T_4(p_2+p_3)^2+T_3T_4p_3^2\over T_{14}}
\right\}\nonumber\\
&&\bigg[\left({(q^++p_4^+)^2\over q^{+2}}+{q^{+2}\over(q^++p_4^+)^2}
\right)
(q+K_1)^\wedge(q+K_2)^\wedge(q+K_3)^\wedge(q+K_4)^\vee
\nonumber\\
&&+{p_1^{+2}p_4^{+2}\over q^{+2}(q^+-p_1^+)^2}
(q+K_1)^\vee(q+K_2)^\wedge(q+K_3)^\wedge(q+K_4)^\wedge
\nonumber\\
&&+{p_2^{+2}p_4^{+2}\over (p_1^++p_2^+-q^+)^{2}(q^+-p_1^+)^2}
(q+K_1)^\wedge(q+K_2)^\vee(q+K_3)^\wedge(q+K_4)^\wedge
\nonumber\\
&&+{p_3^{+2}p_4^{+2}\over (p_1^++p_2^+-q^+)^{2}(q^++p_4^+)^2}
(q+K_1)^\wedge(q+K_2)^\wedge(q+K_3)^\vee(q+K_4)^\wedge\bigg]\\
&\sim&{g^4\over\pi^2}\sum_{q^+}\int {dT_1dT_2dT_3dT_4\over T_{14}}
\delta(T_{14}q^+-T_{24}p^+_1-T_{34}p^+_2-T_4p_3^+)\nonumber\\
&&\exp\left\{
-{T_1T_2p_1^2
+T_1T_3(p_1+p_2)^2+T_1T_4p_4^2
+T_2T_3p_2^2+T_2T_4(p_2+p_3)^2+T_3T_4p_3^2\over T_{14}}
\right\}\nonumber\\
&&\bigg[\left({(q^++p_4^+)^2\over q^{+2}}+{q^{+2}\over(q^++p_4^+)^2}
\right)
\left(K_1^\wedge K_2^\wedge K_3^\wedge K_4^\vee
+{K_1^\wedge K_2^\wedge+ K_2^\wedge K_3^\wedge+K_1^\wedge K_3^\wedge
\over2T_{14}}\right)
\nonumber\\
&&+{p_1^{+2}p_4^{+2}\over q^{+2}(q^+-p_1^+)^2}
\left(K_1^\vee K_2^\wedge K_3^\wedge K_4^\wedge+{K_2^\wedge K_3^\wedge+ 
K_3^\wedge K_4^\wedge+K_2^\wedge K_4^\wedge
\over2T_{14}}\right)
\nonumber\\
&&+{p_2^{+2}p_4^{+2}\over (p_1^++p_2^+-q^+)^{2}(q^+-p_1^+)^2}
\left(K_1^\wedge K_2^\vee K_3^\wedge K_4^\wedge+{K_1^\wedge K_3^\wedge
+ K_3^\wedge K_4^\wedge+K_1^\wedge K_4^\wedge
\over2T_{14}}\right)
\nonumber\\
&&+{p_3^{+2}p_4^{+2}\over (p_1^++p_2^+-q^+)^{2}(q^++p_4^+)^2}
\left(K_1^\wedge K_2^\wedge K_3 ^\vee K_4^\wedge+{K_1^\wedge K_2^\wedge+ K_2^\wedge K_4^\wedge+K_1^\wedge K_4^\wedge
\over2T_{14}}\right)\bigg]
\eea
Here we set $\delta\to0$ because of the absence of UV divergences. In 
particular the $K_i$ may also be taken with $\delta=0$. However
we leave $q^+$ discrete, since the $q^+$ divergences will only cancel
after including the triangle sub-diagrams. Indeed the next task is
to extract these divergences and show the cancellation.

The $q^+$ divergences occur at $q^+=0,p_1^+,-p_4^+, p_1^++p_2^+$.
The range of is $0<q^+<p_1^+p_2^+$, so there are two endpoint
singularities and two interior ones. Although the worst divergences seem
to be linear, we know that those will cancel against terms from the
quartic triangle diagrams, leaving at worst
log divergences. We begin by scaling the variables
$T_i=x_iT_{14}$ and integrating over $T_{14}$:
\bea
\Gamma^{\wedge\wedge\wedge\vee}
&\sim&{g^4\over\pi^2}\sum_{q^+}\int_{x_2+x_3+x_4\leq1} {dx_2dx_3dx_4}
\delta(q^+-x_{24}p^+_1-x_{34}p^+_2-x_4p_3^+)\nonumber\\
&&\bigg[\left({(q^++p_4^+)^2\over q^{+2}}+{q^{+2}\over(q^++p_4^+)^2}
\right)
\left({K_1^\wedge K_2^\wedge K_3^\wedge K_4^\vee\over H^2}
+{K_1^\wedge K_2^\wedge+ K_2^\wedge K_3^\wedge+K_1^\wedge K_3^\wedge
\over2H}\right)
\nonumber\\
&&+{p_1^{+2}p_4^{+2}\over q^{+2}(q^+-p_1^+)^2}
\left({K_1^\vee K_2^\wedge K_3^\wedge K_4^\wedge\over H^2}+
{K_2^\wedge K_3^\wedge+ 
K_3^\wedge K_4^\wedge+K_2^\wedge K_4^\wedge
\over2H}\right)
\nonumber\\
&&+{p_2^{+2}p_4^{+2}\over (p_1^++p_2^+-q^+)^{2}(q^+-p_1^+)^2}
\left({K_1^\wedge K_2^\vee K_3^\wedge K_4^\wedge\over H^2}+
{K_1^\wedge K_3^\wedge
+ K_3^\wedge K_4^\wedge+K_1^\wedge K_4^\wedge
\over2H}\right)
\nonumber\\
&&+{p_3^{+2}p_4^{+2}\over (p_1^++p_2^+-q^+)^{2}(q^++p_4^+)^2}
\left({K_1^\wedge K_2^\wedge K_3 ^\vee K_4^\wedge\over H^2}
+{K_1^\wedge K_2^\wedge+ K_2^\wedge K_4^\wedge+K_1^\wedge K_4^\wedge
\over2H}\right)\bigg]\\
\phantom{\bigg]}H&=&(1-x_2-x_3-x_4)[x_2p_1^2+x_3(p_1+p_2)^2+x_4p_4^2]
+x_2x_3p_2^2+x_2x_4(p_2+p_3)^2+x_3x_4p_3^2
\eea
The endpoint singularities are the easiest to analyze because the
delta function drastically shrinks the range of the $x_i$.
For $q^+\to0$, $x_2,x_3,x_4=O(q^+)$, and, holding the $p_i^2\neq0$,
we have 
\bea
H&\sim& x_2p_1^2+x_3(p_1+p_2)^2+x_4p_4^2=O(q^+)\\
K_1,K_4&=&O(q^+),\qquad K_2\sim{K_{12}\over p_2^+},
\qquad K_3\sim{K_{34}\over p_3^+}
\eea
Then it is simple to extract the divergent behavior:
\bea
&&\hskip-.5in{g^4\over\pi^2}{K_{2}K_{3}p_4^{+2}\over q^{+2}}
\int_{x_2+x_3+x_4\leq1} {dx_2dx_3dx_4}
\delta(q^+-x_{24}p^+_1-x_{34}p^+_2-x_4p_3^+){1\over H}\nonumber\\
&\sim&{g^4\over\pi^2}{K_{12}K_{34}p_4^{+}\over q^{+}p_1^+p_2^+p_3^+}
{p_4^+\over(p_1^++p_2^+)p_4^2+p_4^+(p_1+p_2)^2}\nonumber\\
&&\left[{p_1^+(p_1+p_2)^2\over(p_1^++p_2^+)p_1^2-p_1^+(p_1+p_2)^2}
\ln{p_1^+(p_1+p_2)^2\over(p_1^++p_2^+)p_1^2}
+{p_1^+p_4^2\over p_4^+p_1^2+p_1^+p_4^2}
\ln{p_1^+p_4^2\over(p_1^++p_2^++p_3^+)p_1^2}\right]
\eea
A similar analysis applies for the region $q^+\to p_1^++p_2^+$ with
off-shell external legs, which entails
$x_1,x_2,x_4=O(p_1^++p_2^+-q^+)$:
\bea
H&\sim& x_2p_2^2+x_1(p_1+p_2)^2+x_4p_3^2=O(q^+)\\
K_2,K_3&=&O(p_1^++p_2^+-q^+),\qquad K_1\sim{K_{12}\over p_1^+},
\qquad K_4\sim{K_{34}\over p_4^+}
\eea
\bea
&&\hskip-.5in{g^4\over\pi^2}{K_{1}K_{4}p_4^{+2}\over (p_1^++p_2^+-q)^{2}}
\int_{x_2+x_3+x_4\leq1} {dx_2dx_3dx_4}
\delta(q^+-x_{24}p^+_1-x_{34}p^+_2-x_4p_3^+){1\over H}\nonumber\\
&\sim&{g^4\over\pi^2}
{K_{12}K_{34}p_4^{+}\over (p_1^++p_2^+-q^{+})p_1^+p_2^+p_3^+}
{p_3^+\over(p_1^++p_2^+)p_3^2+p_3^+(p_1+p_2)^2}\nonumber\\
&&\left[{p_2^+(p_1+p_2)^2\over(p_1^++p_2^+)p_2^2-p_2^+(p_1+p_2)^2}
\ln{p_2^+(p_1+p_2)^2\over(p_1^++p_2^+)p_2^2}
+{p_2^+p_3^2\over p_3^+p_2^2+p_2^+p_3^2}
\ln{p_2^+p_3^2\over(p_1^++p_2^++p_4^+)p_2^2}\right]
\eea
Notice that the on shell limit of these expressions is ambiguous.
This happened because they are valid only when $q^+\ll p_i^2$.

To obtain the on-shell scattering of glue by glue, we set $p_i^2=0$
before taking the continuum limit of the $q^+$ sum. This 
corresponds to resolving the infra-red divergences, which are
essentially symptoms of degenerate state perturbation theory,
in the presence of the $q^+$ cutoff. This is the correct procedure
if we commit to $q^+$ discretization as in defining the
theory non-perturbatively, because then, in principle, we should
find the exact energy eigenstates of the theory 
with $q^+$ discrete, and only at the
end take the continuum limit. 

Setting all $p_i^2=0$ drastically simplifies $H$
\bea
H\to H_0=(1-x_2-x_3-x_4)x_3(p_1+p_2)^2+x_2x_4(p_2+p_3)^2
\eea
To study the $q^+$ divergences, we must be careful to not drop terms
in $H_0$ that ensure the convergence of the $x_i$ integrals.
For example in the limit $q^+\to0$, even though $x_2, x_3, x_4 =O(q^+)$,
the integration range includes regions where $x_3\ll x_2x_4$,
so we can't simply drop the second term. Also we must remember
that $H_0$ can be much smaller than $O(q^+)$, so we
cannot
neglect $x_2x_4/H_0^2$ compared to $1/H_0$ as we could
when all legs were off-shell. However it is safe to 
simplify the first term of $H_0$:
\bea
H_0&\approx& x_3(p_1+p_2)^2+x_2x_4(p_2+p_3)^2 \qquad {\rm for}~q^+\sim0
\eea  
It is straightforward to evaluate
\bea
\int_{x_2+x_3+x_4\leq1} {dx_2dx_3dx_4}
\delta(q^+-x_{24}p^+_1-x_{34}p^+_2-x_4p_3^+){1\over H_0}
&\sim&{q^+\over p_4^+p_1^+p_{12}^2}
\left[\ln{q^+(p_1^++p_2^+)p_{23}^2\over-p_4^+p_1^+p_{12}^2}-1\right]\\
\int_{x_2+x_3+x_4\leq1} {dx_2dx_3dx_4}
\delta(q^+-x_{24}p^+_1-x_{34}p^+_2-x_4p_3^+){x_2x_4\over H^2_0}
&\sim&{q^+\over-p_4^+p_1^+p_{12}^2p_{23}^2}\\
\int_{x_2+x_3+x_4\leq1} {dx_2dx_3dx_4}
\delta(q^+-x_{24}p^+_1-x_{34}p^+_2-x_4p_3^+){x_3x_i\over H^2_0}
&\sim& O(q^{+2})\qquad {\rm for}~i=2,4
\eea
Thus, inside the integral we can replace
\bea
K_1^\wedge K_4^\vee+K_1^\vee K_4^\wedge&\to& 
{x_2x_4K_{41}^2\over p_1^+p_4^+}=-x_2x_4p_{23}^2
\eea
and putting everything together obtain
\bea
\Gamma^{\wedge\wedge\wedge\vee}_{\square}
&\sim&{g^4\over\pi^2}
{p_4^{+}K_{12}^\wedge K_{34}^\wedge\over p_1^+p_2^+p_3^+p_{12}^2}
\sum_{q^+}{1\over q^+}
\ln{q^+(p_1^++p_2^+)p_{23}^2\over -p_4^+p_1^+p_{12}^2}=
\Gamma^{\wedge\wedge\wedge\vee}_{0,s}{-g^2\over4\pi^2}\sum_{q^+}{1\over q^+}
\ln{q^+(p_1^++p_2^+)p_{23}^2\over -p_4^+p_1^+p_{12}^2}
\eea
where $\Gamma_{0,s}$ is the direct channel tree amplitude for this
scattering process.
The analysis of the divergence near $q^+=p_1^++p_2^+$ analogously
yields
\bea
\Gamma^{\wedge\wedge\wedge\vee}_{\square}
&\sim&{g^4\over\pi^2}
{p_4^{+}K_{12}^\wedge K_{34}^\wedge\over p_1^+p_2^+p_3^+p_{12}^2}
\sum_{q^+}{1\over p_1^++p_2^+-q^+}
\ln{(p_1^++p_2^+-q^+)(p_1^++p_2^+)p_{23}^2\over -p_3^+p_2^+p_{12}^2}\nonumber\\
&\sim&
\Gamma^{\wedge\wedge\wedge\vee}_{0,s}{-g^2\over4\pi^2}\sum_{q^+}
{1\over p_1^++p_2^+-q^+}
\ln{(p_1^++p_2^+-q^+)(p_1^++p_2^+)p_{23}^2\over -p_3^+p_2^+p_{12}^2}
\eea
The divergences at $q^+=p_1^+,-p_4^+$ are near interior points
of the $q^+$ sum. The leading linear divergence at these
points comes from the factors $(q^+-p_1^+)^{-2},(q^++p_4^+)^{-2}$
which are multiplied by factors continuous at the singular point.
Sub-leading logarithmic divergences can arise when these
factors are expanded about the singular points, and
the first order corrections are discontinuous there. First order corrections
that are continuous at the singular point give rise to factors
$(q^+-p_1^+)^{-1},(q^++p_4^+)^{-1}$ multiplying smooth functions of $q^+$
and the continuum limit on $q^+$ is finite. The leading linear divergence will
be exactly canceled by contributions from the quartic triangle diagrams,
so here we want to extract the sub-leading divergence that is left after
this cancellation. To do this we need to find the discontinuity
of the integrand across these singular points (the leading divergence
cancels in this discontinuity since it is even).

For definiteness, focus on the divergence at $p_1^+$. Discontinuities arise 
from integrating the delta function factor over one of the $x_i$, 
which leads to different boundaries of integration for the remaining 
$x_i$ integrals. For example, integrating over $x_3$ yields
\bea
x_3={q^+-x_2p_1^++x_4p_4^+\over p_1^++p_2^+}
\eea
Then the constraints $x_3>0$ and $x_2+x_3+x_4<1$ reduce to the pair
of inequalities
\bea
x_2p_1^+-x_4p_4^+\leq q^+,\qquad x_2p_2^+-x_4p_3^+\leq p_1^++p_2^+-q^+
\eea
which must be simultaneously satisfied. When these equations are
equalities they define two negative slope lines in the $x_2$-$x_4$
plane. If these lines do not intersect in the first quadrant, 
the region of integration (always in the first quadrant) is bounded by 
the one closest to the origin. If they intersect in the
first quadrant then the boundary is determined by the part
of each line closest to the origin. The intersection point of the two lines
is $(x_2^0,x_4^0)$ where
\bea
x_2^0={q^++p_4^+\over p_1^++p_4^+}=1+{q^+-p_1^+\over p_1^++p_4^+}, \qquad
x_4^0={p_1^+-q^+\over p_1^++p_4^+}=1-{q^++p_4^+\over p_1^++p_4^+}
\eea
This point is in the first quadrant if either $-p_4^+<q^+<p_1^+$
or $p_1^+<q^+<-p_4^+$. In either case we see that the character of the boundary
changes as $q^+$ passes through the singular point, causing
a discontinuity in behavior of the $q^+$ summand. For $q^+$ near $p_1^+$,
the intersection point is near $x_2=1$, $x_3,x_4=0$, so the discontinuity
and hence the logarithmic divergence comes from this corner
of the integration region. Then we can approximate $H_0$ by
\bea
H_0&\approx& x_4(p_2+p_3)^2+x_1x_3(p_1+p_2)^2\nonumber\\
&\approx&x_4(p_2+p_3)^2
+\left({q^+-x_2p_1^++x_4p_4^+\over p_1^++p_2^+}\right)
\left({p_1^++p_2^+-q^+-x_2p_2^++x_4p_3^+\over p_1^++p_2^+}\right)
(p_1+p_2)^2\\
K_1,K_2&=&O(q^+-p_1^+),\qquad K_3\approx {K_{23}\over p_3^+},\quad 
K_4\approx {K_{41}\over p_4^+}
\eea
Since the discontinuity is associated with the boundary of
integration it will be useful to write the approximated integrand as
a total derivative. To this end, note that
\bea
{\bfs v}\cdot\nabla
H_0&=&(p_4^+x_4+p_1^+x_2)(p_2+p_3)^2+{(p_2^++p_3^+)(q^+-x_2p_1^++x_4p_4^+)
\over p_1^++p_2^+}(p_1+p_2)^2\nonumber\\
&\approx& p_1^+(p_2+p_3)^2+{(p_2^++p_3^+)(q^+-x_2p_1^++x_4p_4^+)
\over p_1^++p_2^+}(p_1+p_2)^2\nonumber\\
&\to& p_1^+(p_2+p_3)^2+{(p_2^++p_3^+)((1-x_2)p_1^++x_4p_4^+)
\over p_1^++p_2^+}(p_1+p_2)^2
\eea
for $q^+=p_1^+$.
Here ${\bfs v}\cdot\nabla\equiv p_4^+\partial_2+p_1^+\partial_4$,
and ${\bfs v}\cdot\nabla$ applied to the right side of the
last line vanishes. Thus we can write any function of $H_0$ as
a total derivative:
\bea
f^\prime(H_0)={\bfs v}\cdot\nabla\left({f(H_0)\over 
p_1^+(p_2+p_3)^2+{(p_2^++p_3^+)(q^+-x_2p_1^++x_4p_4^+)
}(p_1+p_2)^2/(p_1^++p_2^+)}\right)
\eea
Then we can write the discontinuity of the amplitude across $q^+=p_1^+$
as
\bea
\Gamma^{\wedge\wedge\wedge\vee}
&\sim&{g^4\over\pi^2}\sum_{q^+}
{K_{23}^\wedge K_{41}^\wedge p_4^{+2}\over p_3^+p_4^+(q^+-p_1^+)^2}
{1\over2}{\rm Disc}\int_{\cal R} {dx_2dx_4\over p_1^++p_2^+}
\left({K_1^\vee K_2^\wedge+ K_1^\wedge K_2^\vee\over H_0^2}+
{1\over H_0}\right)\nonumber\\
&\sim&{g^4\over\pi^2}\sum_{q^+}
{K_{23}^\wedge K_{41}^\wedge p_4^{+2}\over p_3^+p_4^+(q^+-p_1^+)^2}
{1\over2}{\rm Disc}\oint_{\partial{\cal R}} {d\ell}{\bfs v}\cdot{\bfs{\hat n}}
\left({\ln{H_0}-K_1\cdot K_2 H_0^{-1}\over
p_1^+(p_1^++p_2^+)(p_2+p_3)^2}\right)\nonumber
\eea
where we have simplified the denominator in the last line by
dropping terms which vanish for $q^+=p_1^+,x_4=0,x_2=1$, since the 
discontinuity only receives contributions near this point.
For definiteness take $p_1^++p_4^+<0$. Then for $q^+<p_1^+$
the line $x_2p_1^+-x_4p_4^+=q^+$ lies closest to the origin
and forms a part of $\partial{\cal R}$. On this part
${\bfs v}\cdot{\bfs{\hat n}}=0$ so it does not contribute.
The axes $x_4=0$ and $x_2=0$ form the rest of the boundary,
but only the neighborhood of the point $x_2=1,x_4=0$ contributes
to the discontinuity. On the $x_4=0$ axis, ${\bfs v}\cdot{\bfs{\hat n}}=-p_1^+$
and $H_0$ reduces to
\bea
H_0\to \left({q^+-x_2p_1^+\over p_1^++p_2^+}\right)
\left({p_1^++p_2^+-q^+-x_2p_2^+\over p_1^++p_2^+}\right)(p_1+p_2)^2 
\eea
Furthermore, in this region $K_1\cdot K_2 H_0^{-1}\to -1$
so the relevant integral is just
\bea
&&\hskip-.25in-p_1^+\int_{1-\epsilon}^{q^+/p_1^+} {dx_2}
{\ln{H_0}+1\over
p_1^+(p_1^++p_2^+)}=\nonumber\\
&&-{q^+\over p_1^+(p_1^++p_2^+)}
\left(1+\ln{(p_1+p_2)^2
\over (p_1^++p_2^+)^2}\right)+{|p_1^+-q^+|\over p_1^+p_2^+}
\left(\ln{(p_1^++p_2^+)|p_1^+-q^+|\over p_1^+}-1\right)+f(1-\epsilon)
\eea 
In contrast, for $q^+>p_1^+$ the contributing
boundary in the neighborhood of the 
point $x_2=1,x_4=0$ includes not only a segment on the $x_4=0$
axis up to $x_2=(p_1^++p_2^+-q^+)/p_2^+$, but also the
segment of the line $x_2p_2^+-x_4p_3^+=p_1^++p_2^+-q^+$ between
the $x_4=0$ axis and its intersection with the line
$x_2p_1^+-x_4p_4^+=q^+$, which then takes over as boundary,
but contributes nothing because  ${\bfs v}\cdot{\bfs{\hat n}}=0$ on it.
The first part contributes
\bea
&&\hskip-.25in-p_1^+\int_{1-\epsilon}^{(p_1^++p_2^+-q^+)/p_2^+} {dx_2}
\left({\ln{H_0}+1\over
p_1^+(p_1^++p_2^+)}\right)=\nonumber\\
&&-{p_1^++p_2^+-q^+\over p_2^+(p_1^++p_2^+)}
\left(1+\ln{(p_1+p_2)^2
\over (p_1^++p_2^+)^2}\right)+{|p_1^+-q^+|\over p_1^+p_2^+}
\left(\ln{(p_1^++p_2^+)|p_1^+-q^+|\over p_2^+}-1\right)+f(1-\epsilon)
\eea 
With a little rearrangement it is straightforward to show that these
apparently different expressions for $q^+<p_1^+$ and $q^+>p_1^+$
can be written in the unified form
\bea
{|p_1^+-q^+|\over 2p_1^+p_2^+}
\left(\ln{(p_1+p_2)^2|p_1^+-q^+|^2\over p_1^+p_2^+}-1\right)+g(q^+)
\eea
where $g$ and its first derivative are continuous at $q^+=p_1^+$.
To this we must add the contribution (when $q^+>p_1^+$) 
of the segment of the line $x_2p_2^+-x_4p_3^+=p_1^++p_2^+-q^+$.
On this line $H_0=x_4(p_2+p_3)^2=(p_1^++p_2^+-q^+-x_2p_2^+)/(-p_3^+)$
$K_1\cdot k_2\approx0$
and $d\ell{\bfs v}\cdot{\bfs{\hat n}}=dx_2(p_1^+-p_2^+p_4^+/p_3^+)$
and the relevant integral is
\bea
{1\over p_1^+}\int_{x_2^0}^{(p_1^++p_2^+-q^+)/p_2^+}dx_2
(p_1^+-p_2^+p_4^+/p_3^+)\ln H_0={(q^+-p_1^+)\over p_1^+p_2^+}
\left(\ln{p_2^++p_3^+\over(q^+-p_1^+)(p_2+p_3)^2}+1\right)
\eea
which contributes only for $q^+>p_1^+$. But we can write the
same singular behavior as a contribution for both $q^+,p_1^+$
and $q^+>p_1^+$ by enclosing $q^+-p_1^+$ in absolute value
signs and multiplying by $1/2$. Then the complete contribution 
to the singular behavior near $q^+=p_1^+$ to the box diagram
after cancellation of the linear divergence with the quartic triangle diagram
can be written:
 \bea
\Gamma^{\wedge\wedge\wedge\vee}_{\square}
&\sim&
\Gamma^{\wedge\wedge\wedge\vee}_{0,t}{-g^2\over4\pi^2}\sum_{q^+\neq p_1^+}
{1\over 2|p_1^+-q^+|}
\ln{|p_1^+-q^+|(p_2^++p_3^+)p_{12}^2\over p_1^+p_2^+p_{23}^2}
\eea
We see that after the dust has settled, even though the analysis is very
different between end and interior points the final result for the
singular behavior is exactly analogous. Thus we can immediately write
the result for the singular behavior near $q^+=-p_4^+$:
\bea
\Gamma^{\wedge\wedge\wedge\vee}_{\square}
&\sim&
\Gamma^{\wedge\wedge\wedge\vee}_{0,t}{-g^2\over4\pi^2}\sum_{q^+\neq -p_4^+}
{1\over 2|q^++p_4^+|}
\ln{|q^++p_4^+|(p_2^++p_3^+)p_{12}^2\over p_3^+p_4^+p_{23}^2}
\eea
 
\subsection{Quartic Triangle Diagrams}
The quartic triangle diagrams are labeled by the pair of legs 
entering the quartic vertex. We first evaluate
\bea
\Gamma_{12}^{\wedge\wedge\wedge\vee}&=&-2g^4\sum_{q^+}\int dT_1dT_3dT_4
\delta((T_1+T_{34})q^+-T_{34}(p^+_1+p^+_2)-T_4p_3^+)
\int {d^2 q\over(2\pi)^3}\nonumber\\
&&\exp\bigg\{
-{\delta({\boldsymbol k}_0T_1
+{\boldsymbol k}_2T_3+{\boldsymbol k}_3T_4)^2\over
T_{14}(T_{14}+\delta)}-(T_{14}+\delta){\bfs q}^2\bigg\}\nonumber\\
&&\exp\left\{
-{T_1T_3(p_1+p_2)^2+T_1T_4p_4^2+T_3T_4p_3^2\over T_{14}}
\right\}\nonumber\\
&&{p_4^{+2}\over q^+(p_1^++p_2^+-q^+)}\left[{(q^++p_1^+)(p_1^++2p_2^+-q^+)
\over(q^+-p_1^+)^2}+1\right](q+K_3)^\wedge(q+K_4)^\wedge\\
&\sim&-{g^4\over4\pi^2}\sum_{q^+}\int {dT_1dT_3dT_4\over T_1+T_{34}}
\delta((T_1+T_{34})q^+-T_{34}(p^+_1+p^+_2)-T_4p_3^+)\nonumber\\
&&\exp\left\{
-{T_1T_3(p_1+p_2)^2+T_1T_4p_4^2+T_3T_4p_3^2\over T_{14}}
\right\}\nonumber\\
&&{p_4^{+2}\over q^+(p_1^++p_2^+-q^+)}\left[{(q^++p_1^+)(p_1^++2p_2^+-q^+)
\over(q^+-p_1^+)^2}+1\right]{T_1T_3K_{34}^{\wedge2}\over p_3^+p_4^+T_{14}^2}
\nonumber\\
&=&-{g^4\over4\pi^2}\sum_{q^+}\int_{x_3+x_4\leq1} {dx_3dx_4}
\delta(q^+-x_{34}(p^+_1+p^+_2)-x_4p_3^+)\nonumber\\
&&{p_4^{+2}\over q^+(p_1^++p_2^+-q^+)}\left[{(q^++p_1^+)(p_1^++2p_2^+-q^+)
\over(q^+-p_1^+)^2}+1\right]{(1-x_3-x_4)x_3K_{34}^{\wedge2}\over p_3^+p_4^+H}
\eea
with
\bea
H=(1-x_3-x_4)x_3(p_1+p_2)^2+(1-x_3-x_4)x_4p_4^2+x_3x_4p_3^2
\eea
Using the delta function to eliminate 
$x_3=(q^++x_4p_4^+)/(p_1^++p_2^+)$, we find that the upper limit
on $x_4$ is $q^+/(-p_4^+)$ if $q^+<-p_4^+$ and $(p_1^++p_2^+-q^+)/(-p_3^+)$
if $q^+>-p_4^+$. Thus the only $q^+$ singularity is at $q^+=p_1^+$.
Putting the external lines on shell
we find
\bea
\Gamma_{12}^{\wedge\wedge\wedge\vee}
&\to&{g^4\over4\pi^2}{K_{34}^{\wedge2}\over p_3^+p_4^+(p_1^++p_2^+)(p_1+p_2)^2}
\bigg\{\sum_{q^+<-p_4^+}{2p_4^{+}\over (p_1^++p_2^+-q^+)}
\left[{p_1^+(p_1^++p_2^+-q^+)+p_2^+q^+
\over(q^+-p_1^+)^2}\right]
\nonumber\\
&&\hskip1.8in + \sum_{q^+>-p_4^+}{2p_4^{+2}\over p_3^+q^+}
\left[{p_1^+(p_1^++p_2^+-q^+)+p_2^+q^+
\over(q^+-p_1^+)^2}\right]
\bigg\}
\eea
The continuum limit of the $q^+$ sum is singular only due to the factors
$(q^+-p_1^+)^{-2}$. The leading linear divergence is necessary to cancel
the corresponding divergence in the box diagram. As long as $p_1^++p_4^+\neq0$
the potential logarithmic sub-divergence is absent because the $q^+$
discretization implies a principle value prescription.
Next
\bea
\Gamma_{23}^{\wedge\wedge\wedge\vee}&=&-2g^4\sum_{q^+}\int dT_1dT_2dT_4
\delta((T_{12}+T_{4})q^+-T_2p^+_1+T_4p_4^+)
\int {d^2 q\over(2\pi)^3}\nonumber\\
&&\exp\bigg\{
-{\delta({\boldsymbol k}_0T_1
+{\boldsymbol k}_1T_2+{\boldsymbol k}_3T_4)^2\over
T_{14}(T_{14}+\delta)}-(T_{14}+\delta){\bfs q}^2\bigg\}\nonumber\\
&&\exp\left\{
-{T_2T_4(p_2+p_3)^2+T_1T_4p_4^2+T_1T_2p_1^2\over T_{14}}
\right\}\nonumber\\
&&{p_4^{+2}\over (q^++p_4^+)(p_1^+-q^+)}
\left[{(p_2^++q^+-p_1^+)(p_3^+-p_4^+-q^+)
\over(q^+-p_1^+-p_2^+)^2}+1\right](q+K_1)^\wedge(q+K_4)^\wedge\\
&\sim&
-{g^4\over4\pi^2}\sum_{q^+}\int_{x_2+x_4\leq1} {dx_2dx_4}
\delta(q^+-x_{2}p^+_1+x_4p_4^+)\nonumber\\
&&{p_4^{+2}\over (q^++p_4^+)(p_1^+-q^+)}
\left[{(p_2^++q^+-p_1^+)(p_3^+-p_4^+-q^+)
\over(q^+-p_1^+-p_2^+)^2}+1\right]{x_2x_4K_{41}^{\wedge2}\over p_1^+p_4^+H}
\eea
with
\bea
H=x_2x_4(p_2+p_3)^2+(1-x_2-x_4)x_4p_4^2+(1-x_2-x_4)x_2p_1^2
\eea
The delta function implies that $q^+<{\rm max}(p_1^+,|p_4^+|)$.
In the case $p_1^++p_4^+>0$, we use the delta function to eliminate 
$x_2=(q^++x_4p_4^+)/p_1^+$, then the lower limit on $x_4$ is 0, and
we find that the upper limit
on $x_4$ is $q^+/(-p_4^+)$ if $q^+<-p_4^+$ and $(p_1^+-q^+)/(p_1^++p_4^+))$
if $p_1^+>q^+>-p_4^+$. Thus the only $q^+$ singularity is at $q^+=-p_4^+$,
since $q^+$ is prevented from approaching $p_1^++p_2^+$. In the other
case, $p_1^++p_4^+<0$, we eliminate $x_4=(q^+-x_2p_1^+)/|p_4^+|$,
and find the upper limit on $x_2$ to be $q^+/p_1^+$ for $q^+<p_1^+$
and $(|p_4^+|-q^+)/|p_1^++p_4^+|$ for $p_1^+<q^+<|p_4^+|$.
Putting the external lines on shell and assuming $p_1^++p_4^+>0$
we find
\bea
\Gamma_{23}^{\wedge\wedge\wedge\vee}
&\to&-{g^4\over4\pi^2}{K_{41}^{\wedge2}\over p_1^+p_4^+(p_2+p_3)^2}
\bigg\{\sum_{q^+<-p_4^+}{-2|p_4^+|q^+\over p_1^+(q^++p_4^+)(p_1^+-q^+)}
\left[{p_2^+(p_4^++q^+)+p_3^+(p_1^+-q^+)
\over(q^+-p_1^+-p_2^+)^2}\right]
\nonumber\\
&&\hskip.8in + 
\sum_{p_1^+>q^+>-p_4^+}{-2p_4^{+2}\over p_1^+(q^++p_4^+)(p_1^++p_4^+)}
\left[{p_2^+(p_4^++q^+)+p_3^+(p_1^+-q^+)
\over(q^+-p_1^+-p_2^+)^2}\right]
\bigg\}
\eea
The continuum limit of the $q^+$ sum is finite because it dictates a principal
value prescription for the singularity at $q^+=-p_4^+$. The same 
conclusion applies to the case $p_1^++p_4^+<0$.

The remaining two cases of this class of diagrams yield lengthier 
expressions since more than one spin flow is possible.
\bea
\Gamma_{34}^{\wedge\wedge\wedge\vee}&=&-2g^4\sum_{q^+}\int dT_1dT_2dT_3
\delta(T_{13}q^+-T_{23}p^+_1-T_3p_2^+)
\int {d^2 q\over(2\pi)^3}\nonumber\\
&&\exp\bigg\{
-{\delta({\boldsymbol k}_0T_1
+{\boldsymbol k}_1T_2+{\boldsymbol k}_2T_3)^2\over
T_{13}(T_{13}+\delta)}-(T_{13}+\delta){\bfs q}^2\bigg\}\nonumber\\
&&\exp\left\{
-{T_1T_3(p_1+p_2)^2+T_1T_2p_1^2+T_2T_3p_3^2\over T_{13}}
\right\}\nonumber\\
&&\bigg[2{q^+\over p_1^++p_2^+-q^+}
\left({q^+p_4^++(p_1^++p_2^+-q^+)p_3^+\over(p_1^++p_2^+)^2}
+{q^+(p_1^++p_2^+-q^+)+p_3^+p_4^+
\over(q^++p_4^+)^2}\right)\nonumber\\
&&-2{(p_1^++p_2^+-q^+)\over q^+}
{(q^+p_3^++(p_1^++p_2^+-q^+)p_4^+)
\over(p_1^++p_2^+)^2}\bigg](q+K_1)^\wedge(q+K_2)^\wedge\\
&\sim&-{g^4\over4\pi^2}\sum_{q^+}\int_{x_2+x_3\leq1} {dx_2dx_3}
\delta(q^+-x_{23}p^+_1-x_3p_2^+)\nonumber\\
&&\bigg[2{q^+\over p_1^++p_2^+-q^+}
\left({q^+p_4^++(p_1^++p_2^+-q^+)p_3^+\over(p_1^++p_2^+)^2}
+{q^+(p_1^++p_2^+-q^+)+p_3^+p_4^+
\over(q^++p_4^+)^2}\right)\nonumber\\
&&-2{(p_1^++p_2^+-q^+)\over q^+}
{(q^+p_3^++(p_1^++p_2^+-q^+)p_4^+)
\over(p_1^++p_2^+)^2}\bigg]
{(1-x_2-x_3)x_3K_{12}^{\wedge2}\over p_1^+p_2^+H}
\eea
with
\bea
H=(1-x_2-x_3)x_3(p_1+p_2)^2+(1-x_2-x_3)x_2p_1^2+x_2x_3p_2^2
\eea
Using the delta function to eliminate $x_3=(q^+-x_2p_1^+)/(p_1^++p_2^+)$,
we find that the upper limit on $x_2$ is $q^+/p_1^+$ for $q^+<p_1^+$
and $(p_1^++p_2^+-q^+)/p_2^+$ for $q^+>p_1^+$. Putting the external
legs on shell then gives
\bea
\Gamma_{34}^{\wedge\wedge\wedge\vee}&\to&
-{g^4\over4\pi^2}{K_{12}^{\wedge2}\over p_1^+p_2^+(p_1^++p_2^+)(p_1+p_2)^2}
\bigg\{\nonumber\\
&&\sum_{q^+<p_1^+}\bigg[2{q^{+2}\over p_1^+(p_1^++p_2^+-q^+)}
\left({q^+p_4^++(p_1^++p_2^+-q^+)p_3^+\over(p_1^++p_2^+)^2}
+{q^+(p_1^++p_2^+-q^+)+p_3^+p_4^+
\over(q^++p_4^+)^2}\right)\nonumber\\
&&-2{(p_1^++p_2^+-q^+)\over p_1^+}
{(q^+p_3^++(p_1^++p_2^+-q^+)p_4^+)
\over(p_1^++p_2^+)^2}\bigg]\nonumber\\
&&\sum_{q^+>p_1^+}\bigg[2{q^+\over p_2^+}
\left({q^+p_4^++(p_1^++p_2^+-q^+)p_3^+\over(p_1^++p_2^+)^2}
+{q^+(p_1^++p_2^+-q^+)+p_3^+p_4^+
\over(q^++p_4^+)^2}\right)\nonumber\\
&&-2{(p_1^++p_2^+-q^+)^2\over p_2^+q^+}
{(q^+p_3^++(p_1^++p_2^+-q^+)p_4^+)
\over(p_1^++p_2^+)^2}\bigg]\bigg\}
\eea
The only $q^+$ divergence here is due to the factor $(q^++p_4^+)^{-2}$
and as in the $12$ case, the leading divergence cancels a corresponding
divergence in the box diagram and there is no sub-leading divergence because
of the principal value prescription.

The final case is 
\bea
\Gamma_{41}^{\wedge\wedge\wedge\vee}&=&-2g^4\sum_{q^+}\int dT_2dT_3dT_4
\delta(T_{24}q^+-T_{24}p^+_1-T_{34}p_2^+-T_4p_3^+)
\int {d^2 q\over(2\pi)^3}\nonumber\\
&&\exp\bigg\{
-{\delta({\boldsymbol k}_1T_2
+{\boldsymbol k}_2T_3+{\boldsymbol k}_3T_4)^2\over
T_{24}(T_{24}+\delta)}-(T_{24}+\delta){\bfs q}^2\bigg\}\nonumber\\
&&\exp\left\{
-{T_2T_4(p_2+p_3)^2+T_2T_3p_2^2+T_3T_4p_3^2\over T_{24}}
\right\}\nonumber\\
&&\bigg[2{q^++p_4^+\over p_1^+-q^+}
\left({(q^+-p_1^+)p_1^+-(q^++p_4^+)p_4^+\over(p_1^++p_4^+)^2}
+{(q^++p_4^+)(p_1^+-q^+)+p_1^+p_4^+
\over q^{+2}}\right)\nonumber\\
&&-2{p_1^+-q^+\over q^++p_4^+}{(q^+-p_1^+)p_4^+-(q^++p_4^+)p_1^+
\over(p_1^++p_4^+)^2}\bigg](q+K_2)^\wedge(q+K_3)^\wedge\\
&\sim&-{g^4\over4\pi^2}\sum_{q^+}\int_{x_2+x_4\leq1} {dx_2dx_4}
\delta(q^+-p^+_1-(1-x_2)p_2^+-x_4p_3^+)\nonumber\\
&&\bigg[2{q^++p_4^+\over p_1^+-q^+}
\left({(q^+-p_1^+)p_1^+-(q^++p_4^+)p_4^+\over(p_1^++p_4^+)^2}
+{(q^++p_4^+)(p_1^+-q^+)+p_1^+p_4^+
\over q^{+2}}\right)\nonumber\\
&&-2{p_1^+-q^+\over q^++p_4^+}{(q^+-p_1^+)p_4^+-(q^++p_4^+)p_1^+
\over(p_1^++p_4^+)^2}\bigg]
{x_2x_4K_{23}^{\wedge2}\over p_2^+p_3^+H}
\eea
with
\bea
H=x_2x_4(p_2+p_3)^2+(1-x_2-x_4)x_2p_2^2+(1-x_2-x_4)x_4p_3^2
\eea
The $x_2,x_4$ integration in this case closely parallels the
procedure in the $23$ case with the substitutions
$q^+\to p_1^++p_2^+-q^+$, $p_1^+\to p_2^+$, $p_4^+\to p_3^+$.
Then the cases $p_2^++p_3^+<0$, $p_2^++p_3^+>0$ are handled
separately. Just as in that case we find here that the
continuum limit of the $q^+$ sum is finite. For the case
$p_{23}^+<0$, we find
\bea
\Gamma_{41}^{\wedge\wedge\wedge\vee}&=&
-{g^4\over4\pi^2}{2K_{23}^{\wedge2}\over p_2^+p_3^+p_{12}^2}
\bigg\{\sum_{-p_4^+<q^+<p_1^+}{q^++p_4^+\over p_{14}^+}
\bigg[{q^++p_4^+\over p_1^+-q^+}\nonumber\\&&
\hskip-.6in\left({(q^+-p_1^+)p_1^+-(q^++p_4^+)p_4^+\over(p_1^++p_4^+)^2}
+{(q^++p_4^+)(p_1^+-q^+)+p_1^+p_4^+
\over q^{+2}}\right)%\nonumber\\&&
-{p_1^+-q^+\over q^++p_4^+}{(q^+-p_1^+)p_4^+-(q^++p_4^+)p_1^+
\over(p_1^++p_4^+)^2}\bigg]\nonumber\\
&&\hskip-.6in+\sum_{q^+>p_1^+}{p_{12}^+-q^+\over p_2^+}%\nonumber\\&&
\bigg[{q^++p_4^+\over p_1^+-q^+}
\left({(q^+-p_1^+)p_1^+-(q^++p_4^+)p_4^+\over(p_1^++p_4^+)^2}
+{(q^++p_4^+)(p_1^+-q^+)+p_1^+p_4^+
\over q^{+2}}\right)\nonumber\\
&&\hskip.5in-{p_1^+-q^+\over q^++p_4^+}{(q^+-p_1^+)p_4^+-(q^++p_4^+)p_1^+
\over(p_1^++p_4^+)^2}\bigg]\bigg\}
\eea
To summarize this section, 
the $q^+$ divergences of the quartic triangle diagrams are
precisely what is needed to remove the linear divergences in the
box diagrams and they contribute 
no sub-leading logarithmic divergences.

\vskip14pt
\noindent\underline{ Acknowledgments}: 
I would like to acknowledge the valuable assistance of my
collaborators, D. Chakrabarti and J. Qiu, who have helped
correct many errors in these notes. 
This research was supported in part by the Department
of Energy under Grant No. DE-FG02-97ER-41029.

\appendix
\section{Some Useful Integrals}
In the evaluation of the on-shell  triangle diagram, we encounter integrals 
of the form
\bea
\int_{x+y\leq1} dx dy\delta(q^+-(x+y)p^+_1-yp^+_2){\cal I}
\eea
where the integrand is a linear function of $x$ times a linear function of
$\ln xy$, $\ln x(1-x-y)$, or $\ln y(1-x-y)$. By $p^+$ conservation,
two of the momenta $p^+_{1,2,3}$ have one sign and the 
third has the opposite sign. In this section we label momenta so that
$p_1^+>0$ and $p_3^+<0$. If $p_2^+$
is positive do the above integral in its displayed form. If $p_2^+$
is negative rewrite the argument of the delta function in terms
of $p_2^+=-|p_2^+|$ and $p_3^+=-|p_3^+|$, and 
rename $x\leftrightarrow y$, which brings the integral to the form
\bea
\int_{x+y\leq1} dx dy
\delta(q^+-(x+y)|p^+_3|-y|p^+_2|){{\cal I}}
\eea
which reduces it to the first form, with $|p^+_3|$ in the
role of $p_1^+$ and $|p_2^+|$ in the role of $p_2^+$.  
Thus, without loss of generality we
can stipulate that $p_1^+,p_2^+>0$.
Then we do the $y$ integral which sets 
$y=(q^+-xp_1^+)/p^+_{12}$, and also sets
the range of the $x$ integral $0<x<x_m$ where $x_m=q^+/p_1^+$ for
$q^+<p_1^+$ and $x_m=(p^+_{12}-q^+)/p_2^+$ for $q^+>p_1^+$.
Then the following $x$ integrals are needed:
\bea
\hskip-.3in \int dx\ \ln(xy)&=&\left(x_m-{q^+\over p_1^+}\right)
\ln{q^+-x_mp_1^+\over p^+_{12}}
+{q^+\over p_1^+}\ln{q^+\over p^+_{12}}+x_m\ln x_m-2x_m\noindent\\
&=&\cases{\displaystyle
{q^+\over p_1^+}\left(\ln{q^+\over p^+_{12}}+\ln{q^+\over p_1^+}
-2\right)
& for $q^+<p_1^+$\cr
\phantom{[}&\cr
\displaystyle
{p^+_{12}(p_1^+-q^+)\over p_1^+p_2^+}
\ln{q^+-p_1^+\over p_2^+}
+{q^+\over p_1^+}\ln{q^+\over p^+_{12}}&\cr
\displaystyle\hskip1in +{p^+_{12}-q^+\over p_2^+}
\left(\ln{p^+_{12}-q^+\over p_2^+} -2\right)
& for $q^+>p_1^+$\cr}\\
\hskip-.3in \int dx\ x\ln(xy)
&=&{1\over2}\left(x_m^2-{q^{+2}\over p_1^{+2}}\right)
\ln{q^+-x_mp_1^+\over p^+_{12}}
+{q^{+2}\over 2p_1^{+2}}\ln{q^+\over p^+_{12}}+{x_m^2\over2}\ln x_m-
{x_m^2p_1^++q^+x_m\over2p_1^+}\nonumber\\
&=&\cases{\displaystyle
{q^{+2}\over 2p_1^{+2}}\left(\ln{q^+\over p^+_{12}}+\ln{q^+\over p_1^+}
-2\right)
& for $q^+<p_1^+$\cr
\phantom{[}&\cr
\displaystyle
{p^+_{12}(p_1^+-q^+)\over 2p_1^+p_2^+}\left[
{q^+\over p_1^+}+{p^+_{12}-q^+\over p_2^+}\right]
\ln{q^+-p_1^+\over p_2^+}
+{q^{+2}\over 2p_1^{+2}}\ln{q^+\over p^+_{12}}\hskip-.5in&\cr
\displaystyle\quad +{(p^+_{12}-q^+)^2\over 2p_2^{+2}}
\left(\ln{p^+_{12}-q^+\over p_2^+} -1\right)
-{q^+(p^+_{12}-q^+)\over 2p_1^+p_2^{+}}
& for $q^+>p_1^+$\cr}
\eea
\bea
\hskip-.3in \int dx\ \ln x(1-x-y)
&=&\cases{\displaystyle
{q^+\over p_1^+}\left(\ln{q^+\over p_1^+}-2\right)
+{p^+_{12}-q^+\over p_2^+}
\ln{p^+_{12}-q^+\over p^+_{12}}&\cr
\displaystyle\hskip1in -\ 
{p^+_{12}(p_1^+-q^+)\over p_1^+p_2^+}
\ln{p_1^+-q^+\over p_1^+}
& for $q^+<p_1^+$\cr
\phantom{[}&\cr
\displaystyle
{p^+_{12}-q^+\over p_2^+}
\left(\ln{p^+_{12}-q^+\over p^+_{12}}
+\ln{p^+_{12}-q^+\over p_2^+} -2\right)
& for $q^+>p_1^+$\cr}\\
&&\phantom{A}\nonumber\\
\hskip-.3in \int dx\ x\ln(x(1-x-y))&=&%\nonumber\\&&
\cases{\displaystyle
{q^{+2}\over 2p_1^{+2}}\left(\ln{q^+\over p_1^+}-1\right)
+{(p^+_{12}-q^+)^2\over 2p_2^{+2}}
\ln{p^+_{12}-q^+\over p^+_{12}}
-{q^+(p^+_{12}-q^+)\over 2p_1^+p_2^{+}}
\hskip-.5in&\cr
\displaystyle\quad 
-{p^+_{12}(p_1^+-q^+)\over 2p_1^+p_2^+}\left[
{q^+\over p_1^+}+{p^+_{12}-q^+\over p_2^+}\right]
\ln{p_1^+-q^+\over p_1^+}
& for $q^+<p_1^+$\cr
\phantom{[}&\cr
\displaystyle
{(p^+_{12}-q^+)^2\over 2p_2^{+2}}
\left(\ln{p^+_{12}-q^+\over p^+_{12}}+
\ln{p^+_{12}-q^+\over p_2^+} -2\right)
& for $q^+>p_1^+$\cr}
\eea
\bea
\hskip-.3in\int dx\ \ln y(1-x-y)&=&\cases{\displaystyle
{q^+\over p_1^+}\left(\ln{q^+\over p^+_{12}}-2\right)
+{p^+_{12}-q^+\over p_2^+}
\ln{p^+_{12}-q^+\over p^+_{12}}&\cr
\displaystyle\hskip1in -\ 
{p^+_{12}(p_1^+-q^+)\over p_1^+p_2^+}
\ln{p_1^+-q^+\over p_1^+}
& for $q^+<p_1^+$\cr
\phantom{[}&\cr
\displaystyle
{q^+\over p_1^+}\ln{q^+\over p^+_{12}}+{p^+_{12}-q^+\over p_2^+}
\left(\ln{p^+_{12}-q^+\over p^+_{12}} -2\right)
&\cr
\displaystyle\hskip1in -\ 
{p^+_{12}(q^+-p_1^+)\over p_1^+p_2^+}
\ln{q^+-p_1^+\over p_2^+}
& for $q^+>p_1^+$\cr}\\
\hskip-.3in \int dx\ x\ln(y(1-x-y))&=&%\nonumber\\&&
\cases{\displaystyle
{q^{+2}\over 2p_1^{+2}}\left(\ln{q^+\over p^+_{12}}-2\right)
+{(p^+_{12}-q^+)^2\over 2p_2^{+2}}
\ln{p^+_{12}-q^+\over p^+_{12}}
-{q^+(p^+_{12}-q^+)\over 2p_1^+p_2^{+}}
\hskip-.5in&\cr
\displaystyle\quad 
-{p^+_{12}(p_1^+-q^+)\over 2p_1^+p_2^+}\left[
{q^+\over p_1^+}+{p^+_{12}-q^+\over p_2^+}\right]
\ln{p_1^+-q^+\over p_1^+}
& for $q^+<p_1^+$\cr
\phantom{[}&\cr
\displaystyle
{q^{+2}\over 2p_1^{+2}}\ln{q^+\over p^+_{12}}
+{(p^+_{12}-q^+)^2\over 2p_2^{+2}}
\left(\ln{p^+_{12}-q^+\over p^+_{12}} -2\right)
-{q^+(p^+_{12}-q^+)\over 2p_1^+p_2^{+}}
\hskip-.5in&\cr
\displaystyle\quad 
-{p^+_{12}(q^+-p_1^+)\over 2p_1^+p_2^+}\left[
{q^+\over p_1^+}+{p^+_{12}-q^+\over p_2^+}\right]
\ln{q^+-p_1^+\over p_2^+}& for $q^+>p_1^+$ \cr}
\eea

\section{Other Spin configurations}
In the text we analyzed the one-loop three gluon vertex with
spin configuration $\wedge\wedge\vee$
in complete detail. In this appendix we briefly summarize the situation for the
other spin configurations. We maintain the choice of two incoming
particles and one outgoing particle, so $p_1^+,p_2^+>0$ and
$p_3^+<0$. 
\subsection{The remaining triangle diagrams}
For the triangle diagram, the other spin configurations are obtained
by modifying the $A_i$ appearing in Eq.~(\ref{baretriangle}) as described
in the attached footnote, where we called them $A^j_i$ where $j$ labels
the leg with the down spin (so $A^3_i\equiv A_i$). It is straightforward
to work out the consequences of these modifications.

First, the surface terms after the integration by parts become:
\bea
&&\hskip-.25in{\rm Surface~Terms}^{\vee\wedge\wedge}\nonumber\\
&&=-{g^3\over4\pi^2}
{p_1^+\over p_2^+p_3^+}K^\wedge\bigg\{{\cal A}(p_1^2,p_1^+)
+{\cal A}(p_2^2,p_2^+)+{\cal A}(p_3^2,-p_3^+)\nonumber\\
&&\hskip-.25in+\sum_{q^+<p_1^+}
\left[{1\over p^+_{12}-q^+}+{2\over q^+}+{1\over p_1^+-q^+}\right]
+\sum_{q^+>p_1^+}
\left[{2\over p^+_{12}-q^+}+{1\over q^+}+{1\over q^+- p_1^+}\right]
\nonumber\\
&&\hskip-.25in+\sum_{q^+<p_1^+}
\left[{1\over p^+_{12}-q^+}+{1\over q^+}\right]\ln{p_1^2(p_{12}^{+})^2
(p_1^{+}-q^+)\over(p_1+p_2)^2p_1^{+2}(p_{12}^{+}-q^+)}
\\
&&\hskip-.25in+\sum_{q^+>p_1^+}
\left[{1\over p^+_{12}-q^+}+{1\over q^+}\right]\ln{p_2^2(p_1^{+}+p_2^+)^2
(q^+-p_1^{+})\over(p_1+p_2)^2p_2^{+2}q^+}
%\nonumber\\&&
-\sum_{q^+<p_1^+}\ln(\delta e^{\gamma+1}H_<)
\left[{2q^{+2}-2q^+p_1^++4p_1^{+2}
\over p_1^{+3}}\right]\bigg\}\nonumber\\
&&\sim-{g^3\over4\pi^2}
{p_1^+\over p_2^+p_3^+}K^\wedge\bigg\{{\cal A}(p_1^2,p_1^+)
+{\cal A}(p_2^2,p_2^+)+{\cal A}(p_3^2,-p_3^+)
%\nonumber\\&&\hskip-.25in
+\sum_{q^+}
\left[{2\over p^+_{12}-q^+}+{2\over q^+}\right]+
\sum_{q^+\neq p_1^+}{1\over |p_1^+-q^+|}\nonumber\\
&&\hskip-.25in+\sum_{q^+\neq p_1^+}
\left[{1\over q^+}\ln{p_1^2(p_1^{+}+p_2^+)
\over(p_1+p_2)^2p_1^{+}}+{1\over p^+_{12}-q^+}\ln{p_2^2(p_1^{+}+p_2^+)
\over(p_1+p_2)^2p_2^{+}}\right]-\ln{(p^+_{12})^2\over p_1^+p_2^+}
\nonumber\\
&&\hskip-.25in+\int_0^{p_1^+}dq^+
\left[{1\over p^+_{12}-q^+}+{1\over q^+}\right]
\ln{(p_1^{+}+p_2^+)(p_1^{+}-q^+)
\over p_1^{+}(p_1^{+}+p_2^+-q^+)}
\\&&\hskip-.25in
+\int_{p_1^+}^{p^+_{12}}dq^+
\left[{1\over p^+_{12}-q^+}+{1\over q^+}\right]\ln{(p_1^{+}+p_2^+)
(q^+-p_1^{+})\over p_2^{+}q^+}
%\nonumber\\&&\hskip-.25in
-\int_0^{1}du\ln(\delta e^{\gamma+1}u(1-u)p_1^2)
\left[2u^{2}-2u+4\right]\bigg\}\nonumber
\label{surfaceterms1}
\eea
\bea
&&\hskip-.25in{\rm Surface~Terms}^{\wedge\vee\wedge}\nonumber\\
&&=-{g^3\over4\pi^2}
{p_2^+\over p_1^+p_3^+}K^\wedge\bigg\{{\cal A}(p_1^2,p_1^+)
+{\cal A}(p_2^2,p_2^+)+{\cal A}(p_3^2,-p_3^+)\nonumber\\
&&\hskip-.25in+\sum_{q^+<p_1^+}
\left[{1\over p^+_{12}-q^+}+{2\over q^+}+{1\over p_1^+-q^+}\right]
+\sum_{q^+>p_1^+}
\left[{2\over p^+_{12}-q^+}+{1\over q^+}+{1\over q^+- p_1^+}\right]
\nonumber\\
&&\hskip-.25in+\sum_{q^+<p_1^+}
\left[{1\over p^+_{12}-q^+}+{1\over q^+}\right]\ln{p_1^2(p_1^{+}+p_2^+)^2
(p_1^{+}-q^+)\over(p_1+p_2)^2p_1^{+2}(p_1^{+}+p_2^+-q^+)}
\\
&&\hskip-.25in+\sum_{q^+>p_1^+}
\left[{1\over p^+_{12}-q^+}+{1\over q^+}\right]\ln{p_2^2(p_1^{+}+p_2^+)^2
(q^+-p_1^{+})\over(p_1+p_2)^2p_2^{+2}q^+}
\nonumber\\&&
-\sum_{q^+>p_1^+}\ln(\delta e^{\gamma+1}H_>)
\left[{2(p^+_{12}-q^{+})^2-2(p^+_{12}-q^+)p_2^++4p_2^{+2}
\over p_2^{+3}}\right]\bigg\}\nonumber\\
&&\sim-{g^3\over4\pi^2}
{p_2^+\over p_1^+p_3^+}K^\wedge\bigg\{{\cal A}(p_1^2,p_1^+)
+{\cal A}(p_2^2,p_2^+)+{\cal A}(p_3^2,-p_3^+)
%\nonumber\\&&\hskip-.25in
+\sum_{q^+}
\left[{2\over p^+_{12}-q^+}+{2\over q^+}\right]+
\sum_{q^+\neq p_1^+}{1\over |p_1^+-q^+|}\nonumber\\
&&\hskip-.25in+\sum_{q^+\neq p_1^+}
\left[{1\over q^+}\ln{p_1^2(p_1^{+}+p_2^+)
\over(p_1+p_2)^2p_1^{+}}+{1\over p^+_{12}-q^+}\ln{p_2^2(p_1^{+}+p_2^+)
\over(p_1+p_2)^2p_2^{+}}\right]-\ln{(p^+_{12})^2\over p_1^+p_2^+}
\nonumber\\
&&\hskip-.25in+\int_0^{p_1^+}dq^+
\left[{1\over p^+_{12}-q^+}+{1\over q^+}\right]
\ln{(p_1^{+}+p_2^+)(p_1^{+}-q^+)
\over p_1^{+}(p_1^{+}+p_2^+-q^+)}
\\&&\hskip-.25in
+\int_{p_1^+}^{p^+_{12}}dq^+
\left[{1\over p^+_{12}-q^+}+{1\over q^+}\right]\ln{(p_1^{+}+p_2^+)
(q^+-p_1^{+})\over p_2^{+}q^+}
%\nonumber\\&&\hskip-.25in
-\int_0^{1}du\ln(\delta e^{\gamma+1}u(1-u)p_2^2)
\left[2u^{2}-2u+4\right]\bigg\}\nonumber
\label{surfaceterms2}
\eea
Note that the divergent factor multiplying the tree vertex are
spin independent.

Next we quote the analogs of Eq.~(\ref{gammatriangle-}) for the other spin
configurations:
\bea
\Gamma_{\triangle-}^{\vee\wedge\wedge}&=&
{\rm Surface~Terms}-{g^3\over4\pi^2}
{p_1^+\over p_2^+p_3^+}K^\wedge
\sum_{q^+}\int_0^{x_{max}} {dx\over p^+_{12}}\nonumber\\
&&\bigg\{
-\left[2+{p_2^+\over q^+}
+{p_1^+\over (p^+_{12}-q^+)}\right]I_1
%\nonumber\\&&
+{1\over q^{+2}}\left[{q^{+4}\over p_1^{+2}(q^+-p^+_1)^{2}}
+1+{(p^+_1-q^+)^2\over p_1^{+2}}+{2q^+\over p^+_{12}}\right]I_3\nonumber\\&&
+{1\over(p^+_{12}-q^+)^2}\left[{p^{+2}_2\over (q^+-p^+_1)^{2}}+1
+{2(p^+_{12}-q^{+})\over p^+_{12}}\right]I_2\bigg\}\\
\Gamma_{\triangle-}^{\wedge\vee\wedge}&=&
{\rm Surface~Terms}-{g^3\over4\pi^2}
{p_2^+\over p_3^+p_1^+}K^\wedge
\sum_{q^+}\int_0^{x_{max}} {dx\over p^+_{12}}\nonumber\\
&&\bigg\{
-\left[2+{p_2^+\over q^+}
+{p_1^+\over (p^+_{12}-q^+)}\right]I_1
%\nonumber\\&&
+{1\over q^{+2}}\left[{p_1^{+2}\over (q^+-p^+_1)^{2}}
+1+{2q^+\over p^+_{12}}\right]I_3\nonumber\\&&
+{1\over(p^+_{12}-q^+)^2}\left[{(p^+_{12}+q^+)^4
\over p^{+2}_2(q^+-p^+_1)^{2}}+{(q^+-p_1^+)^2\over p_2^{+2}}+1
+{2(p^+_{12}-q^{+})\over p^+_{12}}\right]I_2\bigg\}
\label{gammatriangle-1}
\eea 
The extraction of the $q^+$ divergences for these spin configurations 
parallels the discussion in the text. The $I_1$ term does not contain
$q^+$ divergences as in the text. Again, as in the text, 
the worst (linear) divergences are for $q^+\sim p_1^+$ and arise
only from the ``First Terms'' of $I_{2,3}=xp_1^+p_2^+ +{\hat I}_{2,3}$.
This linear divergence is canceled by a term in the corresponding
swordfish diagram as in the text:
\bea
({\rm First Term})^{\vee\wedge\wedge}&=&-{g^3\over8\pi^2}
{p_1^+\over p_2^+p_3^+}{K^\wedge\over p^+_{12}}\bigg\{
\sum_{q^+<p_1^+}
{p_2^+\over p_1^{+}}\bigg(\left[{q^{+4}\over p_1^{+2}(q^+-p^+_1)^{2}}
+1+{(p^+_1-q^+)^2\over p_1^{+2}}+{2q^+\over p^+_{12}}\right]\nonumber\\
&&
+{q^{+2}\over (p^+_{12}-q^+)^2}
\left[{p^{+2}_2\over (q^+-p^+_1)^{2}}+1
+{2(p^+_{12}-q^{+})\over p^+_{12}}\right]\bigg)\nonumber\\
&&
+\sum_{q^+>p_1^+}
{p_1^+\over p_2^+}\bigg({(p^+_{12}-q^+)^2\over q^{+2}}
\left[{q^{+4}\over p_1^{+2}(q^+-p^+_1)^{2}}
+1+{(p^+_1-q^+)^2\over p_1^{+2}}+{2q^+\over p^+_{12}}\right]\nonumber\\
&&
+\left[{p^{+2}_2\over (q^+-p^+_1)^{2}}+1
+{2(p^+_{12}-q^{+})\over p^+_{12}}\right]\bigg)\bigg\}\\
({\rm First Term})^{\wedge\vee\wedge}&=&-{g^3\over8\pi^2}
{p_2^+\over p_3^+p_1^+}{K^\wedge\over p^+_{12}}
\bigg\{\sum_{q^+<p_1^+}
{p_2^+\over p_1^+}\bigg(\left[{p_1^{+2}\over (q^+-p^+_1)^{2}}
+1+{2q^+\over p^+_{12}}\right]\nonumber\\&&
+{q^{+2}\over(p^+_{12}-q^+)^2}\left[{(p^+_{12}+q^+)^4
\over p^{+2}_2(q^+-p^+_1)^{2}}+{(q^+-p_1^+)^2\over p_2^{+2}}+1
+{2(p^+_{12}-q^{+})\over p^+_{12}}\right]\bigg)\nonumber\\
&&+\sum_{q^+>p_1^+}
{p_1^+\over p_2^+}\bigg({(p^+_{12}-q^+)^2\over q^{+2}}
\left[{p_1^{+2}\over (q^+-p^+_1)^{2}}
+1+{2q^+\over p^+_{12}}\right]\nonumber\\&&
+\left[{(p^+_{12}+q^+)^4
\over p^{+2}_2(q^+-p^+_1)^{2}}+{(q^+-p_1^+)^2\over p_2^{+2}}+1
+{2(p^+_{12}-q^{+})\over p^+_{12}}\right]\bigg)\bigg\}
\eea

\subsection{The remaining swordfish diagrams}
We list here the amplitudes for the other spin configurations of 
the swordfish diagrams.
\bea
\Gamma^{\wedge\vee\wedge}_{\rm SF3}&=&-{g^3\over8\pi^2}\sum_{q^+}
{1\over p^+_{12}}\bigg\{{p^+_{12}-q^+\over q^+}\left[
-3-{(p_1^++q^+)(q^+-p_1^+-2p_2^+)\over(p_1^+-q^+)^2}\right]\nonumber\\
&&\hskip-.25in+\left[{p^+_{12}-q^+\over q^+}
+{ q^+\over p^+_{12}-q^+}\right]
\left[1-{(p^+_{12}-2q^+)(p_2^+-p_1^+)
\over(p^+_{12})^2}\right]\bigg\}(x_3k_2+(1-x_3)k_0)^\wedge
\nonumber\\
&=&-{g^3\over8\pi^2}{1\over p^+_{12}}\sum_{q^+}
\bigg\{{y_3\over 1-y_3}\left[
-3-{(y_3+1-\eta)(y_3-1-\eta)\over(\eta-1+y_3)^2}\right]\nonumber\\
&&\hskip.25in+\left[{1-y_3\over y_3}
+{ y_3\over 1-y_3}\right]
\left[1-{(2y_3-1)(1-2\eta)}
\right]\bigg\}(y_3k_0+(1-y_3)k_2)^\wedge\\
\phantom{\int}\nonumber\\
\Gamma_{SF1}^{\wedge\vee\wedge}&=&-{g^3\over8\pi^2}{1\over p_1^+}
\sum_{q^+<p_1^+}
\bigg\{{1-x_1\over x_1}\left[
-3-{(p^+_{12}+q^+)(q^+-p^+_{12})\over(p^+_{12}-q^+)^2}\right]
\nonumber\\&&
+\left[{1-x_1\over x_1}+{x_1\over1-x_1}\right]\left[
1-{(2q^+-p_1^+)(p_1^++2p_2^+)\over p_1^{+2}}\right]\bigg\}
(x_1k_1+(1-x_1)k_0)^\wedge\nonumber\\
&=&-{g^3\over8\pi^2}{1\over p_1^+}
\sum_{q^+<p_1^+}
\bigg\{{y_1\over 1-y_1}\left[
-3-{(y_1+1-\eta^{-1})(y_1-1-\eta^{-1})\over(\eta^{-1}-1+y_1)^2}\right]
\nonumber\\&&
+\left[{1-y_1\over y_1}+{y_1\over1-y_1}\right]\left[
1-{(2y_1-1)(1-2\eta^{-1})}\right]\bigg\}
(y_1k_0+(1-y_1)k)^\wedge
\eea
\bea
\Gamma_{SF2}^{\wedge\vee\wedge}&=&\hskip-6pt
-{g^3\over8\pi^2}\sum_{q^+>p_1^+}
{p_2^+\over(q^+-p_1^+)(p^+_{12}-q^+)}\left[
1-{(q^+-2p_1^+-2p_2^+)(q^+-2p_1^+)\over q^{+2}}\right]
(x_2k_1+(1-x_2)k_2)^\wedge\nonumber\\
&=&\hskip-6pt-{g^3\over8\pi^2}{1\over p_2^+}\sum_{q^+>p_1^+}
{1\over x_2(1-x_2)}\left[
1-{(x_2+(1-\eta)^{-1})(x_2+(1-\eta)^{-1}-2)\over (x_2-(1-\eta)^{-1})^2}\right]
(x_2k_1+(1-x_2)k_2)^\wedge
\\
\phantom{\int}\nonumber\\
\Gamma^{\vee\wedge\wedge}_{\rm SF3}&=&
-{g^3\over8\pi^2}{1\over p^+_{12}}\sum_{q^+}
\bigg\{{ q^+\over p^+_{12}-q^+}\left[
-3-{(p_1^++q^+)(q^+-p_1^+-2p_2^+)\over(p_1^+-q^+)^2}\right]\nonumber\\
&&\hskip-.25in+\left[{p^+_{12}-q^+\over q^+}
+{ q^+\over p^+_{12}-q^+}\right]
\left[1-{(p^+_{12}-2q^+)(p_2^+-p_1^+)
\over(p^+_{12})^2}\right]\bigg\}(x_3k_2+(1-x_3)k_0)^\wedge
\nonumber\\
&=&-{g^3\over8\pi^2}{1\over p^+_{12}}\sum_{q^+}
\bigg\{{ x_3\over 1-x_3}\left[
-3-{(x_3+\eta)(x_3+\eta-2)\over(\eta-x_3)^2}\right]\nonumber\\
&&\hskip-.25in+\left[{ x_3\over 1-x_3}+{1- x_3\over x_3}
+\right]
\left[1-{(2x_3-1)(2\eta-1)}\right]\bigg\}
(x_3k_2+(1-x_3)k_0)^\wedge
\\%\eea
\phantom{\int}\nonumber\\
\Gamma_{SF1}^{\vee\wedge\wedge}&=&
-{g^3\over8\pi^2}{1\over p_1^+}\sum_{q^+<p_1^+}
{1\over x_1(1-x_1)}\left[
1-{(p^+_{12}+q^+)(q^+-p^+_{12})\over(p^+_{12}-q^+)^2}\right]
(x_1k_1+(1-x_1)k_0)^\wedge\nonumber\\
&=&-{g^3\over8\pi^2}{1\over p_1^+}\sum_{q^+<p_1^+}
{1\over x_1(1-x_1)}\left[
1-{(x_1+\eta^{-1})(x_1+\eta^{-1}-2)\over(\eta^{-1}-x_1)^2}\right]
(x_1k_1+(1-x_1)k_0)^\wedge
\\
\phantom{\int}\nonumber\\
\Gamma_{SF2}^{\vee\wedge\wedge}&=&-{g^3\over8\pi^2}\sum_{q^+>p_1^+}
{1\over p_2^+}\bigg\{{q^+-p_1^+\over p^+_{12}-q^+}\left[
-3-{(q^+-2p_1^+-2p_2^+)(q^+-2p_1^+)\over q^{+2}}\right]\nonumber\\
&&\hskip-.25in +\left[{q^+-p_1^+\over p^+_{12}-q^+}
+{p^+_{12}-q^+\over q^+-p_1^+}\right]\left[1
-{(2p^+_{12}-2q^+)(2p^+_{12})\over p_2^{+2}}
\right]\bigg\}
(x_2k_1+(1-x_2)k_2)^\wedge\nonumber\\
&=&-{g^3\over8\pi^2}{1\over p_2^+}\sum_{q^+>p_1^+}
\bigg\{{y_2\over1- y_2}\left[
-3-{(y_2+1-(1-\eta)^{-1})(y_2-1-(1-\eta)^{-1})\over 
(y_2-1+(1-\eta)^{-1})^2}\right]\nonumber\\
&&\hskip-.25in +\left[{1-y_2\over y_2}+{y_2\over 1-y_2}\right]\left[1
-{(2y_2-1)(1-2(1-\eta)^{-1})}
\right]\bigg\}
(y_2k_2+(1-y_2)k)^\wedge
\eea
In these formulas the $x_i$'s are defined as before and the $y_i\equiv 1-x_i$.

Notice the different spin cases involve summands that can be
obtained from one another through simple substitution rules. In terms of 
the rescaled variables $z=x_i$ or $y_i$ we only encounter two forms:
\bea
F_1(z,\xi)&=&{1\over z(1-z)}\left[
1-{(z+\xi)(z+\xi-2)\over(\xi-z)^2}\right]
(zk_1+(1-z)k_2)^\wedge\\
F_2(z,\xi)&=&\bigg\{{ z\over 1-z}\left[
-3-{(z+\xi)(z+\xi-2)\over(\xi-z)^2}\right]\nonumber\\&&\hskip+.5in
+\left[{ z\over 1-z}+{1- z\over z}
\right]
\left[1-{(2z-1)(2\xi-1)}\right]\bigg\}(zk_1+(1-z)k_2)^\wedge
\eea
where $k_1,k_2$ are any pair of $k,k_2,k_0$
and $\xi$ is one of the variables $\eta,1-\eta,\eta^{-1},1-\eta^{-1},
(1-\eta)^{-1},1-(1-\eta)^{-1}$. It is convenient to decompose
each of these expressions into partial fractions:
\bea
F_1&=&{2k_1\over(1-z)(1-\xi)}+{2k_2\over z\xi}+4{\xi k_1+(1-\xi)k_2
\over(z-\xi)^2}+2{\xi k_1-(1-\xi)k_2\over(z-\xi)\xi(1-\xi)}\\
F_2&=&{2\xi^2k_1\over(1-z)(1-\xi)}+{2\xi k_2\over z}+4\xi^2{\xi k_1+(1-\xi)k_2
\over(z-\xi)^2}+2\xi{\xi(5-4\xi)k_1+(1-\xi)(3-4\xi)k_2\over(z-\xi)(1-\xi)}
+{\cal P}(z)\quad
\eea
where ${\cal P}$ is a quadratic polynomial in $z$. These decompositions
are useful in the evaluations of the singular contributions
discussed in the text.
%%%%%%%%%%
\section{Scalar Box}
The one-loop scalar box diagram is given by
\bea
\int {d^4q\over(2\pi)^4}{1\over(q-k_0)^2(q-k_1)^2(q-k_2)^2(q-k_3)^2}
\eea
It has infra-red divergences when two or more
neighboring external lines are on-shell
$(k_{i+1}-k_i)^2=0$, when the propagator connecting two on-shell
vertices carries zero momentum, causing the integrand to
behave as
\bea
{1\over(q-k_i)^2}{1\over2(q-k_i)\cdot(k_{i+1}-k_i)}
{1\over2(q-k_i)\cdot(k_{i-1}-k_{i})}
\eea
which causes a logarithmic divergence in 4 dimensions.
The discretization of  $p^+$ in the context of light-cone quantization
doesn't regulate these divergences. To see this, go to the Galilei
frame where one of the on-shell particles has ${\bfs p}=0$. Then
the two propagators hooked to this vertex become
\bea
{1\over{\bfs q}^2-2q^+q^-}{1\over{\bfs q}^2-2(q^+-p_1^+)q^-}
\eea 
which show a log divergence near $({\bfs q},q^-)=0$.
If the massless particles were instead gluons, this vertex would
supply an additional factor of ${\bfs q}$ and the divergence  
would be regulated by discretizing $p^+$.

The previous paragraph shows that the box infra-red divergences
are identical to certain triangle infra-red divergences. We can
use this fact to define a regulated scalar box integral by subtracting from
the box integrand a sum of triangle integrands times the IR 
limit of the fourth (non-diverging propagator). This amounts to
supplying a numerator factor
\bea
{\cal N} =1 -{(q-k_0)^2+(q-k_2)^2\over(p_1+p_2)^2}-
{(q-k_1)^2+(q-k_3)^2\over(p_1+p_4)^2}
\eea
Introducing Schwinger parameters and completing the square 
in the exponential leads to
\bea
\Gamma_{4,S}&=&\int_0^\infty dT_1dT_2dT_3dT_4\int {d^4q\over(2\pi)^4}{\cal N}
\exp\left\{{-T_{14}(q-K)^2-{T_1T_3(p_1+p_2)^2
+T_2T_4(p_1+p_4)^2\over T_{14}}}\right\}\\
{\cal N}&=&2{p_{12}^2+p_{14}^2\over p_{12}^2p_{14}^2}
\left[{T_1T_3p_{12}^2+T_2T_4p_{14}^2\over T_{14}^2}-(q-K)^2\right]
\eea
where $K=(T_2p_1+T_3(p_1+p_2)-T_4p_4)/T_{14}$. Changing variables
to $x_i=T_i/T_{14}, T=T_{14}, \sum x_i=1$, and integrating $q,T$
yields
\bea
\Gamma_{4,S}&=&-{p_{12}^2+p_{14}^2\over 8\pi^2p_{12}^2p_{14}^2}
\int_{x_2+x_3+x_4\leq1} dx_2dx_3dx_4
{1\over x_1x_3(p_1+p_2)^2+x_2x_4(p_1+p_4)^2}\nonumber\\
&=&-{p_{12}^2+p_{14}^2\over 8\pi^2p_{12}^2p_{14}^2}\int_0^\infty dt
{1\over (1+t)(p_{12}^2t-p_{14}^2)}\ln{p_{12}^2t\over p_{14}^2}
=-{1\over 16\pi^2p_{12}^2p_{14}^2}
\left[\ln^2{p_{12}^2\over p_{14}^2}+{\pi^2}\right]
\eea
\section{More Box Integrals}
We list here some integrals over Feynman parameters that
arise in the box diagrams. We use the shorthand notation
$d^3x=\prod_{i=1}^4 dx_i \delta(1-\sum_i x_i)$.
\bea
L &=&\int d^3x \ln(x_1x_3A+x_2x_4B)=-{11\over18}+{B\ln B+A\ln A\over
6(A+B)}+{AB\over 12(A+B)^2}\left(\pi^2+\ln^2{A\over B}\right)\\
L_1 &=& \int d^3x {1\over x_1x_3A+x_2x_4B}=
{1\over 2(A+B)}\left(\pi^2+\ln^2{A\over B}\right)\\
L_{1A}&=&\int d^3x {(x_1,x_3)\over x_1x_3A+x_2x_4B}={\ln(A/B)\over2(A+B)} 
+{B\over 4(A+B)^2}\left(\pi^2+\ln^2{A\over B}\right)\\
L_{1B}&=&\int d^3x {(x_2,x_4)\over x_1x_3A+x_2x_4B}={\ln(B/A)\over2(A+B)} 
+{A\over 4(A+B)^2}\left(\pi^2+\ln^2{A\over B}\right)\\
L_A&=&\int d^3x {x_1x_3\over x_1x_3A+x_2x_4B}= {1\over6(A+B)}+{B\ln(A/B)\over
3(A+B)^2}+{B(B-A)\over 12(A+B)^3}\left(\pi^2+\ln^2{A\over B}\right)\\
L_B&=&\int d^3x {x_2x_4\over x_1x_3A+x_2x_4B}= {1\over6(A+B)}+{A\ln(B/A)\over
3(A+B)^2}+{A(A-B)\over 12(A+B)^3}\left(\pi^2+\ln^2{A\over B}\right)\\
L_C &=& \int d^3x {(x_1x_2,x_2x_3,x_3x_4,x_4x_1)
\over x_1x_3A+x_2x_4B}\nonumber\\
&=& -{1\over6(A+B)}+{(A-B)\ln(A/B)\over
6(A+B)^2}+{AB\over 6(A+B)^3}\left(\pi^2+\ln^2{A\over B}\right)\\
L_{2B}&=&\int d^3x {(x_2^2,x_4^2)\over x_1x_3A+x_2x_4B}
= {1-\ln(A/B)\over6(A+B)}-{A\ln(A/B)\over
3(A+B)^2}+{A^2\over 6(A+B)^3}\left(\pi^2+\ln^2{A\over B}\right)\\
L_{2A} &=& \int d^3x {(x_1^2,x_3^2)\over x_1x_3A+x_2x_4B}
= {1-\ln(B/A)\over6(A+B)}
-{B\ln(B/A)\over
3(A+B)^2}+{B^2\over 6(A+B)^3}\left(\pi^2+\ln^2{A\over B}\right)\\
L_{AB}&=&\int d^3x {x_1x_2x_3x_4\over (x_1x_3A+x_2x_4B)^2}\nonumber\\
&=&{1\over2(A+B)^2}+{(B-A)\ln(A/B)\over
2(A+B)^3}+{A^2+B^2-4AB\over 12(A+B)^4}\left(\pi^2+\ln^2{A\over B}\right)\\
L_{AA}&=&\int d^3x {x_1^2x_3^2\over (x_1x_3A+x_2x_4B)^2}\nonumber\\
&=&
{A-2B\over6A(A+B)^2}+{B(5A-B)\ln(A/B)\over
6A(A+B)^3}+{B(2B-A)\over 6(A+B)^4}\left(\pi^2+\ln^2{A\over B}\right)\\
L_{CA}&=&\int d^3x {(x_1x_2,x_2x_3,x_3x_4,x_4x_1)
x_1x_3\over (x_1x_3A+x_2x_4B)^2}\nonumber\\
&=& 
{B-2A\over6A(A+B)^2}-{(5B-A)\ln(A/B)\over
6(A+B)^3}+{B(2A-B)\over 6(A+B)^4}\left(\pi^2+\ln^2{A\over B}\right)\\
L_{2AB}&=&\int d^3x {(x_1^2,x_3^2)x_2x_4\over (x_1x_3A+x_2x_4B)^2}\nonumber\\
&=& 
{A+4B\over6B(A+B)^2}+{(A-5B)\ln(B/A)\over
6(A+B)^3}+{B(B-2A)\over 6(A+B)^4}\left(\pi^2+\ln^2{A\over B}\right)
\eea

\end{document}